\documentclass[amsmath,amssymb,pra,aps,showpacs,superscriptaddress,twocolumn]{revtex4-1}
\pdfoutput=1
\usepackage{amsmath,amsfonts,amssymb,amsthm,graphics,graphicx,epsfig,bbm}
\usepackage[colorlinks=true,citecolor=blue,linkcolor=blue,urlcolor=blue]{hyperref}
\usepackage[usenames]{color}
\usepackage{graphicx}
\usepackage{subfigure}
\usepackage{amsmath}
\usepackage{epsfig}
\usepackage{dcolumn}
\usepackage{bm}
\usepackage{color}
\usepackage{epstopdf}
\usepackage{amssymb}
\usepackage{amstext}
\usepackage{latexsym}
\usepackage{hyperref}
\usepackage{amsfonts}
\usepackage{psfrag}
\usepackage{xcolor}
\usepackage[normalem]{ulem}
\usepackage{dsfont}
\usepackage{txfonts}
\usepackage{cases}

\graphicspath{{./Images/}}

\newcommand{\ket}[1]{\ensuremath{\left\vert #1 \right\rangle}}

\newcommand{\an}[2]{\ensuremath{\hat{#1}^{\protect\phantom{\dagger}}_{#2}}}
\newcommand{\cn}[2]{\ensuremath{\hat{#1}^\dagger_{#2}}}
\newcommand{\nn}[2]{\ensuremath{\hat{n}^{#1}_{#2}}}

\newcommand{\expU}[1]{\ensuremath{e^{#1}}}
\newcommand{\abs}[1]{\left|#1\right|}
\newcommand{\pdif}[2]{\ensuremath{\frac{\partial#1}{\partial#2}}}

\newcommand{\subfigimg}[3][,]{%
	\setbox1=\hbox{\includegraphics[#1]{#3}}% Store image in box
	\leavevmode\rlap{\usebox1}% Print image
	\rlap{\hspace*{2pt}\raisebox{\dimexpr\ht1-0.5\baselineskip}{{\bfseries \large\textsf{#2}}}}% Print label
	\phantom{\usebox1}% Insert appropriate spcing
}

\newcommand{\subfigimgraised}[3][,]{%
	\setbox1=\hbox{\includegraphics[#1]{#3}}% Store image in box
	\leavevmode\rlap{\usebox1}% Print image
	\rlap{\hspace*{2pt}\raisebox{\dimexpr\ht1-0.\baselineskip}{{\bfseries \large\textsf{#2}}}}% Print label
	\phantom{\usebox1}% Insert appropriate spcing
}

\newcommand{\rev}[1]{#1}
\newcommand\numberthis{\addtocounter{equation}{1}\tag{\theequation}}

\newcommand{\idg}[1]{{\bfseries #1)}}

\newcommand{\densdescr}[4]{The propagation of the density excitation in time. The upper curves show the transmitted density wave into the drain lead (integrated between site \ensuremath{#1} and \ensuremath{#2}), and the lower curves the incoming and reflected wave in the source lead (\ensuremath{#3} and \ensuremath{#4}). The background density is subtracted.}

\begin{document}
	
\title{Andreev-reflection and Aharonov-Bohm dynamics in atomtronic circuits}

\author{Tobias Haug}
\affiliation{Centre for Quantum Technologies, National University of Singapore,
3 Science Drive 2, Singapore 117543, Singapore}
\author{Rainer Dumke}
\affiliation{Centre for Quantum Technologies, National University of Singapore, 3 Science Drive 2, Singapore 117543, Singapore}
\affiliation{Division of Physics and Applied Physics, Nanyang Technological University, 21 Nanyang Link, Singapore 637371, Singapore}
\affiliation{MajuLab, CNRS-UNS-NUS-NTU International Joint Research Unit, UMI 3654, Singapore}
\author{Leong-Chuan Kwek}
\affiliation{Centre for Quantum Technologies, National University of Singapore,
	3 Science Drive 2, Singapore 117543, Singapore}
\affiliation{MajuLab, CNRS-UNS-NUS-NTU International Joint Research Unit, UMI 3654, Singapore}
\affiliation{Institute of Advanced Studies, Nanyang Technological University,
	60 Nanyang View, Singapore 639673, Singapore}
\affiliation{National Institute of Education, Nanyang Technological University,
	1 Nanyang Walk, Singapore 637616, Singapore}

\author{Luigi Amico}
\affiliation{Centre for Quantum Technologies, National University of Singapore,
3 Science Drive 2, Singapore 117543, Singapore}
\affiliation{MajuLab, CNRS-UNS-NUS-NTU International Joint Research Unit, UMI 3654, Singapore}
\affiliation{Dipartimento di Fisica e Astronomia, Via S. Sofia 64, 95127 Catania, Italy}
\affiliation{CNR-MATIS-IMM \&   INFN-Sezione di Catania, Via S. Sofia 64, 95127 Catania, Italy}
\affiliation{LANEF {\it 'Chaire d'excellence'}, Universit\`e Grenoble-Alpes \& CNRS, F-38000 Grenoble, France}

%\author{Rainer Dumke}
%\affiliation{Centre for Quantum Technologies, National University of Singapore, 3 Science Drive 2, Singapore 117543, Singapore}
%\affiliation{Division of Physics and Applied Physics, Nanyang Technological University, 21 Nanyang Link, Singapore 637371, Singapore}

\date{\today}

\begin{abstract}
We study the quantum transport through two specific atomtronic circuits: a Y-junction  and a ring-shaped condensate pierced by an effective magnetic flux. We demonstrate that for bosons, the two circuits display Andreev-like reflections. For the Y-junction, the transport depends on the coupling strength of the Y-junction. 
For the ring-shaped condensate, the transport crucially depends on the particle statistics. For interacting bosons we find that the Aharonov-Bohm interference effect is absent. By breaking the translational invariance of the ring, the flux dependence can be restored. A complementary view of the problem is obtained  through a specific non-equilibrium quench protocol. We find that the steady-state is independent of the flux, however the actual time-dynamics depends on the flux. The dynamics of the full closed system can be fitted with an approximated open system approach. For all the protocols we studied, we find striking differences in the dynamics of the Bose-Hubbard model and the Gross-Pitaevskii equation. 
\end{abstract}

%\pacs{03.65.Yz, 03.67.Lx, 42.50.Ct, 42.50.Pq}
 \maketitle

\section{Introduction}
Atomtronics  seeks to realize circuits of cold-atoms guided with  laser light beams or magnetic means\cite{seaman2007atomtronics,amico2005quantum,Amico_NJP}.  Key aspects of this emerging  field in quantum technology are the charge-neutrality and  the coherence properties of the fluid  flowing in the circuits, the bosonic/fermionic statistics that  carriers may have,  the tunable  particle-particle interaction, the versatility of the operating conditions of the circuit elements both in shape and time. In this way, atomtronic circuits import  the reduced decoherence, flexibility and controllability of cold atoms  quantum technology to define  new quantum devices and simulators exploiting the properties of  coherent atomic matter waves\cite{Amico_Atomtronics}.  A clearly interesting domain in which atomtronics can play an important  role  is  provided by mesoscopic physics\cite{kulik2012quantum,fazio2003new,nazarov2012quantum}.  Important chapters of the field like  persistent currents in mesoscopic normal or superconducting rings, transport through quantum dots and more complex heterostructures could be taken as inspiration  and explored with a new twist. 
With this logic, ring-shaped  condensates  interrupted  by  one or several weak  links and pierced by an effective magnetic flux\cite{dalibard2011colloquium},  have been studied: the Atomtronic Quantum Interference Device (AQUID) in analogy with the SQUIDs of  mesoscopic
superconductivity \cite{wright2013driving,Ramanathan2011,Ryu2013,eckel2014hysteresis,yakimenko2015,eckel2014hysteresis,hallwood2006macroscopic,solenov2010metastable,amico2014superfluid,aghamalyan2015coherent,aghamalyan2016atomtronic,Mathey_Mathey2016,haug2018readout}.   In this context, the single impurity problem for mesoscopic ring condensates  was demonstrated to be characterized by an unexpected and non trivial behaviour\cite{cominotti2014optimal,cominotti2015optimal}.  Persistent currents have been also studied to address the  vortex configuration in a mesoscopic circuit made of two two coupled ring condensates\cite{haug2018mesoscopic}.
%In particular,  for AQUIDs with  weak barriers and  weak atom-atom interaction, hysteresis effects were evidenced\cite{}. 
%Several problems have been explored already\cite{Amico_focus}. Persistent currents in toroidal or ring-shaped confining potential were studied both experimentally and theoretically\cite{}

In this paper we study Andreev scattering and the Aharonov-Bohm (AB) effect  in specific atomtronic circuits. These effects have been of defining  importance for the understanding of quantum transport in mesoscopic structures. Andreev scattering is inherent in the transport in heterostructures of  quantum electronics: Because of the pairing interaction, when one electron propagates from a normal to a superconducting material, one hole, instead of an electron, is reflected  back to the normal lead. 
Aharonov-Bohm oscillations  were extensively studied both in normal and superconducting  mesoscopic circuits: An electronic fluid confined to a ring-shaped wire pierced by a magnetic flux is  the typical configuration employed  to study the Aharonov-Bohm effect. In this way, a matter-wave interferometer is realized:
%The persistent current in ring systems is a hallmark feature of the interplay between charge and magnetic field. 
The current through the ring-shaped quantum system displays characteristic oscillations depending on the imparted magnetic flux\cite{gefen1984quantum,buttiker1984quantum,webb1985observation,nitzan2003electron}.  Neutral particles with magnetic moments display similar interference effects\cite{AharonovCasher}.

In the last few years cold atoms technologies allowed to explore quantum transport with enhanced flexibility and control of the system. The source to drain dynamics with cold atom systems was pioneered by the Esslinger group\cite{brantut2012conduction,krinner2015observation,husmann2015connecting,krinner2017two}.

Both Andreev-like scattering and AB effect have a counterpart  in atomtronic circuits. Andreev reflections can occur at the interface of two  bosonic condensates: If the density wave excitation in a one-dimensional condensate is transmitted from the first to second condensate, a hole (an excitation with negative amplitude) is reflected back into the first condensate\cite{daley2008andreev,zapata2009andreev,watabe2008reflection}; AB oscillations are expected to occur in ring-shaped matter-wave circuits in artificial gauge fields. 

In this paper, we study Andreev reflections and AB effect in specific bosonic circuits connected with leads: Y-junctions and AB rings.
%\rev{Within Luttinger liquid theory, Andreev-reflection have been predicted at the interface of three bosonic condensates strongly coupled together at a single point, akin to a Y-junction\cite{tokuno2008dynamics}. Within the same theory, it was shown that in a bosonic ring coupled to leads the Aharonov-Bohm effect is absent. It remains an open question what happens to these effects beyond the low energy approximation of Luttinger liquids or when condensates are not strongly coupled together.}
\rev{ A field theoretic study  indicates that Andreev reflection at the lead-circuit interface can occur;  intriguingly,  no AB oscillations were predicted to be  displayed  in the transport through bosonic rings\cite{tokuno2008dynamics}. These results were obtained within the well-known field-theoretic constraints. In particular, the field theory  describes the asymptotics of low-energy excitations. In addition, specific limitations in the physical parameters of the circuit (like the strength of the leads-circuit coupling) must be fulfilled in order such approach can be applied. \rev{Here, we mention that part of such constraints have been waived in a  very recent study, in which, however,  the leads were treated in a idealised regime\cite{haug2017aharonov}. In fact, it results that the latter aspect put severe constraints on the possible dynamical regimes that can be established in the system.}

In the present paper, we  explore the parameter regimes and dynamics beyond these limitations: We analyse  the non-equilibrium dynamics of excitations near and far from the ground state of the system and the full range of lead-circuit interface coupling. These features turned out to be clearly important for the analysis of this problem: } \rev{We find that Andreev reflection and AB effect strongly depend on the system parameters. In particular, we find that Andreev reflections can be suppressed by tuning the lead-circuit coupling. The AB effect is absent for all interaction strengths. To get further insight on the physics behind, we add impurities into the ring to perturb the phase dynamics locally. Indeed we find that the AB effect can reemerge because of the impurities. } 

%The Aharonov-Bohm effect  in bosonic condensates has been studied very  recently: A Bose-Einstein condensate propagating out of equilibrium from source to drain along a mesoscopic ring-shaped laser light potential, pierced by an effective magnetic flux. It was found that the  system experiences a subtle crossover between  physical regimes dominated by pronounced  interference patterns and  others in which the Aharonov-Bohm effect is effectively washed out\cite{haug2017aharonov}. 

In this manuscript, 
%we study the  Andreev-like scattering and the Aharonov-Bohm  effect in Bose condensates for a wide parameter range in two specific atomtronic circuit elements: a Y-junction and a ring condensate attached to leads.
we study the dynamics of the propagation  of an incoming density 'bump' created in the source lead, which moves through both  Y-junction and ring-shaped lattices.   The leads are modeled through  finite length bosonic lattice chains. \rev{This setup allows us to tune the energies of the incoming density wave and study different regimes of propagation, from low energy excitations to ones with high amplitude.} %Furthermore, the coupling strength of lead and system can be freely tuned.}
We analyze how the device properties, atom-atom interaction and  initial conditions affect the time evolution, the transmission and reflection coefficients of the atomtronic setup.
For the same setups, we study  the transport dynamics induced by a quench protocol to view the complementary dynamics when the system is strongly perturbed. %The quench dynamics of the full system can be fitted with an open system approach of the reduced system.

The paper is outlined as follows.
In Sec.\ref{models}, we explain the configuration and protocol we employ to study  the atomtronic circuits. 
In Sec. \ref{YExc} and Sec. \ref{RingSimple},    both the propagation of density and quench dynamics are reported for  the Y-junction and Aharonov-Bohm matter-wave interferometer respectively. Sec.\ref{discussion} is devoted to the discussion of the results. For the analysis, we employ the Density Matrix Renormalization Group (DMRG) to calculate ground states, \rev{time dependent Density Matrix Renormalization Group (tDMRG) to calculate the time dynamics} and the Lindblad master equation (as detailed in the Appendix). For comparison with the bosonic case, the transport dynamics of fermions is discussed in Appendix.

\section{Models and protocols}\label{models}
\begin{figure}[htbp]
	\centering
	\subfigure{\includegraphics[width=0.4\textwidth]{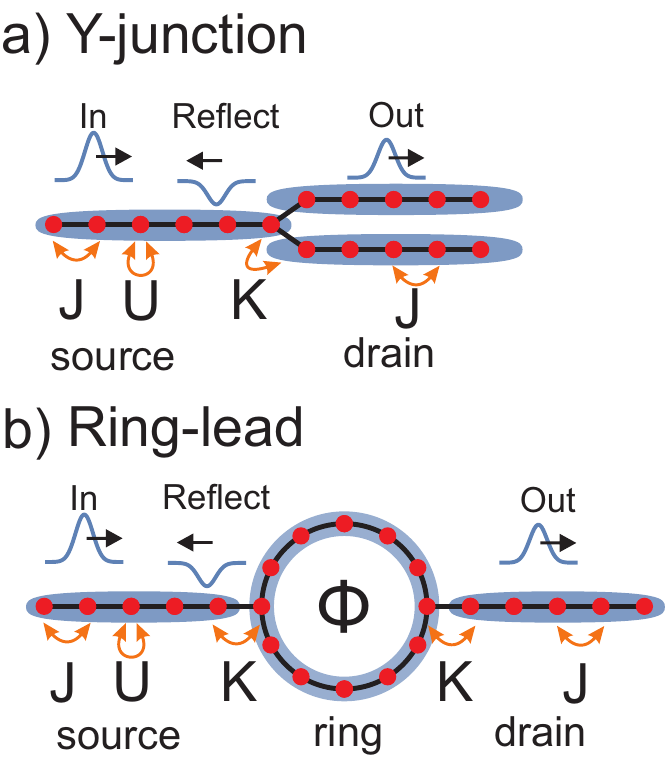}}
	\caption{Sketch of the two configurations \idg{a} Y-junction and \idg{b} ring-lead. The red dots denote sites, the black links tunneling between sites. $J$ is the tunneling strength inside leads and ring. $K$ is the coupling between different leads and ring. $U$ denotes the on-site interaction, which is the same on all sites, and $\Phi$ the flux of the ring. We study excitations on top of a atom condensate. An excitation incoming through the source lead is transmitted and reflected by the Y-junction and the ring. }
	\label{Sketch}
\end{figure}

We consider two setups: A Y-junction, and a lead-ring system. A sketch of both systems is presented in Fig.\ref{Sketch}. We model atomtronic circuits with the Bose-Hubbard model. 

{\bfseries Y-junction--}
The Y-junction is a system consisting of three one-dimensional chains, which are coupled together at a single point. Such systems have been proposed and realized experimentally\cite{kevrekidis2003guidance,andersen2018physical,Boshier_integrated}.
The Hamiltonian for the Y-junction is ${\mathcal{H}_\text{S}+\mathcal{H}_\text{D}+\mathcal{H}_\text{I}}$, with the source lead Hamiltonian (analogue for the two drain leads) 
\begin{equation}\label{HamiltonSource}
\mathcal{H}_\text{S}=-\sum_{j=1}^{L_\text{S}-1}\left(J\cn{s}{j}\an{s}{j+1} + \text{H.C.}\right)+\sum_{j=1}^{L_\text{S}}\frac{U}{2}\nn{s}{j}(\nn{s}{j}-1)\;,
\end{equation}
where $\an{s}{j}$  and $\cn{s}{j}$  are the annihilation and creation operator at site $j$ in the source lead, ${\nn{s}{j}=\cn{s}{j}\an{s}{j}}$ is the particle number operator of the source, $J$ is the intra-lead hopping, $L_\text{S}$ the number of source lead sites and $U$ is the on-site interaction between particles. \rev{The Hamiltonian $\mathcal{H}_\text{D}$ for the two drain leads have similar Hamiltonians, where we replace the index $s$ with respective $d$ (for first drain) and $f$ (second drain), and define the drain length $L_\text{D}$. We chose interaction $U$ globally the same everywhere in the system in source, system and drain. We define $N_\text{p}$ as the number of atoms in the total system, which is a conserved quantity.}%For bosons, the annihilation and creation operators commute: $\left[\an{s}{i},\cn{s}{j}\right]=\delta_{i,j}$.

The coupling Hamiltonian between the source lead and the two drain leads is
\begin{equation}
\mathcal{H}_\text{I}=-K\cn{s}{1}\left(\an{d}{1} + \an{f}{1}\right)+ \text{H.C.}\;,
\end{equation}
where $K$ is the coupling strength between source and drain leads. 

{\bfseries Leads-ring--}
To study the Aharonov-Bohm effect, we introduce a ring system. It is coupled to two leads (source and drain) symmetrically at two opposite sites of the ring. 
The ring-lead Hamiltonian is $\mathcal{H}_\text{R}+\mathcal{H}_\text{S}+\mathcal{H}_\text{D}+\mathcal{H}_\text{I}$. The ring Hamiltonian is
\begin{equation}
\mathcal{H}_\text{R}=-\sum_{j=1}^{L_\text{R}}\left(J\expU{i2\pi\Phi/L}\cn{a}{j}\an{a}{j+1} + \text{H.C.}\right)+\frac{U}{2}\sum_{j=1}^{L_\text{R}}\nn{a}{j}(\nn{a}{j}-1)\;,
\end{equation}
where $\an{a}{j}$ and $\cn{a}{j}$ are the annihilation and creation operator at site $j$ in the ring, $L_\text{R}$ the number of ring sites, ${\nn{a}{j}=\cn{a}{j}\an{a}{j}}$ is the particle number operator of the ring, $J$ is the intra-ring hopping and $\Phi$ is the total flux through the ring. Periodic boundary conditions are applied for the ring with ${\cn{a}{L}=\cn{a}{0}}$. In the following, we set ${J=1}$, and all values of $U$, $K$ are given in units of $J$.
The source and drain Hamiltonian are analogue to the Y-junction as defined in Eq.\ref{HamiltonSource}.
The coupling Hamiltonian between leads and ring is
\begin{equation}
\mathcal{H}_\text{I}=-K\left(\cn{a}{0}\an{s}{0} + \cn{a}{L_\text{R}/2}\an{d}{0}+ \text{H.C.}\right)\;.
\end{equation}

The current through the Y-junction is defined as
\begin{equation}\label{Ycurrent}
j_\text{Y}=-iK\cn{s}{0}\an{d}{0} + \text{H.C.} \;,
\end{equation}
and for the current into the ring (source current) and from the ring into the drain (drain current) 
\begin{align*}\label{ringcurrent}
j_\text{source}=&-iK\cn{a}{0}\an{s}{0} + \text{H.C.}\\
j_\text{drain}=&-iK\cn{a}{L_\text{R}/2}\an{d}{0} + \text{H.C.}\;.\numberthis
\end{align*}

We also consider the addition of impurities into the ring. We define this as a potential offset in Hamiltonian at two symmetric sites in the middle of the ring  \begin{equation}\mathcal{H}_\text{R,impurity}=\Delta(\nn{a}{L_\text{R}/4}+\nn{a}{3L_\text{R}/4} )\; ,
\end{equation}
where $\Delta$ is the strength of the impurity.

\rev{In this paper, we study three different models, which describe different on-site interaction $U$ regimes of above systems:}

{\bfseries 1. Hard-core Bose-Hubbard--}
In the limit of very strong interaction ${U\rightarrow \infty}$, only zero or one boson is allowed per site. This is the so called Hard-core boson limit. In a strictly one-dimensional system, it can be mapped to non-interacting fermions with the Jordan-Wigner transformation. In our quasi one-dimensional system as a ring-lead system or Y-junction, this mapping is not possible since it is not one-dimensional and the Jordan-Wigner transformation introduces non-local terms. The hard-core Bose-Hubbard model is studied in Figs.\ref{YDensitysmall},\ref{YDensityFermionBosonComp},\ref{YCurrentComparison},\ref{RingOnlyHardcore},\ref{RingFermionBosonComparison},\ref{RingDensityComp},\ref{TwoImpRingNew},\ref{RingDensityfullBoson},\ref{RingDensityComparison}.

{\bfseries 2. Bose-Hubbard model--}
\rev{For finite $U$, several atoms can occupy the same site. This Bose-Hubbard model (BHM) is a non-integrable model, requiring numerical solution even in strictly one dimension. When $U$ is not too small, interaction penalizes the occupation of several atoms per site. Thus, we can restrict the local Hilbert space to only a few atoms without loosing accuracy and speed up numerical simulation. The model is studied in Figs.\ref{YDensityHubbardsmall},\ref{RingBH},\ref{RingDensityComp}.}

{\bfseries 3. Gross-Pitaevskii--}
\rev{In the limit of many particles, low interaction and many lattice sites} we can replace the operators with complex numbers ${\an{a}{j}\rightarrow\psi_a(j)}$, ${\cn{a}{j}\rightarrow\psi^*_a(j)}$ and ${\nn{a}{j}\rightarrow|\psi_a(j)|^2}$ \cite{oelkers2007ground}. The result is a lattice version of the non-linear Gross-Pitaevskii equation, which we refer to as discrete Gross-Pitaevskii equation (dGPE). For example, the equation of motions for the source lead at site $j$ in the Y-junction are
\begin{align*}
-i\partial_t\psi_s(j)=&-J\left(\psi_s(j-1)+\psi_s(j+1)\right)+P_s(j)\psi_s(j)\\
&-K(\psi_d(1)+\psi_f(1))\delta_{j,1}+g|\psi_s(j)|^2\psi_s(j)\;,\numberthis
\end{align*}
where $\psi_s(j)$ ($\psi_d(j)$,$\psi_f(1)$) is the wavefunction of the source (drain) at site $j$ and $g=UN_\text{p}$ the non-linear interaction term. \rev{The number of particles $N_\text{p}$ is just a scaling factor in this model.} The other equation of motions for ring or drains follow in a similar way. The sum of the absolute square of the wavefunctions is normalized to one. The model is studied in Figs.\ref{YDensitysmallGPE},\ref{YDensitybigPosGPE},\ref{YDensitybigNegGPE},\ref{RingDensityComp},\ref{RingDensitydarkGPE},\ref{RingDensitybrightGPE}.

%The Gross-Pitaevskii equation (GPE) describes the flow of a Bose-Einstein condensate in the limit of many particles and low interaction. In the limit of many lattice sites, the discretized GPE yields the same result as the continous version.
\rev{In the limit of many lattice sites, we can replace the nearest-neighbor coupling terms by second-order derivatives and get the continuous Gross-Pitaevskii equation (cGPE). We use this equation to get analytic understanding of numerical results of the dGPE.}
In a continuous 1D system, the cGPE is given by
\begin{equation}
i\partial_t\Psi(x,t)=\left[\frac{1}{2}(-i\partial_x-A(x,t))^2+V(x)+g\abs{\Psi(x,t)}^2\right]\Psi(x,t)\;,
\end{equation}
where $\Psi(x,t)$ is the condensate wavefunction, $V(x)$ the potential, $g$ the coefficient for the atom-atom interaction and $A(x,t)$ the effective magnetic field. For a ring, we can relate it to the flux through a ring as introduced in the BHM model ${A=\frac{2\pi\Phi}{L}}$, where $L$ is the length of the ring.
In the following, we assume that there is potential ${V(x)=0}$ and the effective magnetic field is constant ${A(x)=\text{const}}$. For a wavefunction ansatz $\Psi(x,t)=\sqrt{n(x,t)}\expU{i\phi(x,t)}$, the cGPE can be rewritten in terms of the density $n(x,t)$ and the velocity of the condensate $v(x,t)=-\partial_x\phi(x,t)$ \cite{pethick2008bose}
\begin{align}
\partial_t n+\partial_x(n(v-A))&=0\label{ConservationMass}\\
\partial_t v + \partial_x\left(gn+\frac{1}{2}(v-A)^2-\frac{1}{2\sqrt{n}}\partial_x^2\sqrt{n}\right)&=0\,.\label{EquationGPE}
\end{align}
The first equation is the conservation of mass, while the second is a hydrodynamic equation with an additional term, which represents the ``quantum pressure" (Eq.\ref{EquationGPE}, third term). \rev{In the regimes normally considered for cold atom experiments,} we can neglect the quantum pressure term. Now, we expand the equation in terms of small excitations on top of the static condensate with $n=n_0+\delta n$ and $v=v_0+\delta v$. We assume that $n_0$ and $v_0$ is constant in space, and that the density and velocity variation $\delta n$, $\delta v$ is very small.
For the linearized equation of motion, we find that both $\delta_n$ and $\delta_v$ are decoupled dispersion-less wave equations
\begin{align*}
\partial_t^2\delta n&=\left[gn_0-(A-v_0)^2\right]\partial_x^2\delta n +2(A-v_0)\partial_t\partial_x\delta n\; \\
\partial_t^2\delta v&=\left[gn_0-(A-v_0)^2\right]\partial_x^2\delta v +2(A-v_0)\partial_t\partial_x\delta v\;,\numberthis\label{LinEq}
\end{align*}
where ${A-v_0}$ causes a direction-dependent shift in the speed of sound and is explained further in Sec.\ref{RingSimple}.

%Next, we introduce two different ways of studying transport in these system.
\rev{To study the dynamics, we use two different approaches. We either let a small excitation propagate from the leads through the system, or strongly quench the leads:}

{\bfseries 1. Excitation propagation--}
To study the propagation of a density excitation through our setups, we prepare the system in the ground state of the full Hamiltonian with initially a small local potential offset in the lead Hamiltonian. This will create a localized density bump in the source lead.
We add the following Hamiltonian for the offset potential to the source Hamiltonian
\begin{equation}
\mathcal{H}_P=-\epsilon_\text{D}\sum_{j=1}^{L_\text{S}}\exp\left(-\frac{(j-d)^2}{2\sigma^2}\right)\nn{s}{j}\;,
\end{equation}
where $d$ is the distance of the initial excitation to the junction, and $\sigma$ is the width of the potential offset, which we set to ${\sigma=2}$ unless specified otherwise.
At the start of the time evolution the offset potential is instantaneously switched off. The density bump will propagate as an excitation in both positive and negative direction. In this paper, we are only interested in the forward direction, and disregard the excitation in backward direction.

{\bfseries 2. Reservoir quench--}
In this configuration, all particles are loaded initially into the ground state of the  source lead, which is disconnected from the rest of the system. The rest of the system is initially empty. The ground state is obtained through DMRG.  Then, at the start of the time-evolution, the coupling terms with the rest of the circuits are suddenly switched on, and atoms start propagating into them. \rev{The dynamics is calculated with tDMRG. As the difference in particle number between source and rest of the system is large, the resulting evolution is highly non-equilibrium and cannot be treated as small perturbation of the ground state within some quasi-particle description.}

{\bfseries Transmission and reflection coefficients--}
To evaluate the dynamics, we calculate the total density of the incoming wave by taking the first $a$ sites of the source lead at a specific time $t_\text{in}$ when the density waves has entered this region, and subtracting from it the density at time ${t=0}$ before the wave has entered the region
\begin{equation}\label{EqInc}
N_\text{inc}=\sum_{i\in {a\text{ sites of source}}}\left[n_i(t_\text{in})-n_i(0)\right]\;. 
\end{equation}
Here, $n_i(t)$ is the expectation value of the density at the $i$-th site of the system at time $t$.
We find the transmission coefficient by dividing the change in atom number in the drain density by the total density of the incoming wave
\begin{equation}\label{EqTrans}
T=\frac{\sum_{i\in\text{drain}}\left[n_i(t)-n_i(0)\right]}{N_\text{inc}}
\end{equation}
and the reflection coefficient as 
\begin{equation}\label{EqRef}
R=1-T\;.
\end{equation} 
%In the following, all parameters for coupling $K$, interaction $U$ and $g$ are given in units of the hopping strength ${J=1}$.

{\bf{Experimental realization--}}
\rev{Our proposed model can be realized with currently available cold atom techniques. The Bose-Hubbard model has been successfully realized for cold atoms trapped in lattices in many experiments\cite{bloch2008many}. Transport through one-dimensional systems attached to reservoirs has been demonstrated\cite{brantut2012conduction,krinner2015observation,husmann2015connecting,krinner2017two,eckel2016contact}. Using magnetic traps, spatial light modulators and digital mirror devices (DMD) nearly any type of two-dimensional trapping potential  can be generated\cite{mcgloin2003applications,gaunt2012robust,gauthier2016direct}. In particular, magnetic traps have been utilized to create Y-junctions\cite{Boshier_integrated} and light shaping techniques to create lattice rings with weak links\cite{amico2014superfluid}.}

%\section{Results}
%\label{results}
%In section \ref{YExc}, we show the results for the Y-junction. 
%In section \ref{RingSimple}, we show results for propagation through ring.

\section{Propagation of excitations in Y-junctions}\label{YExc}
{\bfseries Linearized equations--}
First, we study the linearized equations of motion of the cGPE in Eq.\ref{LinEq}. For a small excitation, they yield dispersion-less waves traveling with velocity ${c=\sqrt{gn_0}}$ (without magnetic field). In general, the solution of the equations can be written as a non-dispersive wave, with $\delta n=f(t-\frac{x}{c})$, where $f(y)$ is any function, $x$ the position and $t$ time. From the continuity equation, we can then derive the phase of the excitation ${\delta v=c\frac{\delta n}{\delta n +n_0}}$. Next, we derive the transmission and reflection at the Y-junction (here ${A=0}$, ${v_0=0}$). At the junction, we have the following components: incoming wave $\delta n_\text{in}$, reflected wave $\delta n_\text{r}$ and the transmitted waves $\delta n_\text{t}$ in the two drain leads. Due to symmetry, the transmission into each drain wire is the same. We demand that at the junction of source and drain the density and velocity are the same
\begin{align*}
\delta n_\text{in}+\delta n_\text{r}=&\delta n_\text{t} 
\delta v_\text{in}+\delta v_\text{r}=&\delta v_\text{t} \;.
\end{align*} 
Also, the continuity equation has to be fulfilled at the junction of source and the two drains:
\begin{equation}
n_\text{source}v_\text{source}=2n_\text{drain}v_\text{drain}\;.
\end{equation}
The factor two arises as we have two drain leads. Inserting the equations for density and velocity, we get
\begin{equation}
(n_0+\delta n_\text{in}+\delta n_\text{r})(\delta v_\text{in}-\delta v_\text{r})=2(n_0+\delta n_\text{t})\delta v_\text{t}\;.
\end{equation}
For small velocity and density excitations, we can approximate $\delta v\approx c\frac{\delta n}{n_0}$. Then, by inserting the conditions, we find for the transmitted and reflected density (same for velocity variation $\delta v$)
\begin{align*}\label{transmission}
\delta n_\text{t}=&\frac{2}{3}\delta n_\text{in}\\
\delta n_\text{r}=&-\frac{1}{3}\delta n_\text{in}\;.\numberthis
\end{align*}
The total transmission into the two arms of the Y-junction is then ${2\delta n_\text{t}=\frac{4}{3}\delta n_\text{in}}$.

The transmission and reflection into each lead can easily be generalized to a junction with $N$ drains
\begin{align*}\label{transmissionN}
\delta n_\text{t}=&\frac{2}{N+1}\delta n_\text{in}\\
\delta n_\text{r}=&-\frac{N-1}{N+1}\delta n_\text{in}\;.\numberthis
\end{align*}

{\bfseries Full dynamics--}
\begin{figure}[htbp]
	\centering
	%\subfigure{\includegraphics[width=0.24\textwidth]{{chiralcurrentFluxN1DTwistDataN10000}.pdf}}\hfill
	%\subfigure{\includegraphics[width=0.24\textwidth]{{chiralcurrentTwistN1DTwistN10000}.pdf}}
	\subfigimgraised[width=0.24\textwidth]{a}{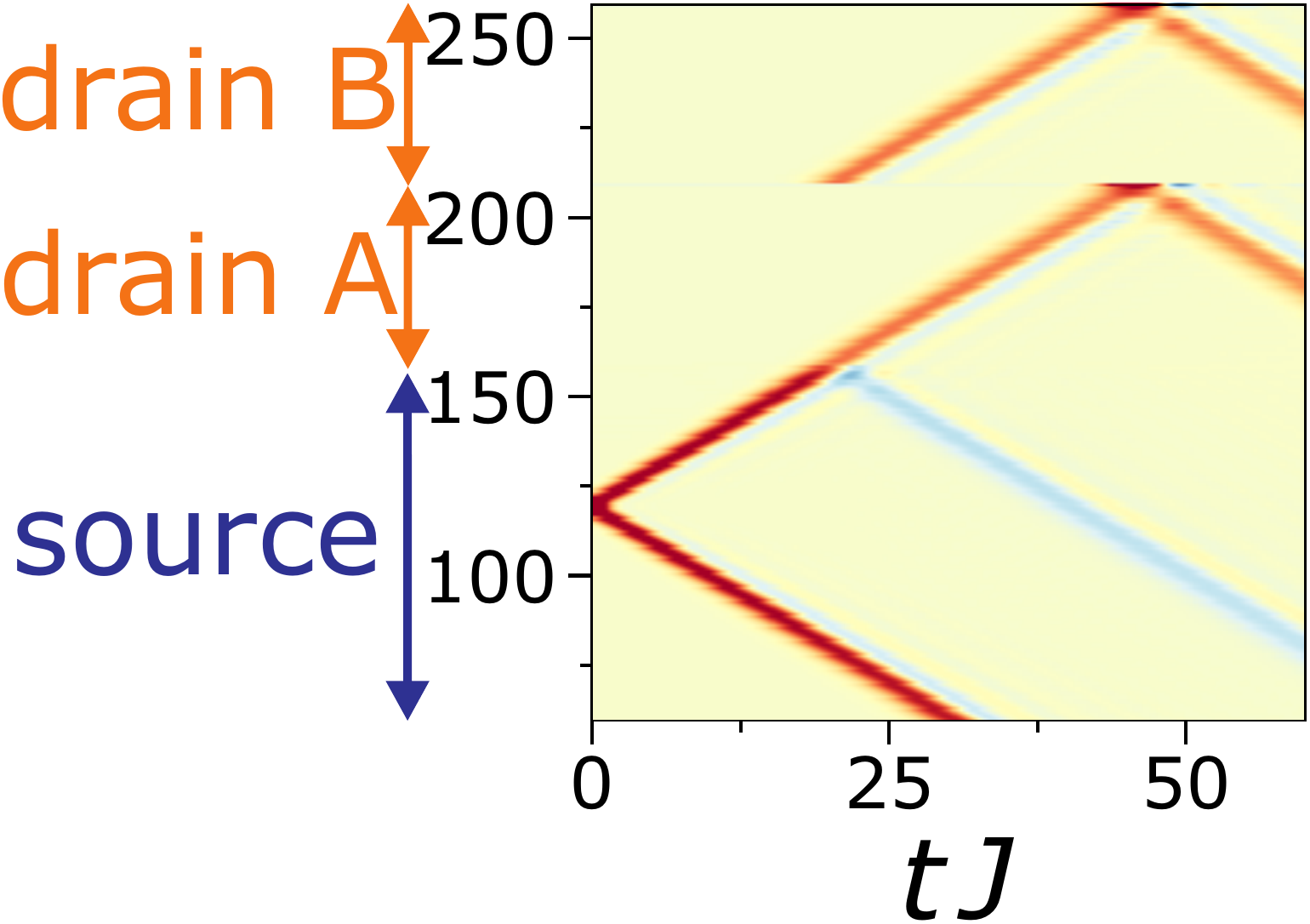}\hfill
	\subfigimg[width=0.24\textwidth]{b}{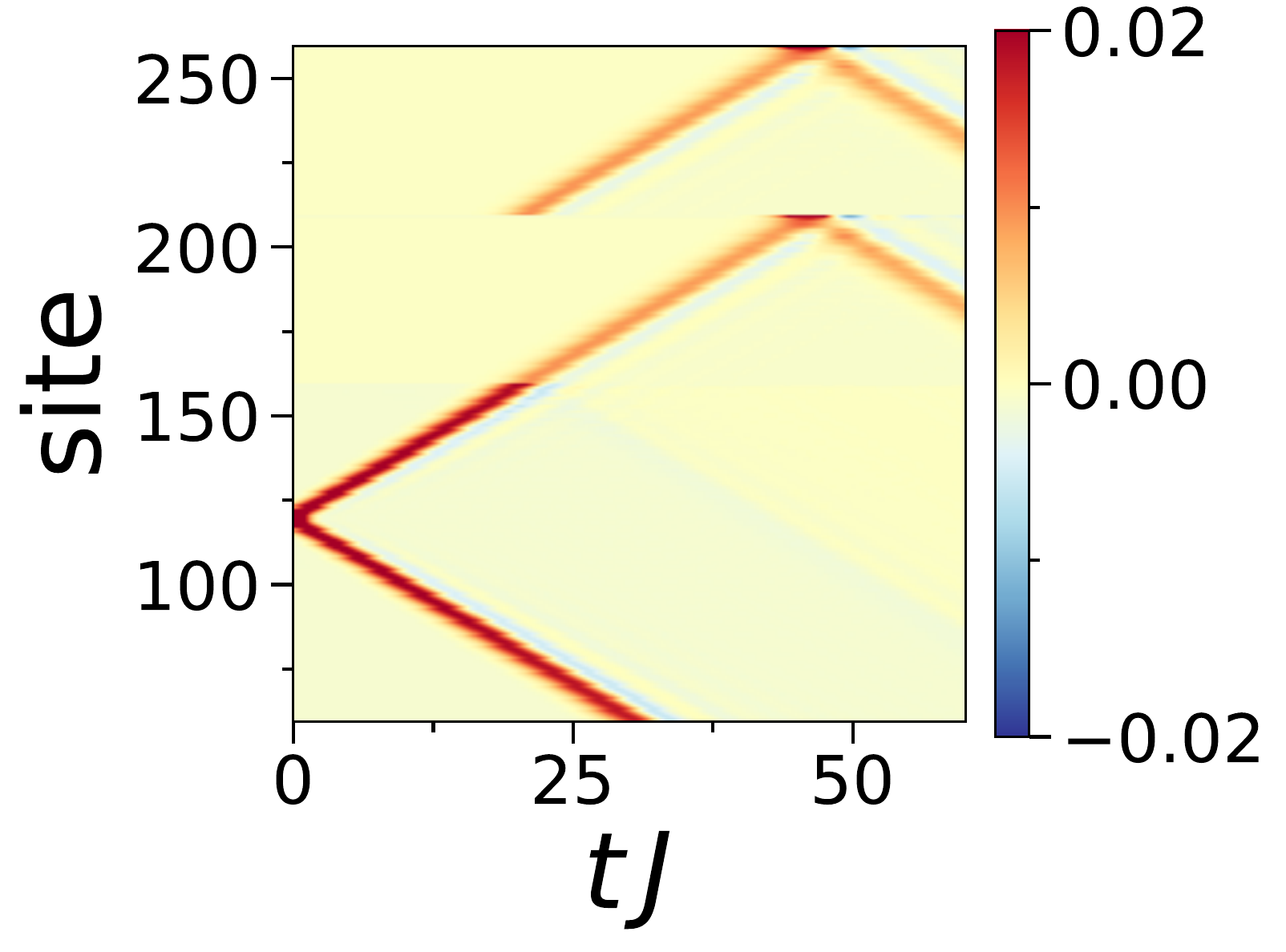}
	\subfigimg[width=0.24\textwidth]{c}{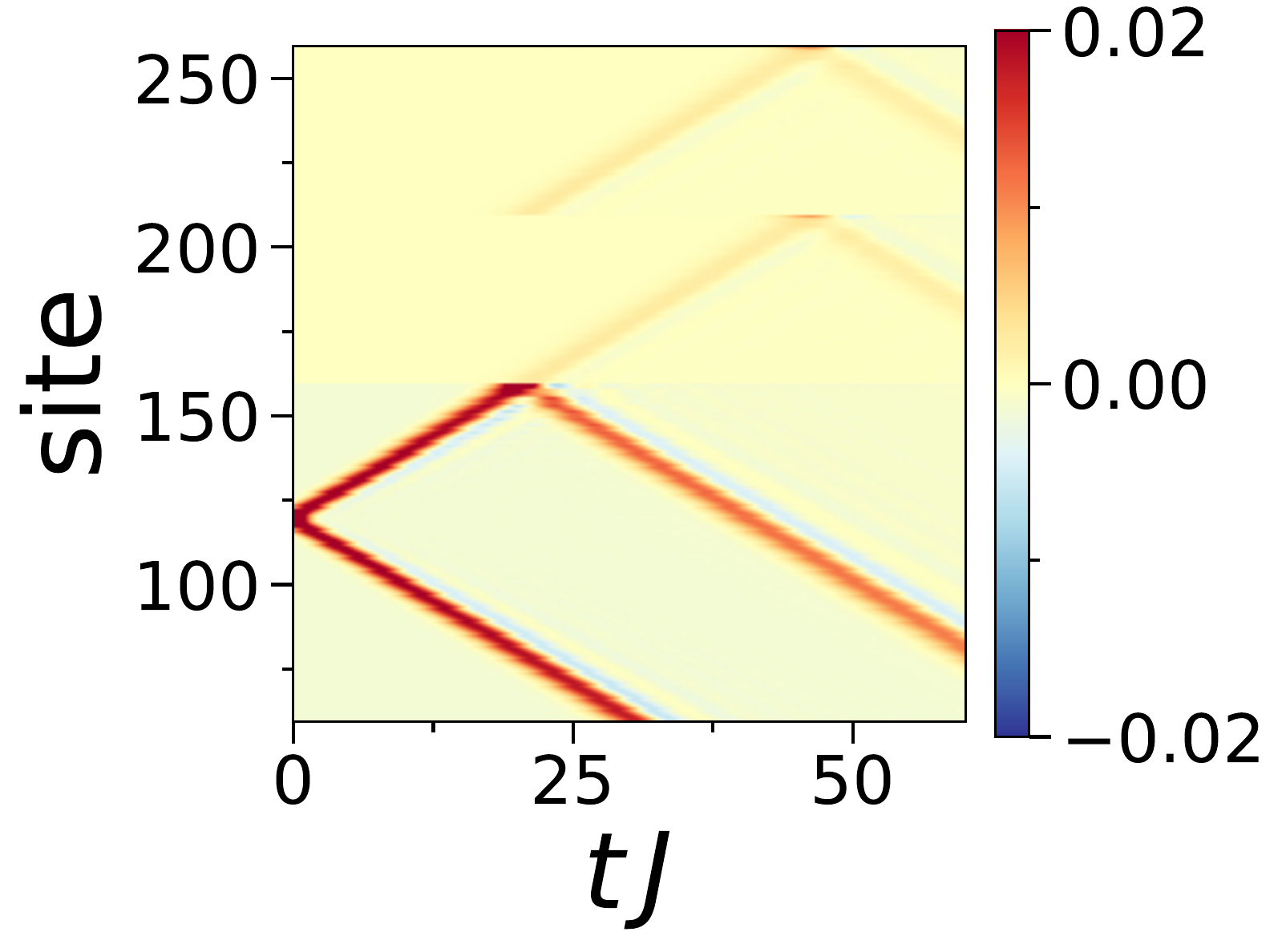}\hfill
	\subfigimg[width=0.24\textwidth]{d}{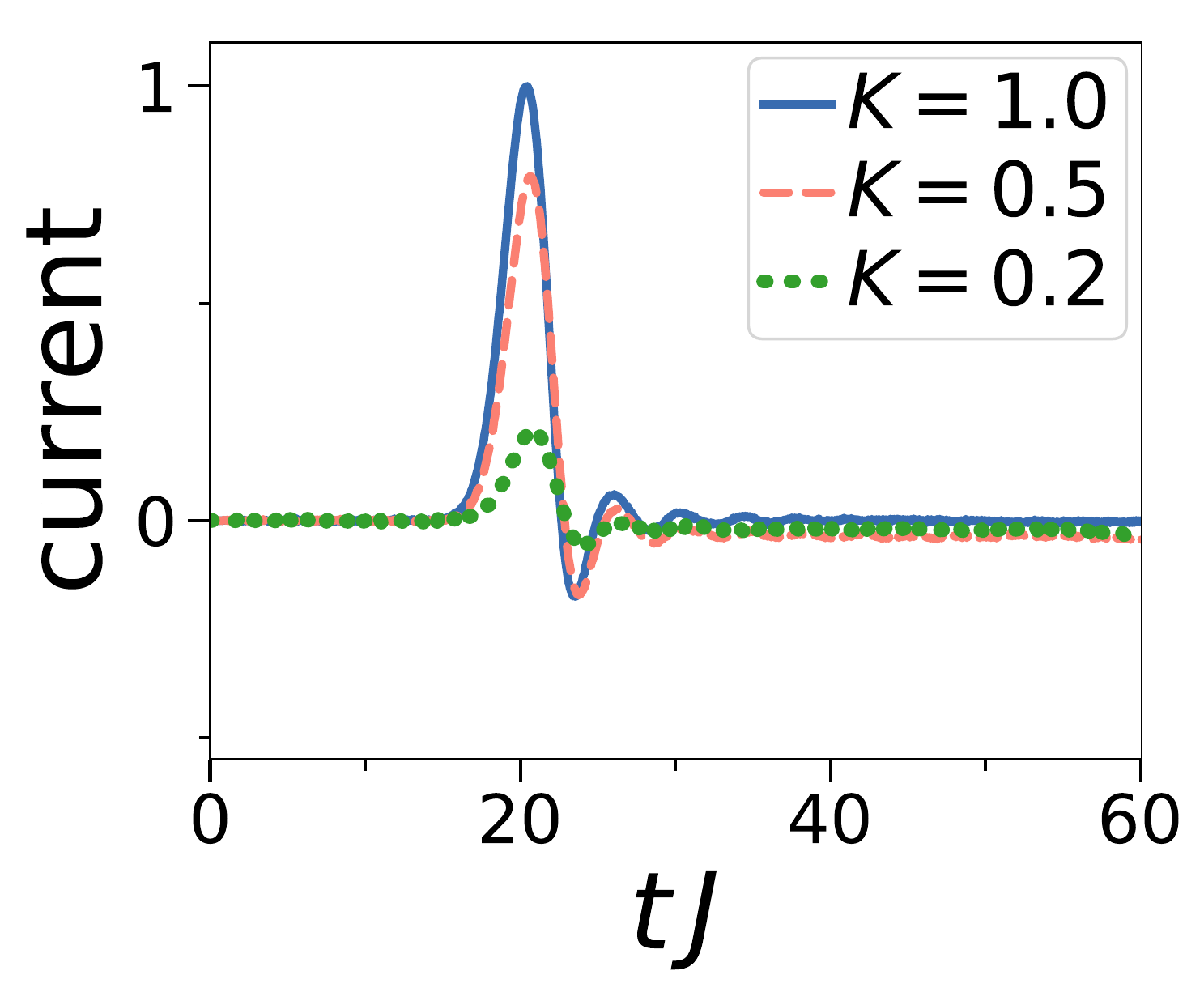}\\
	\subfigimg[width=0.32\textwidth]{e}{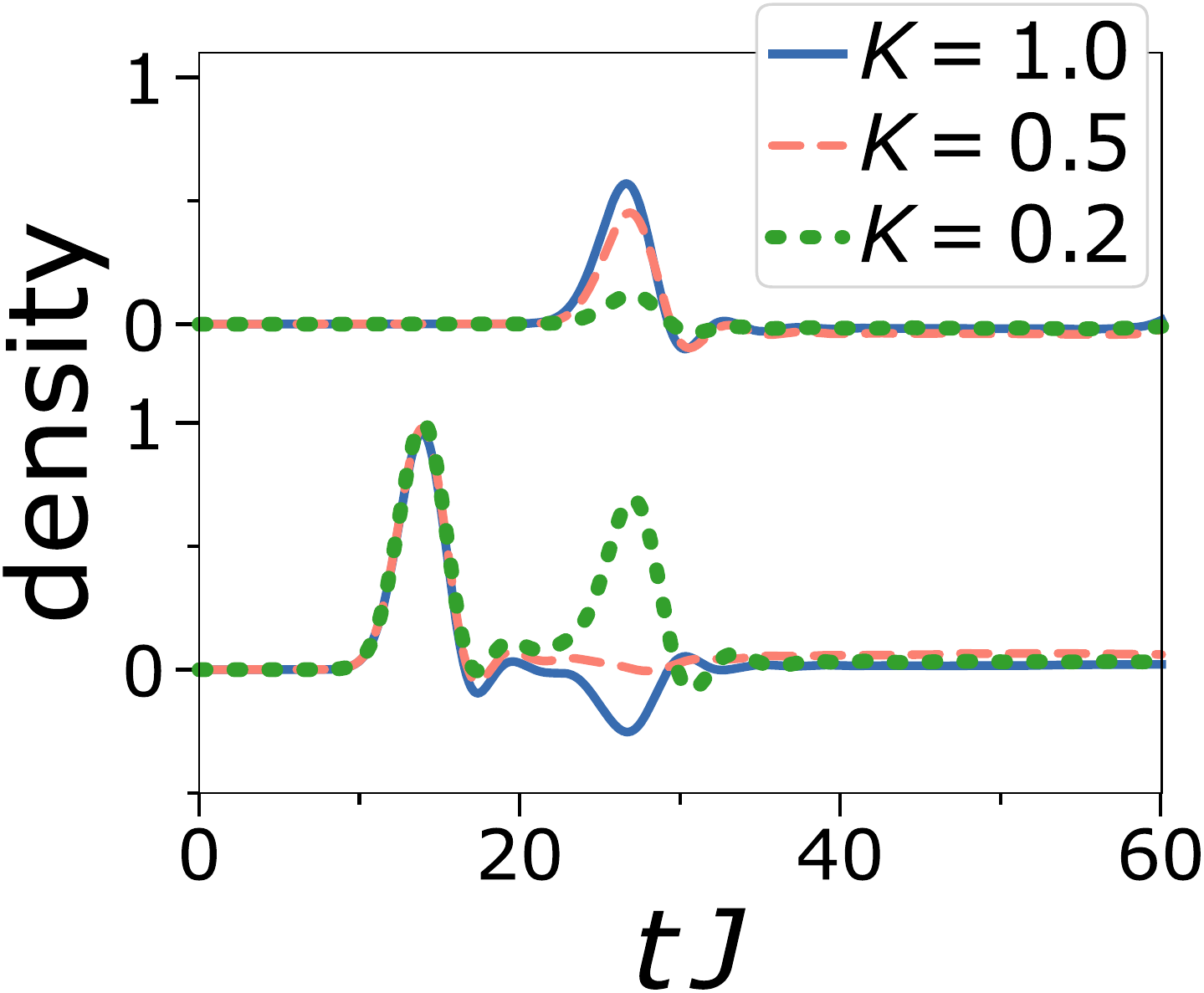}
	\caption{Propagation of small excitation in a Y-junction for the hard-core boson model. The source lead has length ${L_\text{S}=160}$, the drain lead each ${L_\text{D}=50}$, the particle number ${N=130}$, initial distance of the excitation to the junction ${d=40}$ and ${\epsilon_\text{D}=0.3}$. The source lead is from site 1 to 160, the first drain lead from 160 to 210, and the second one from site 210 to 260. The coupling at the junction (site 160) is \idg{a} ${K=1}$, \idg{b} ${K=0.5}$, \idg{c} ${K=0.2}$. \idg{d} Current through the junction (Eq.\ref{Ycurrent}) in time. \idg{e} \densdescr{170}{175}{145}{150}  For ${K=1}$ (solid) we observe a negative reflection (Andreev-like), ${K=0.5}$ (dashed) nearly no reflection, ${K=0.2}$ (dots) a large positive reflection amplitude. 
	The table below shows the transmission and reflection coefficients, calculated at ${t=31/J}$ with Eq.\ref{EqInc}-\ref{EqRef} (${t_\text{in}=15}$, ${a=30}$). 
	%We find small, but long-lived currents, which change the coefficients when calculated at later times. However, these currents are related to a atom imbalance between drain and source, and could be removed by a constant global potential offset in the source.
}
	%\begin{tabular}{l|*{3}{c}}
	%	& ${K=1}$ & ${K=0.5}$ & ${K=0.2}$ \\
	%	\hline
	%	transmission & $1.333$ & $0.818$ & $0.13$   \\
	%	reflection   & $-0.333$ & $0.182$ & $0.87$
	%\end{tabular}
	\begin{tabular}{l|*{3}{c}}
	& ${K=1}$ & ${K=0.5}$ & ${K=0.2}$ \\
	\hline
	transmission & $1.332$ & $0.947$ & $0.207$   \\
	reflection   & $-0.332$ & $0.053$ & $0.793$
\end{tabular}
	\label{YDensitysmall}
\end{figure}
Here we present our numerical results for the Y-junction. First, we concentrate of the limiting cases of infinitely strong on-site interaction with hard-core bosons in Fig.\ref{YDensitysmall}. (Spinless fermions in Fig.\ref{YDensitysmallFermion} as a reference in the supplemental material). 

{\bfseries Hard-core bosons.}
\rev{In Fig.\ref{YDensitysmall}a-c, we study the propagation for different values of lead coupling $K$. In the source lead an initial excitation bump is prepared. At ${t=0}$, the potential offset is quenched, and the excitation starts moving in forward and backward direction. We ignore the backward propagating part of the wave. The forward moving part of the wave propagates from the source through the junction to the two drain leads.  We find that the wave at the junction (site 160) is both transmitted and reflected. }
For the reflection amplitude,  we find three characteristic reflection regimes, which are controlled by the junction coupling $K$.
First, we look at the reflection peak as seen in Fig.\ref{YDensitysmall}e) at time ${t J=27}$. 
In the strong coupling regime ${K=1}$, we see a negative (Andreev-like) reflection amplitude peak.

For the intermediate coupling regime ${K\approx0.5}$ the back reflection amplitude is very small, and the reflected wave consists of a small, first positive and then negative part, of nearly equal weight.

Finally, for the weak coupling regime with $K$ small, we find a large positive back-reflection and small transmission.

%Interestingly, we find that both in strong and intermediate regime, the total transmission is close to unity, however the underlying process is quite different (Andreev-like reflection or perfect transmission). Here, we would like to point out the strong similarities with a power line. In a power line, at the junction the resistance of the line has to be matched in order to avoid unwanted reflection. This is called impedance matching. In our Y-junction, the impedance matching is done via the coupling strength $K$. By tuning $K$, we can change the amount of reflection and transmission, and reduce the reflection amplitude to nearly zero at ${K=0.5}$. 

In the table below Fig.\ref{YDensitysmall}, we plot the total transmitted and reflected density at time ${t=31/J}$ (calculated using Eq.\ref{EqInc}-\ref{EqRef}). This gives us the transmission coefficient and reflection coefficient of the density wave packet.

\rev{The evolution of the density has two components: The dynamics of the main wavepacket, clearly visible in transmission and reflection as peaks. This is the dynamics we want to study. 
Additionally, there is a small, but non-decaying background current between source and drain, which is an artifact of our procedure to generate the density excitation. This small current is most likely result of finite size of the leads and how the excitation are initialized: We prepare the initial wavepacket by a potential offset on the source side only, which breaks the symmetry of the bulk states between source and drain. This may generate a phase difference between the leads and subsequently this small current flows. We find that this current does not decay in our numerical simulation. Depending on the parameters, we observe currents in either direction: Source-drain or drain-source. %As we want to study the dynamics of the wavepacket only, we want to neutralize the effect of this small background current. This current could be eliminated by a measuring it at long times, long after the excitation has passed, and subtracting the current from the dynamics. 
%We choose a different route to minimize this current: 
To avoid the impact of this background current, we calculate the reflection coefficient at ${t=31/J}$, immediately after the main wavepacket has propagated into the drain and the oscillations in the source-drain current have rung down. At these early times, the background current has no significant impact yet on the transmission. Calculating the transmitted density at later times would yield different transmission coefficients, since the background current would disturb the result.}
%However, we find that the transmission and reflection coefficient can be different when calculated at longer times, at which the initial wave has already passed the junction.
%For example, at ${K=0.5}$, we find that the reflected wave has nearly zero amplitude, and a very small reflection coefficient when it is calculate at ${t=40/J}$. This is the contribution from the density wave. However, if the coefficient is calculated at later times, the reflection increases. This is caused by a small, long-lived current from drain into source even after the wave has passed. This current contributes to the reflection coefficient when it is calculated for later times. However, this small current is related to an atom imbalance of source and drain due to our quenched potential offset, and could be eliminated by a static potential offset in the source. We avoid this complication by calculating the reflection coefficient at ${t=40/J}$.
%For ${K\le0.5}$ we get a positive reflection, and for ${K>0.5}$ a negative reflection. Exactly at ${K=0.5}$, there is no reflection, which means that the junction is impedance matched. 

\begin{figure*}[htbp]
	\centering
	\subfigimg[width=0.25\textwidth]{a}{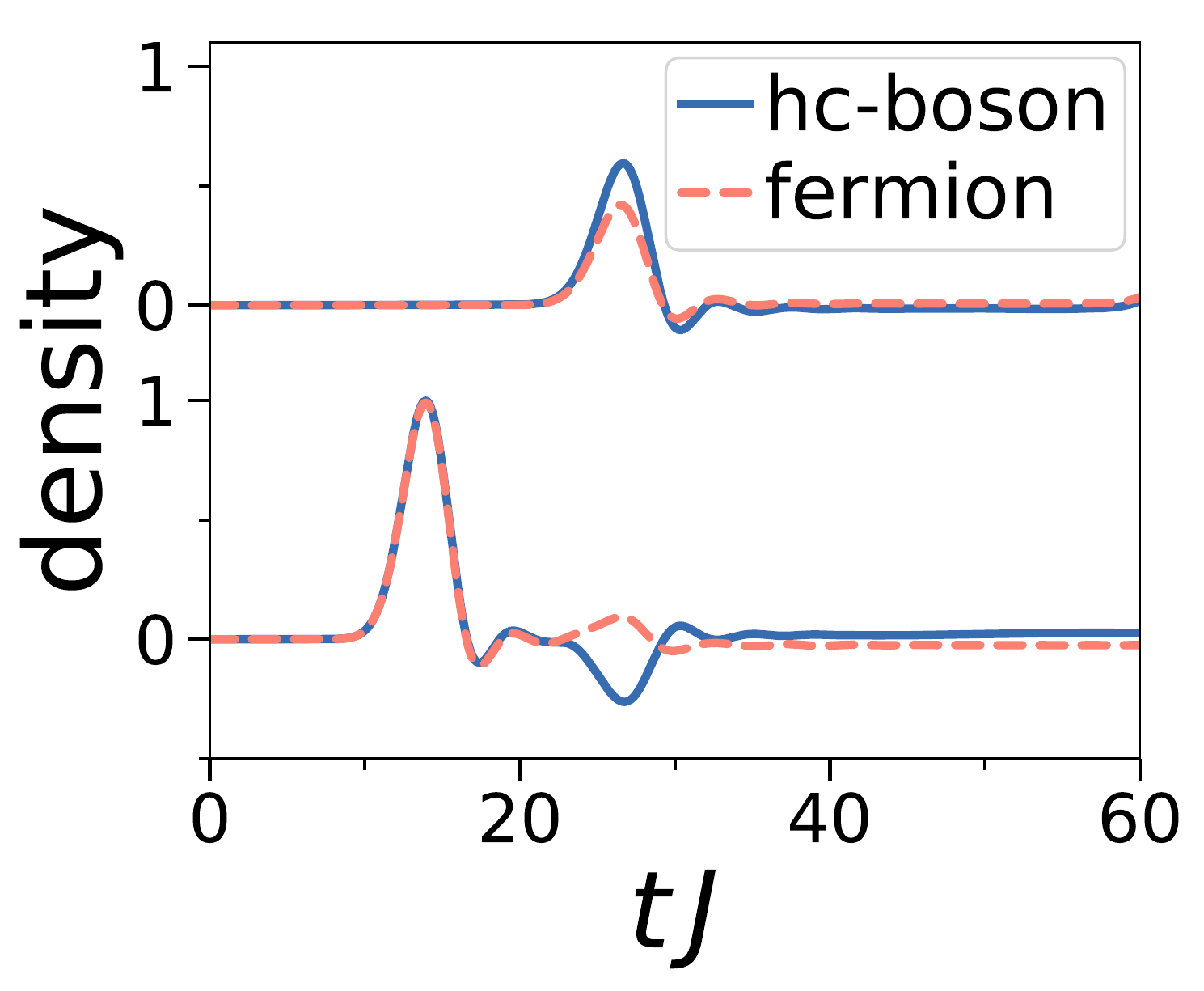}
	\subfigimg[width=0.25\textwidth]{b}{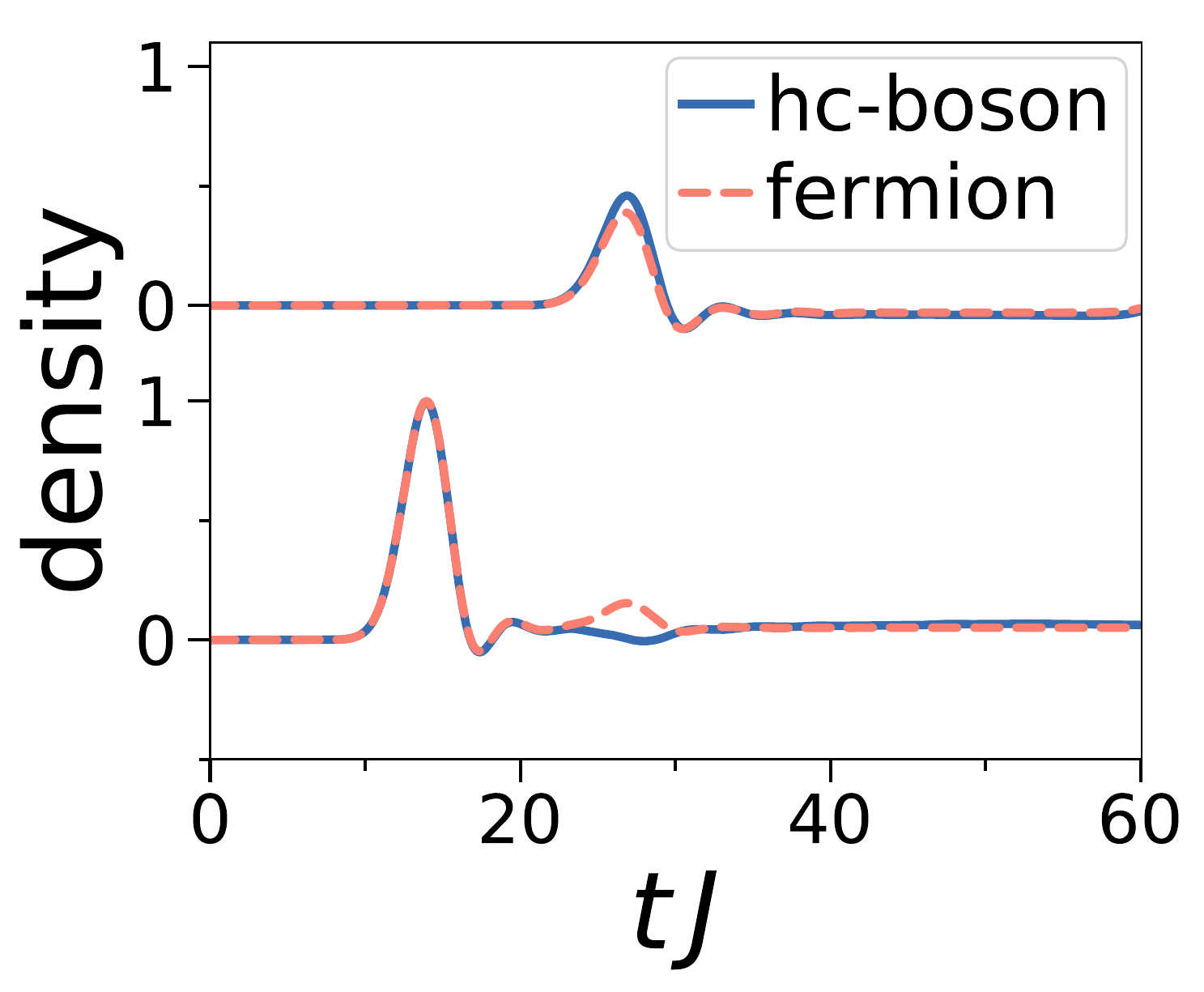}
	\subfigimg[width=0.25\textwidth]{c}{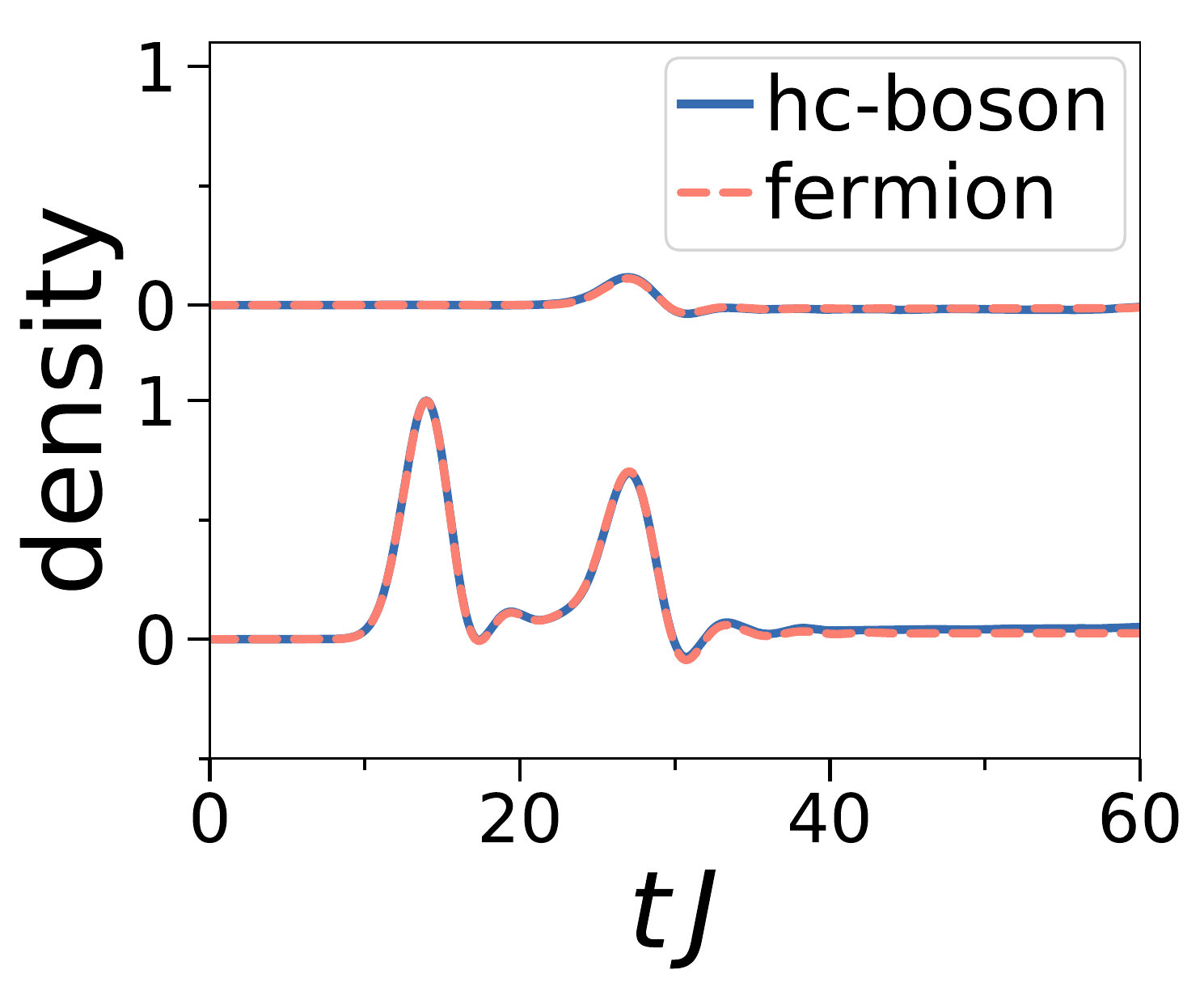}
	\caption{Comparison between hard-core bosons and spinless fermions for the propagation of a small excitation. The source lead has length ${L_\text{S}=160}$, the drain lead each ${L_\text{D}=50}$, the particle number ${N=130}$, ${d=40}$ and ${\epsilon_\text{D}=0.3}$. \idg{a-c} \densdescr{170}{175}{145}{150}. The coupling at the junction is \idg{a} ${K=1}$, \idg{b} ${K=0.5}$, \idg{c} ${K=0.2}$. 	The table below compares the transmission and reflection coefficients for hard-core bosons and fermions, calculated at ${t=31/J}$ with Eq.\ref{EqInc}-\ref{EqRef} (${t_\text{in}=15}$, ${a=30}$) . 
	}
	\label{YDensityFermionBosonComp}
		\begin{tabular}{l|*{3}{c}}
		& ${K=1}$ & ${K=0.5}$ & ${K=0.2}$ \\
		\hline
		transmission hard-core bosons& $1.332$ & $0.947$ & $0.207$\\
		transmission fermions & $1.061$ & $0.772$ & $0.199$\\
	\end{tabular}
\end{figure*}

Next, we compare the behavior of hard-core bosons with spinless fermions. In Fig.\ref{YDensityFermionBosonComp}, we plot the propagating density wave for transmission and reflection. For strong-coupling, hard-core bosons show a clear Andreev-reflection, while spinless fermions do not. \rev{For fermions we find the above mentioned background current, which is more substantial than for bosons; this is why the transmission is slightly above 1.} 
We find that for weak coupling, hard-core bosons and spinless fermions produce nearly the same result. For Y-junction, the Jordan-Wigner transformation cannot map spinless fermions and hard-core bosons since there is no notion of ordering. For weak coupling, the one-dimensional source and drain chains become disconnected, effectively restoring the mapping.

\begin{figure}[htbp]
	\centering
	\subfigimg[width=0.23\textwidth]{a}{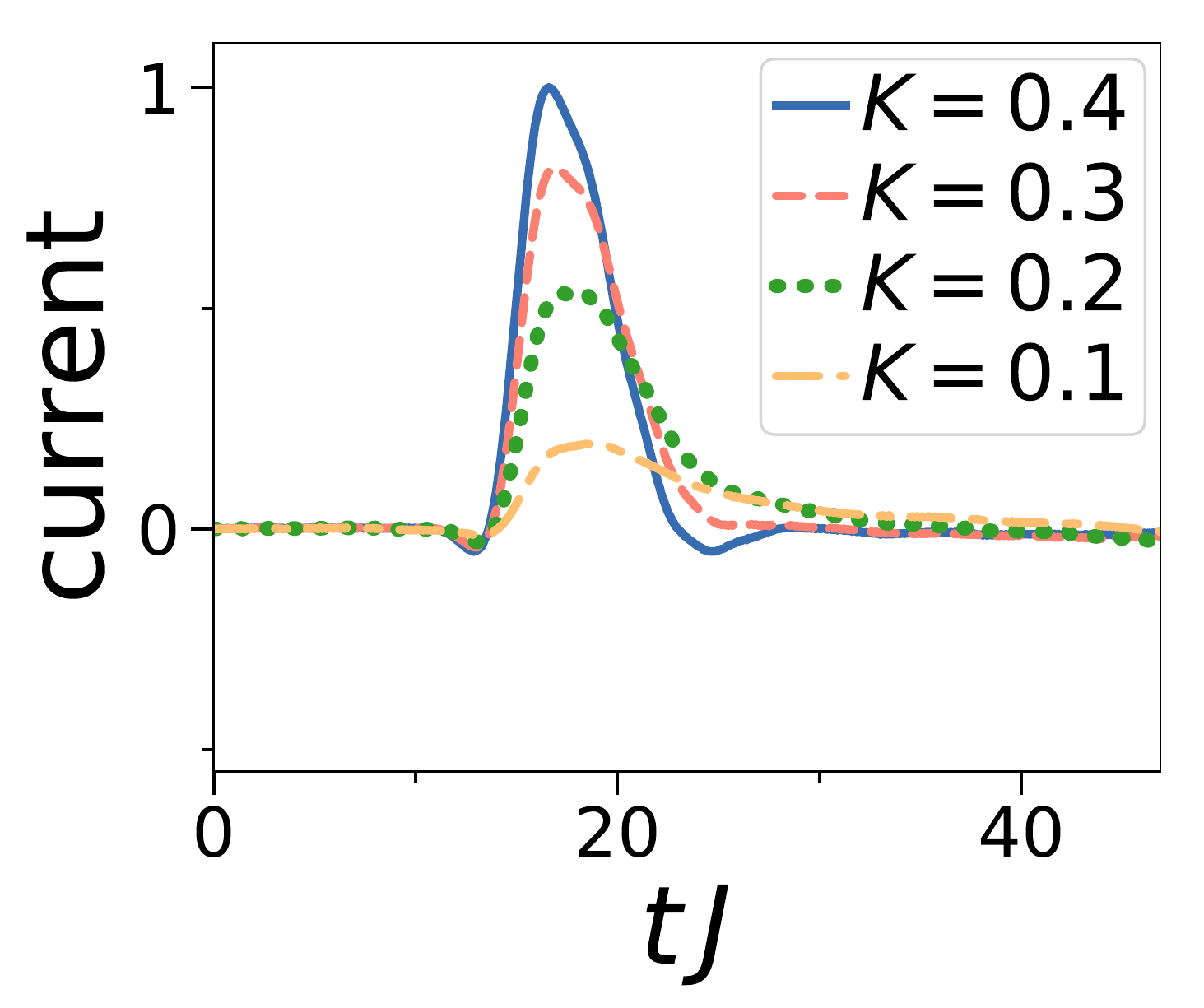}\hfill
	\subfigimg[width=0.25\textwidth]{b}{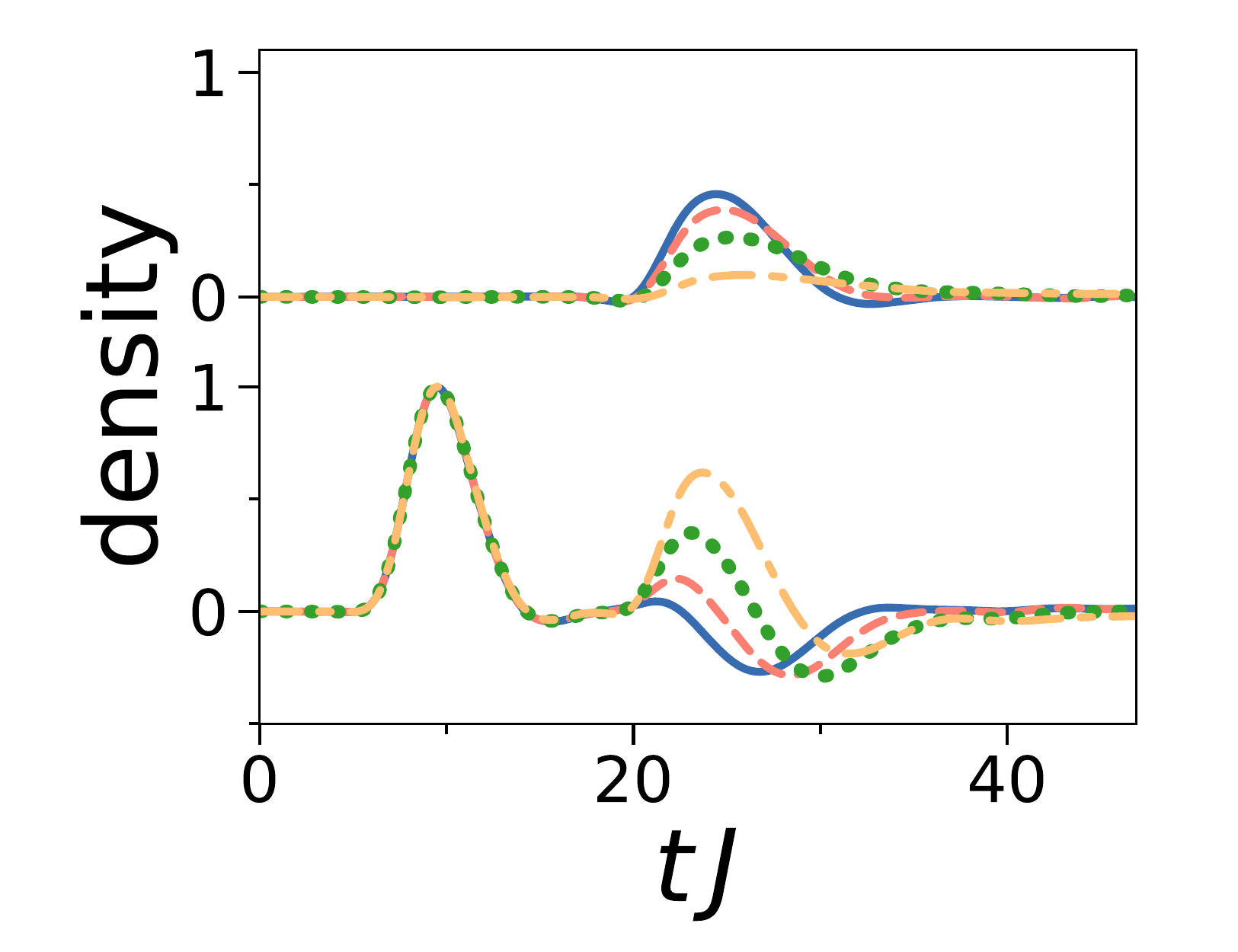}
	\caption{Propagation of small excitation in a Y-junction for the Bose-Hubbard model for ${U=5}$. The source lead has length ${L_\text{S}=80}$, the drain lead each ${L_\text{D}=40}$, the particle number ${N=80}$ and ${d=30}$. The source lead is from site 1 to 80, the first drain lead from 80 to 120, and the second one from site 120 to 160. \idg{a} Current at the junction in time. \idg{b} \densdescr{80}{85}{65}{70} For ${K=0.4}$ we observe a negative reflection (Andreev-like), ${K=0.1}$ (dots) a mostly positive reflection amplitude. In between for ${K=0.3}$ and ${K=0.2}$, the reflection changes in time from positive to negative. The maximum number of atoms per site is restricted to 4.
	}
	
	\label{YDensityHubbardsmall}
\end{figure}

{\bfseries Bose-Hubbard model.}
Next, we relax the hard-core condition, and investigate to the Bose-Hubbard model with finite $U$. Here, more than one particle is allowed per site. In Fig.\ref{YDensityHubbardsmall}, we plot the current and density in time for different junction couplings $K$. The transition from strong to weak coupling is different compared to the hard-core model. The dynamics of strong and weak behaves similarly. However, in the intermediate regime the reflection amplitude is not flat as in the hard-core limit, but has first a positive and then a negative part. 
Note that for finite $U$, the effective coupling of the leads is renormalized \cite{kane1992transport,cominotti2014optimal} with interaction. As a result, the transmission coefficient increases with interaction. 
The coupling regime is a function of both interaction and coupling $K$ (e.g. for half-filling and for finite $U$, we find that ${K=0.5}$ is the strong-coupling regime, while for hard-core bosons ${U=\infty}$, it is in the intermediate coupling regime). 

We note that our results are general for both positive and negative sign of the initial density wave $\epsilon_\text{D}$ as long as the amplitude of $\epsilon_\text{D}$ is small (the regime we considered so far). For larger values we find small differences in propagation speed (further details see supplementary materials section \ref{PosNeg}).
	
\begin{figure}[htbp]
	\centering

	\subfigimg[width=0.24\textwidth]{a}{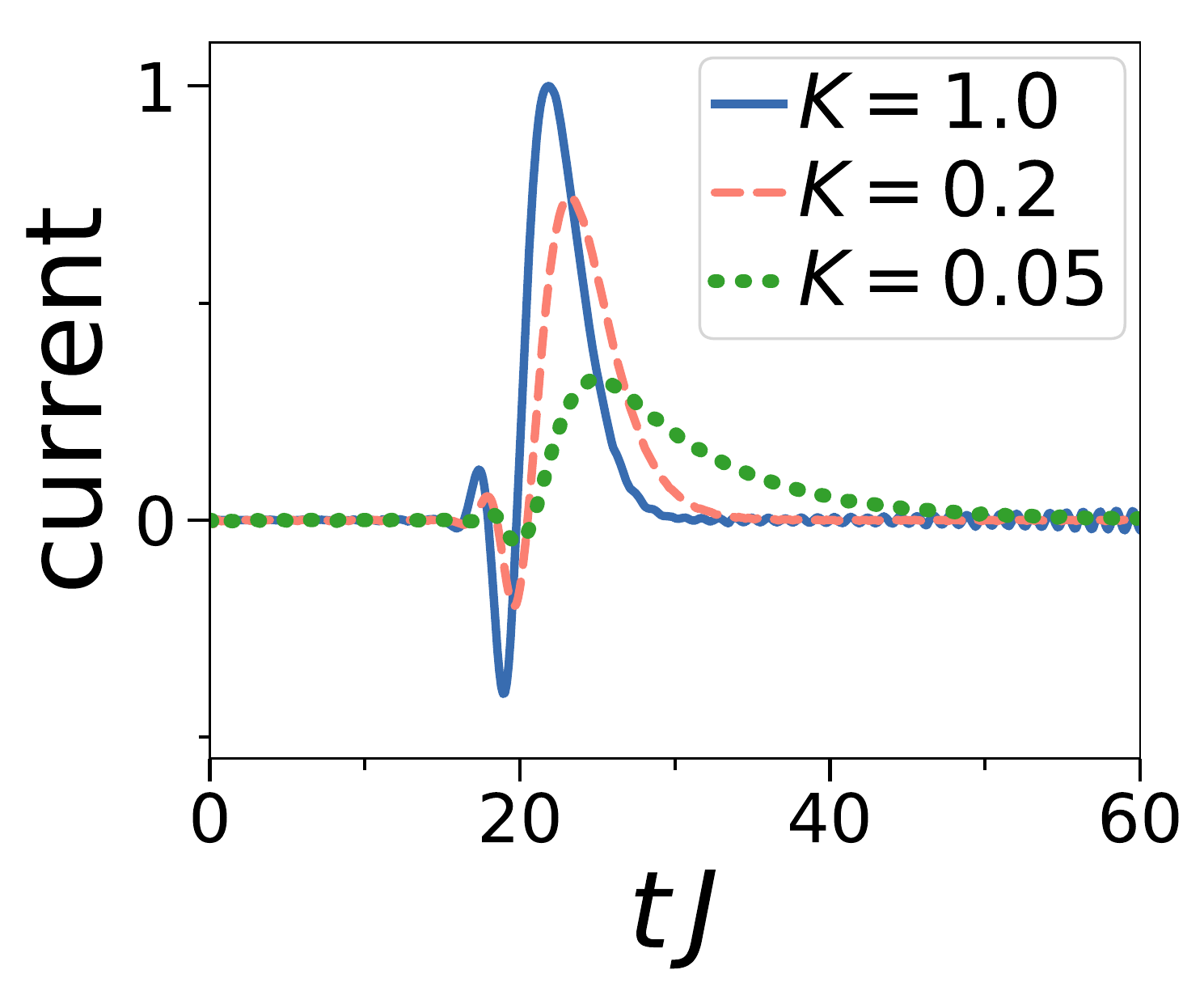}\hfill
	\subfigimg[width=0.24\textwidth]{b}{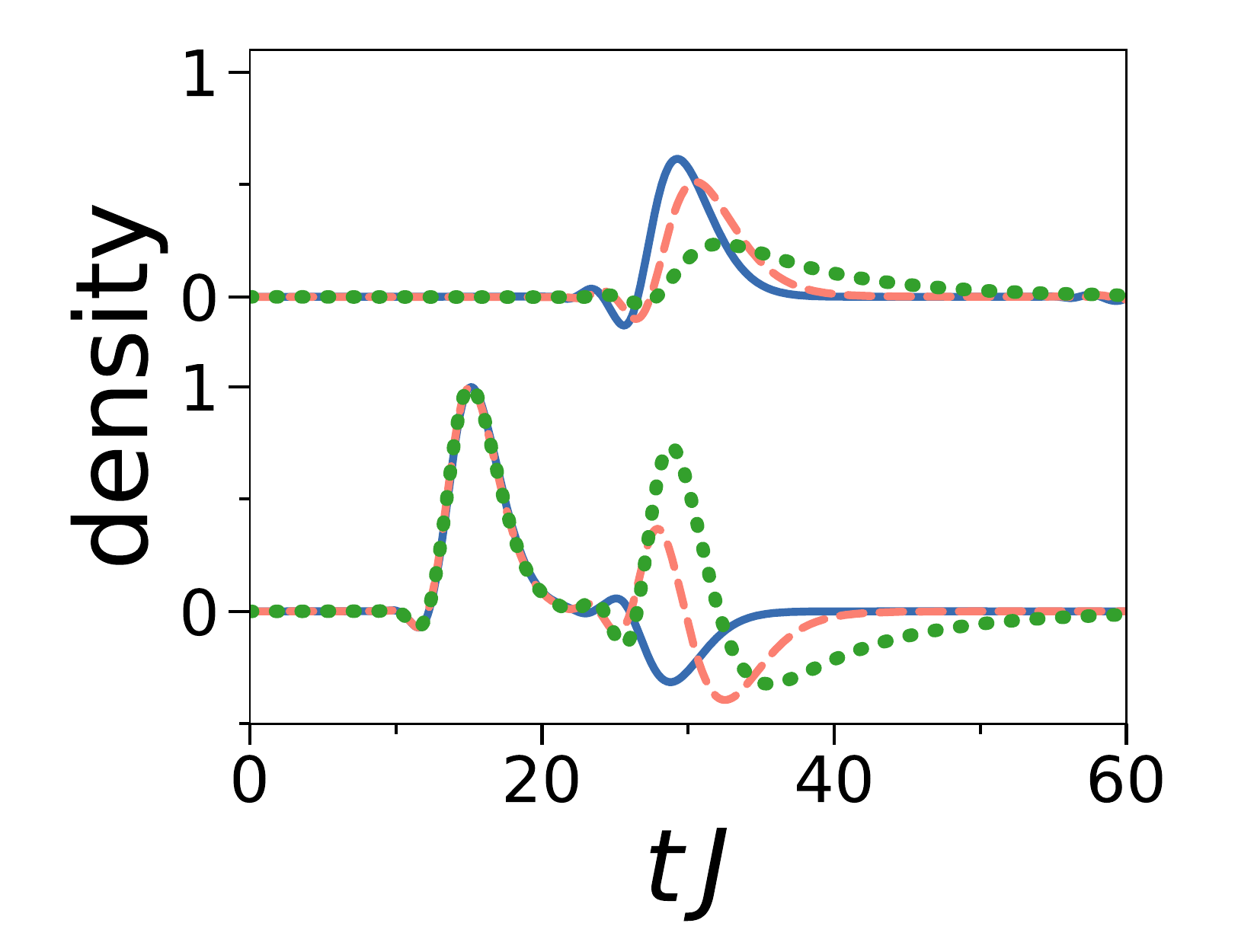}
	\caption{Propagation of small excitation in a Y-junction for the discrete Gross-Pitaevskii model (dGPE). The source lead has length ${L_\text{S}=160}$, the drain lead each ${L_\text{D}=50}$, the interaction strength ${g=360}$, initial potential offset ${\epsilon_\text{D}=\pm0.1}$ and ${d=40}$. \idg{a} the current through the junction in time \idg{b} \densdescr{170}{175}{145}{150} For ${K=1}$ (solid) we observe a negative reflection (Andreev), ${K=0.2}$ (dashed) reflection with both positive and negative contribution, ${K=0.05}$ (dots) a large positive reflection amplitude. Note that the interaction $g$ renormalizes the coupling $K$ between the chains, and the type of reflection changes depending on $g$. The velocity of the excitation also depends on $g$.
		The table below shows the transmission and reflection coefficients, calculated at ${t=60/J}$ with Eq.\ref{EqInc}-\ref{EqRef} (${t_\text{in}=15}$, ${a=30}$).
	}
	\begin{tabular}{l|*{3}{c}}
		& ${K=1}$ & ${K=0.2}$ & ${K=0.05}$ \\
		\hline
		transmission & $1.331$ & $1.33$ & $1.317$   \\
		reflection   & $-0.331$ & $-0.33$ & $-0.317$ 
	\end{tabular}
	\label{YDensitysmallGPE}
\end{figure}

{\bfseries Gross-Pitaevskii model.}
Now, we turn our attention towards the weakly-interacting regime with many atoms. This limit is described the dGPE. The results are plotted in Fig.\ref{YDensitysmallGPE}.
For the dGPE and strong coupling, we find similar dynamics with Andreev-reflection as in the Bose-Hubbard model. 
For small $K$ we find an oscillating behavior of the intermediate coupling regime: The reflection is a density wave with initial positive and then negative amplitude density. For smaller couplings $K$, the initial positive reflecting part of the wave gets larger and the negative part gets a smaller amplitude, however it becomes very broad in time. 
Even for very small $K$, the reflected wave shows this oscillating behavior of the intermediate regime. Only at exactly ${K=0}$ we find a purely positive reflection which characterizes the weak-coupling regime. Thus, we conclude there is no weakly coupled regime in the dGPE limit.

We find for the dGPE that the transmission and reflection coefficient is independent of the coupling $K$ as shown in the table below Fig.\ref{YDensitysmallGPE}. The values for any $K$ corresponds to the ones we observed for strong coupling in the Bose-Hubbard model. Our numerical values match the analytic result of Eq.\ref{transmission} (-1/3 reflection, 4/3 transmission). This is in stark contrast to the hard-core and finite $U$ Bose-Hubbard model, where total transmitted and reflected density depends on the junction coupling $K$. %\rev{It has been shown that for repulsive bosons in a 1D system  the conductance is independent of barriers and re-normalizes to the result of zero barrier height\cite{kane1992transport} (repulsive bosons correspond to the case of attractive fermions within Luttinger liquid theory). The dGPE total transmission indeed matches this prediction for Y-junctions (total transmission is always the same value as if there is no barrier). However, we note that for the Bose-Hubbard model we find a dependence on total transmission with $K$.}

\begin{figure*}[htbp]
	\centering
	%\subfigure{\includegraphics[width=0.24\textwidth]{{chiralcurrentFluxN1DTwistDataN10000}.pdf}}\hfill
	%\subfigure{\includegraphics[width=0.24\textwidth]{{chiralcurrentTwistN1DTwistN10000}.pdf}}
	\subfigimgraised[width=0.3\textwidth]{a}{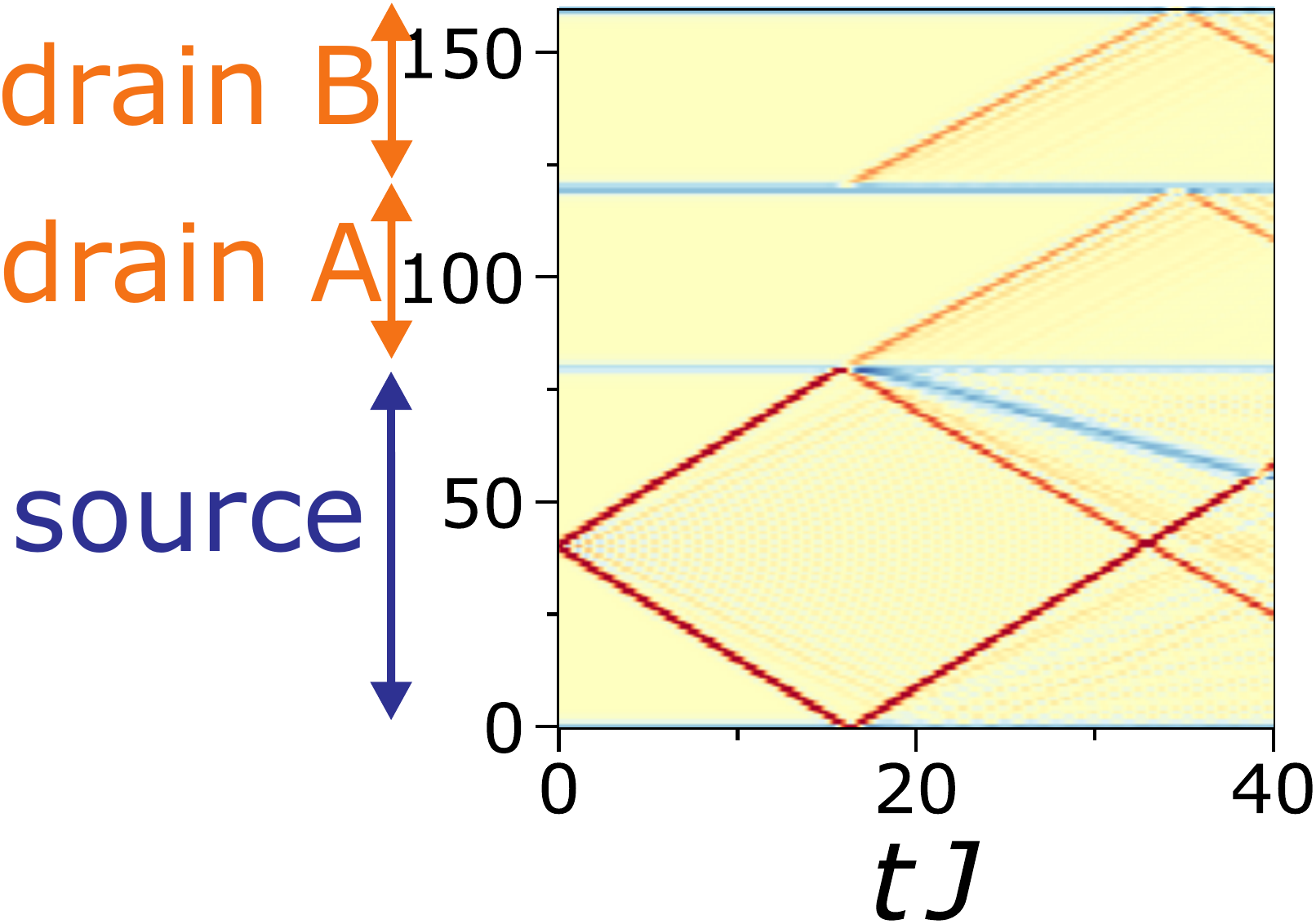}
	\subfigimg[width=0.26\textwidth]{b}{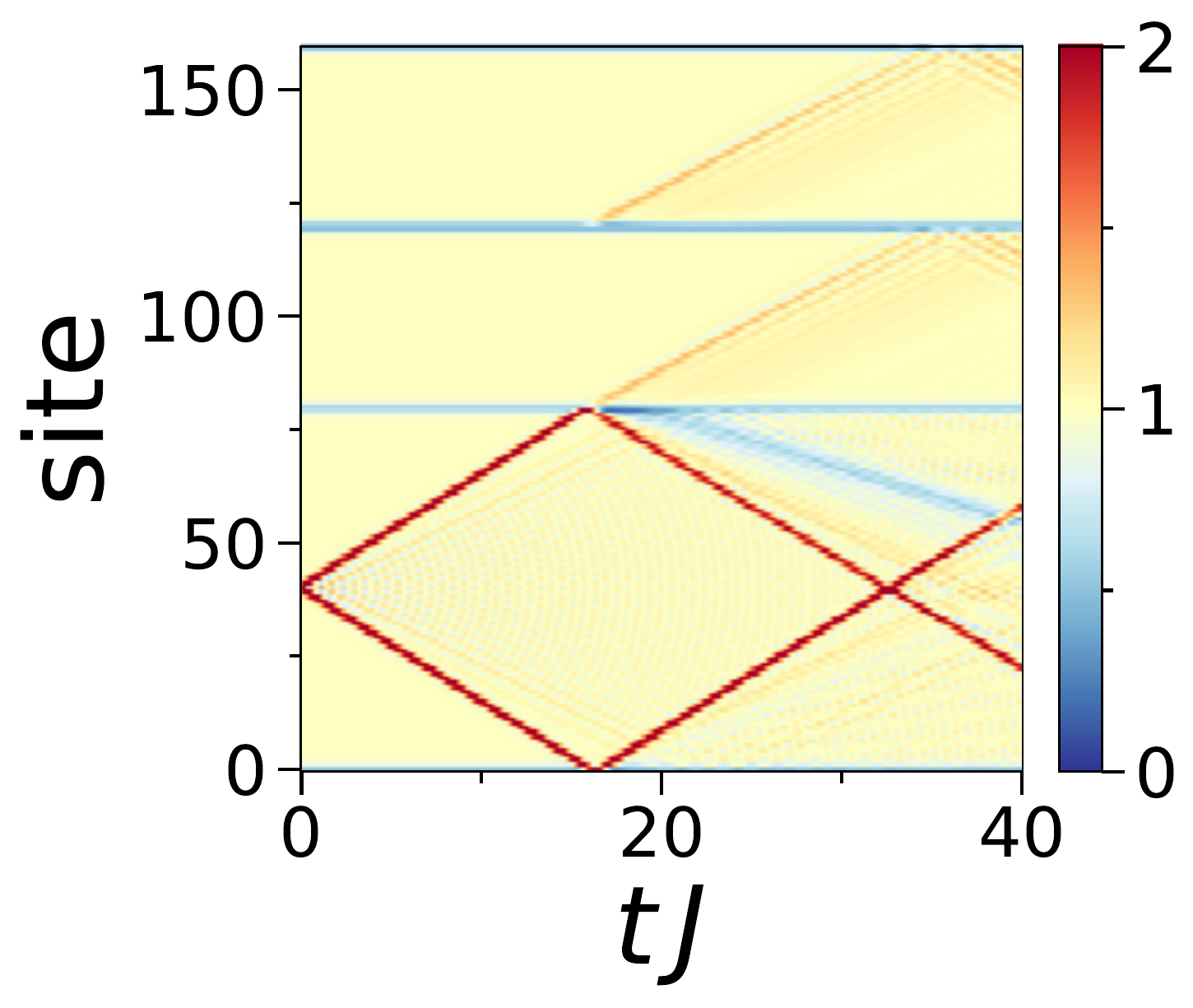}
	\subfigimg[width=0.26\textwidth]{c}{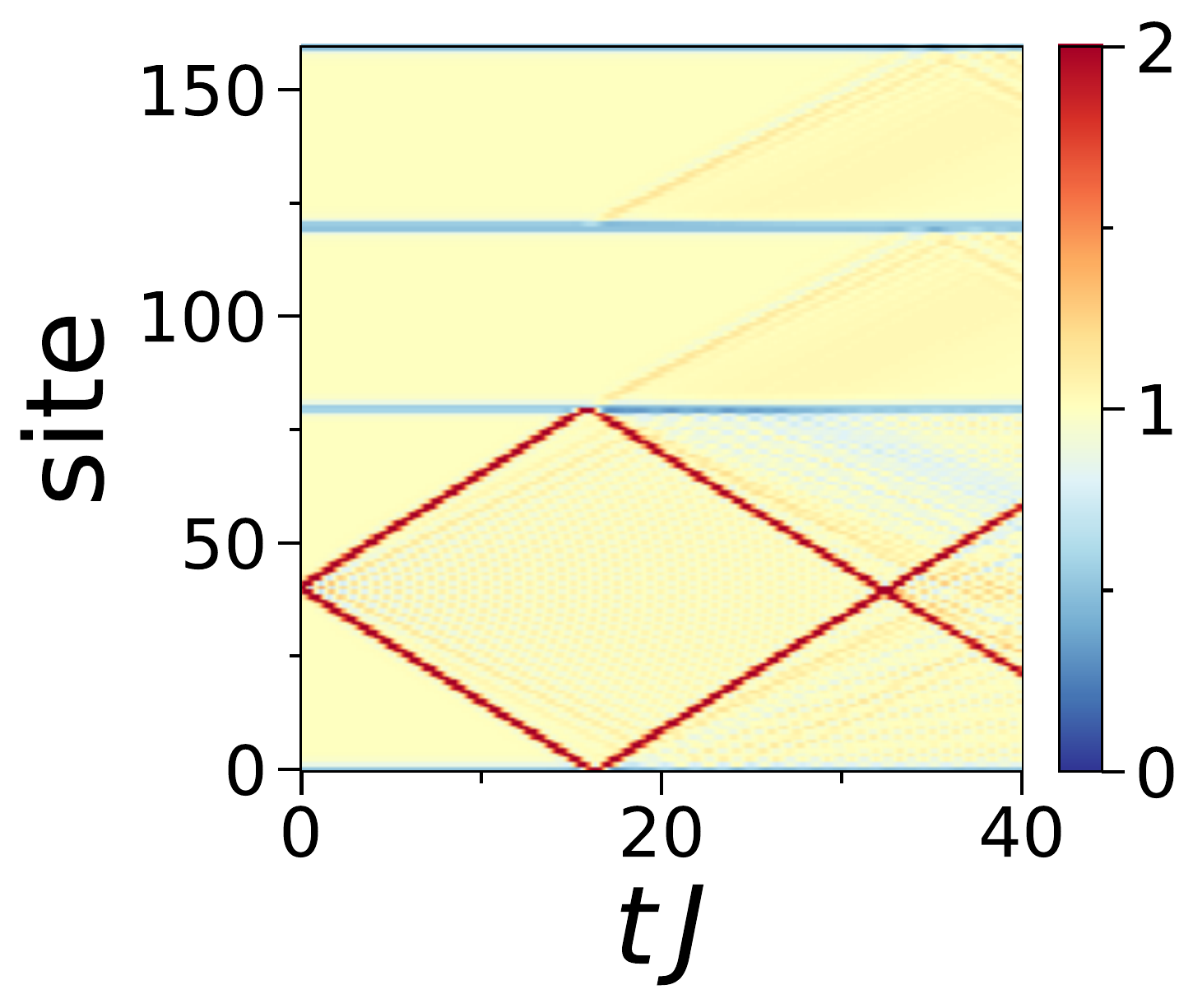}

	\caption{Propagation of large positive excitation in a Y-junction for the discrete Gross-Pitaevskii model (dGPE). The source lead has length ${L_\text{S}=80}$, the drain lead each ${L_\text{D}=40}$, interaction strength ${g=220}$ and ${d=40}$. \idg{a} ${K=0.3}$, \idg{b} ${K=0.2}$, \idg{c} ${K=0.1}$. The initial potential offset is ${\epsilon_\text{D}=5}$, the width ${\sigma=0.01}$.}
	\label{YDensitybigPosGPE}
\end{figure*}

Next, we study excitations with large amplitudes using the dGPE. In Fig.\ref{YDensitybigPosGPE}, we plot a large positive excitation for different coupling strengths. We find that there are two reflections, one positive and one negative. The negative reflection moves slower than the positive one. The ratio between positive and negative reflection changes with coupling strength. For strong coupling, the negative dominates, while for weak coupling the positive reflection is larger. The positive reflection is reflected when the incoming wave arrives (Fig.\ref{YDensitybigPosGPE}a). However, the negative reflection is stuck in the junction for a while and is delayed, before it is reflected back (Fig.\ref{YDensitybigPosGPE}b). For very small $K$, the negative reflection becomes broader, and also the reflection is delayed more (Fig.\ref{YDensitybigPosGPE}c). We find similar dynamics for other values of interaction $g$.

\begin{figure}[htbp]
	\centering
	\subfigimgraised[width=0.24\textwidth]{a}{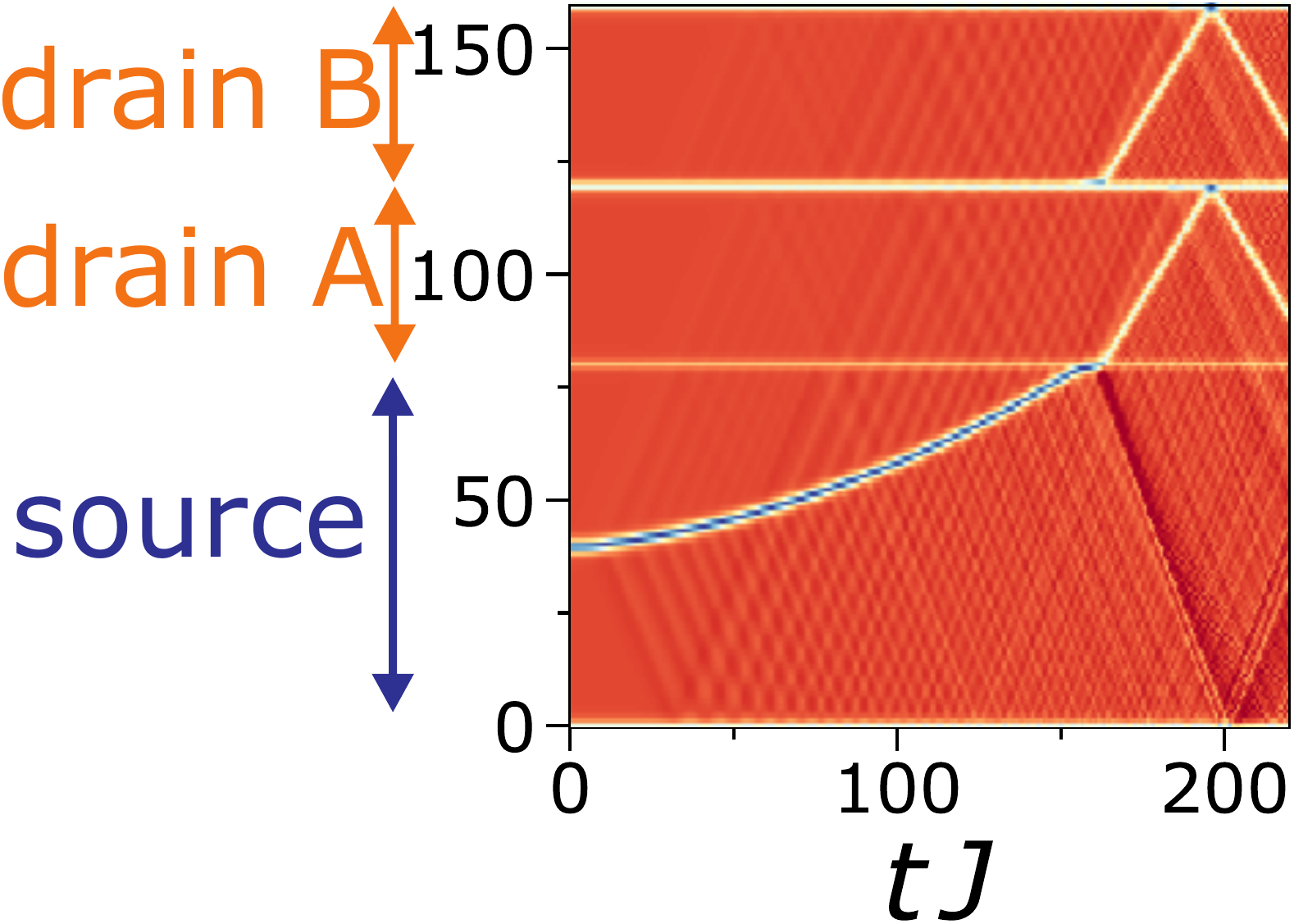}\hfill
	\subfigimg[width=0.24\textwidth]{b}{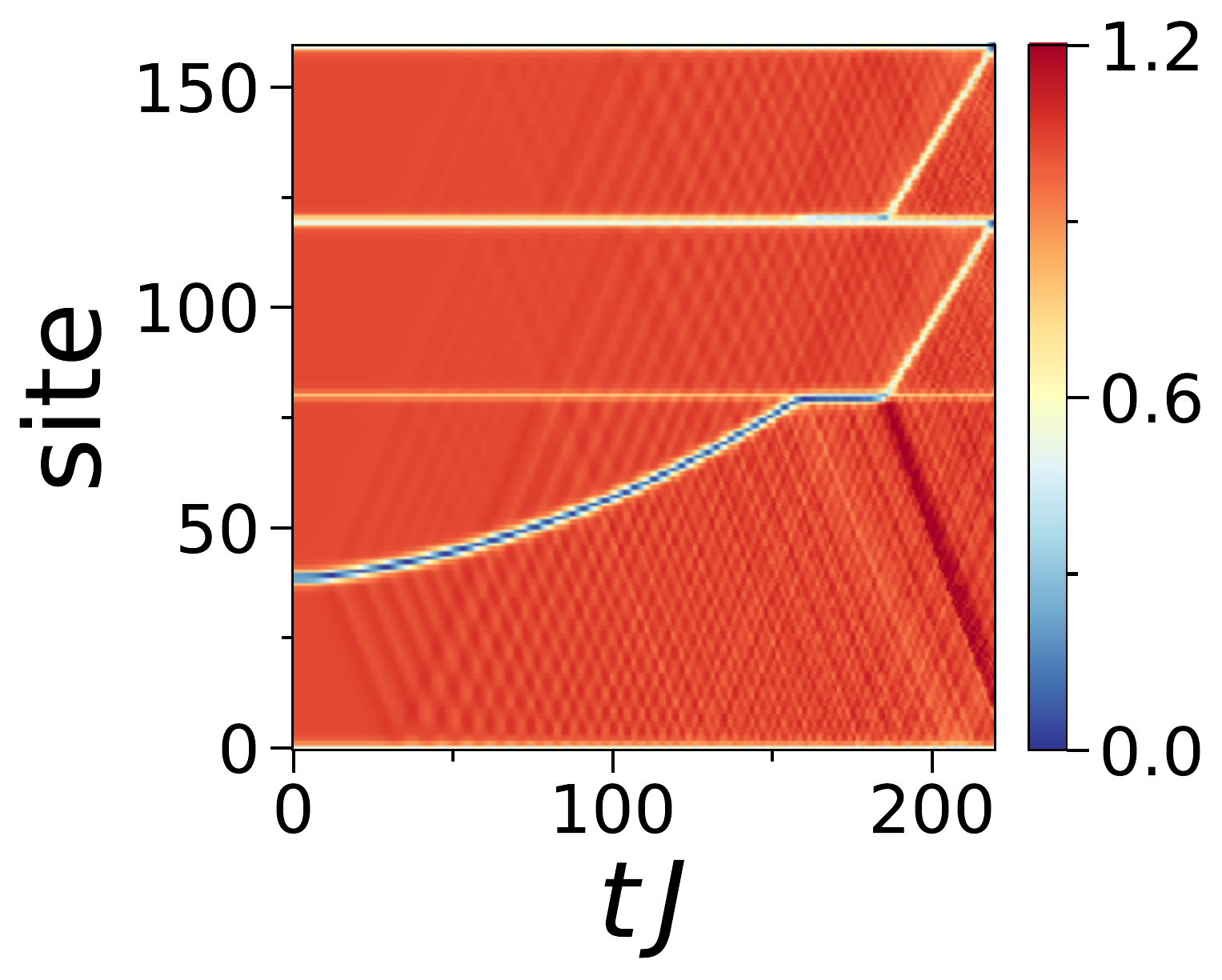}\\
	\subfigimg[width=0.24\textwidth]{c}{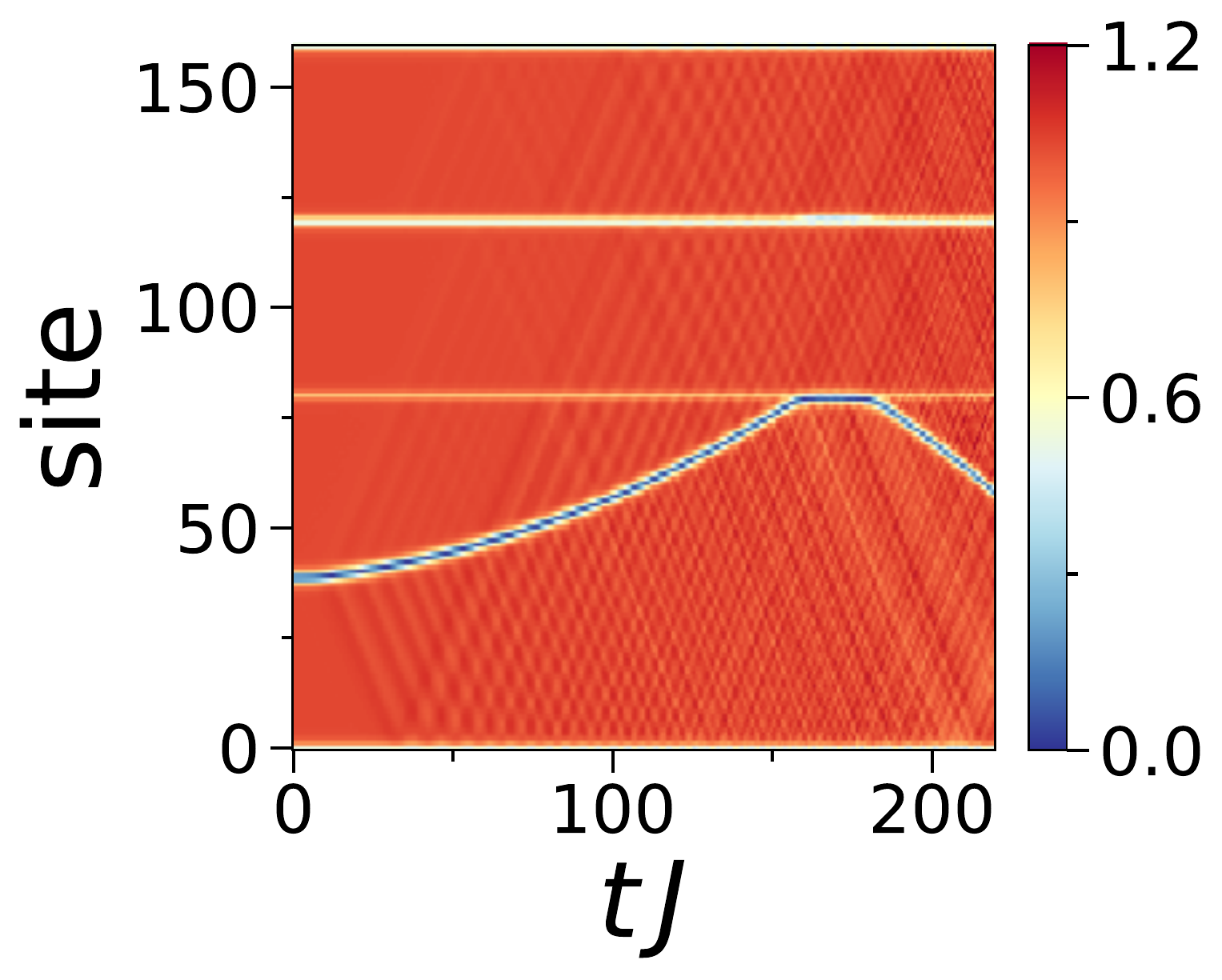}\hfill
	\subfigimg[width=0.24\textwidth]{d}{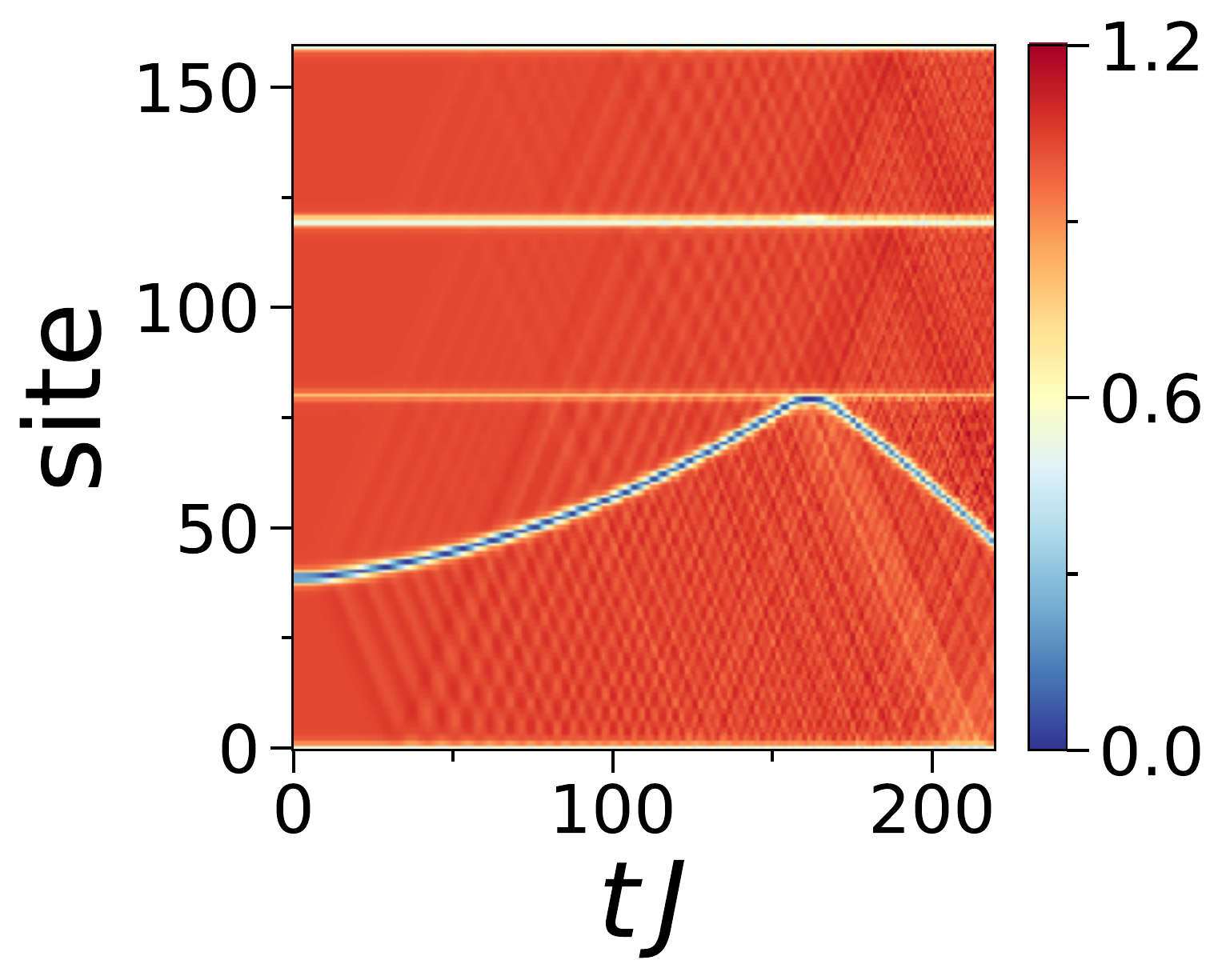}
	\caption{Propagation of large negative excitation in a Y-junction for the discrete Gross-Pitaevskii model (dGPE). The source lead has length ${L_\text{S}=80}$, the drain lead each ${L_\text{D}=40}$, the interaction strength ${g=220}$ and ${d=40}$. The coupling at the junction (site 80) \idg{a} ${K=0.47}$ \idg{b} ${K=0.46055}$, \idg{c} ${K=0.46054}$, \idg{d} ${K=0.45}$. The initial potential offset is ${\epsilon_\text{D}=-0.01}$, the width ${\sigma=0.01}$. \rev{To generate a gray soliton we calculate the ground state of the system, then apply a phase shift of $\pi$ at site 40.}}
	\label{YDensitybigNegGPE}
\end{figure}

In Fig.\ref{YDensitybigNegGPE}, we study large negative excitations with the dGPE. We use an additional step to generate this excitation. We apply a phase shift of $\Pi$ across the excitation. This will generate a gray soliton. The soliton is not completely stable, we observe that its speed increases over time as its density depression decreases. 
The transmission and reflection depends on coupling $K$. We find a critical $K$, below which the gray soliton is totally reflected. Above, the soliton is transmitted. For the transmitting case, we also observe a (positive) Andreev-like reflection. The behavior changes abruptly. Close to the critical coupling strength, the soliton has a residing time in the junction, before being either transmitted or reflected (see Fig.\ref{YDensitybigNegGPE}b,c). 

\begin{figure}[htbp]
	\centering
	\subfigimg[width=0.3\textwidth]{a}{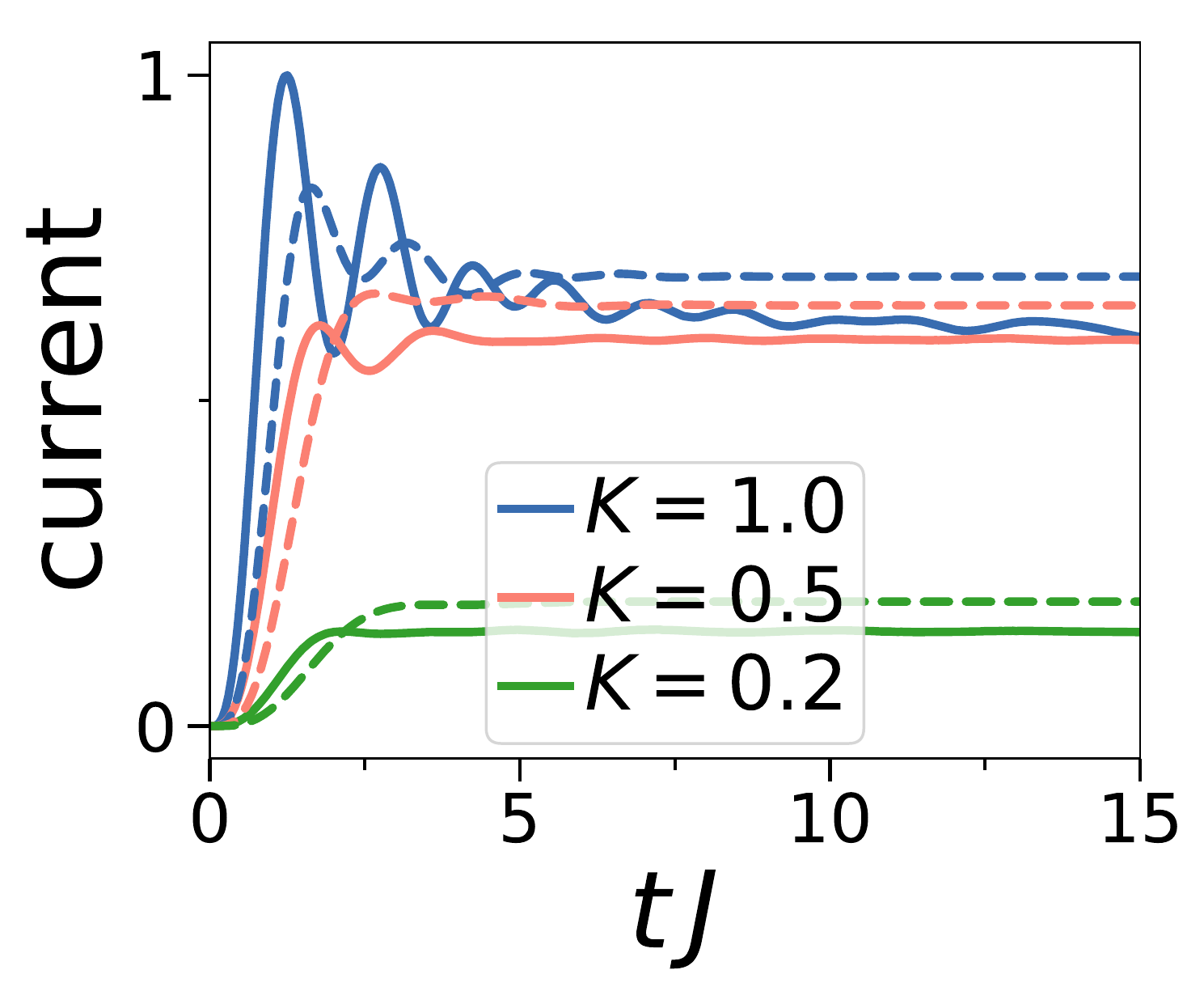}
	\caption{Quench dynamics in the Y-junction. Hard-core bosons are initially prepared in the ground state of the source lead of a Y-junction. Comparison of the expectation value of the current  through the junction for different coupling strengths $K$ with two different methods: Closed system with full modeling of junction and lead (solid) and open system with leads modeled with Markovian bath (dashed). For DMRG, the source (drain) lead has length ${L_\text{S}=31}$ (${L_\text{D}=30}$), the particle number ${N=15}$. The atoms are prepared in the ground state of the uncoupled source (the first 30 sites), then the coupling is switched on instantaneously at ${t=0}$. For the open system approach, the source and drain leads consist of 2 site each. The Lindblad operator couple to the first source site. The Lindblad parameters are ${\Gamma=1.5}$ and ${r=0.65}$. The graph shows different values of coupling $K$.}
	\label{YCurrentComparison}
\end{figure}

{\bfseries Quench of Y-junction}
In this subsection, we study a non-equilibrium quench where the atoms are initially loaded into the source, and then \rev{the atoms are suddenly released into the two drains of the Y-junction}. We calculate the ground state of atoms in the source, without coupling to the rest of the system. Then, we suddenly switch on the coupling $K$ at ${t=0}$. 
We use two different approaches: First, we simulate the full Y-junction including leads through DMRG. In the second approach, we trace out the leads, and simulate the junction itself as open system coupled to a atom reservoir, driving the junction. The open system equations are presented in the supplemental material Sec.\ref{opensystem}. 
We investigate the expectation value of the current through the junction (Eq.\ref{Ycurrent}) for different junction couplings $K$ for the hard-core Bose-Hubbard model. The current for both numerical methods is plotted in Fig.\ref{YCurrentComparison}. Both methods produce similar dynamics and steady-states.
We find that the steady-state current does not change significantly by changing the junction coupling for ${K\ge0.5}$. However, we find that the oscillations in the dynamics are much smaller for ${K=0.5}$ compared to ${K=1}$.  
\rev{We conjecture the following: After atoms are injected from the source into the drain, they can tunnel back with a certain probability. This probability increases with the tunneling rate $K$. This causes the oscillating dynamics. For weaker junction couplings $K$ the back-propagation is suppressed (since atoms simply propagate further into the drain) and the oscillations do not appear. %For smaller ${K=0.2}$, both current and oscillations diminish greatly.
}

\section{Propagation of excitations through a ring}\label{RingSimple}
As a next step, we study the propagation through a ring-lead system.

{\bfseries Linearized equations--}
First, we study the linearized equation Eq.\ref{LinEq} of the cGPE again, this time for a ring with magnetic field. When the magnetic field $A$ and  the velocity $v_0$ coincide ${A=v_0}$, this is the wave equation, describing the dispersion-less propagation with the speed of sound ${c=\sqrt{gn_0}}$. With ${A\ne v_0}$, the propagation of an excitation in forward and backward direction becomes asymmetric. For a generic wave ${\delta n=f(t\pm\frac{x}{c})}$, we find that for ${\abs{A-v_0}\ll \sqrt{gn_0}}$, the propagation velocity 
\begin{equation}\label{EqFluxVelocity}
c_\pm=\sqrt{gn_0}\pm (A-v_0)\;.
\end{equation}
For larger $(A-v_0)$, an initial excitation will move in forward and backward direction with asymmetric amplitude as well as different velocity.
The quantity $(A-v_0)\propto j$ is proportional to the persistent current $j$ of the condensate. Thus, the difference in excitation velocity of forward and backward direction can be used to measure the persistent current of the condensate\cite{StringariDoppler2016}.
When two wavefunctions are added, we take the absolute square to get the density. The complex phase of the added wavefunctions can give rise to constructive or destructive interference. The flux in a closed loop influences the complex phase, giving rise to the Aharonov-Bohm effect.
The linearized equation for the density does not have this property. Two real-valued density and velocity excitations are simply added up. There is no complex phase that is influenced by flux. Therefore, we conclude that for small excitations traveling on top of a condensate there is no Aharonov-Bohm effect with interference. 
The effective magnetic field only influences the propagation velocity, however it does not cause any interference pattern. 
However, beyond the small excitation approximation, the dynamics of an isolated wavepacket shows the Aharonov-Bohm effect.

\begin{figure}[htbp]
	\centering
	\subfigimg[width=0.24\textwidth]{a}{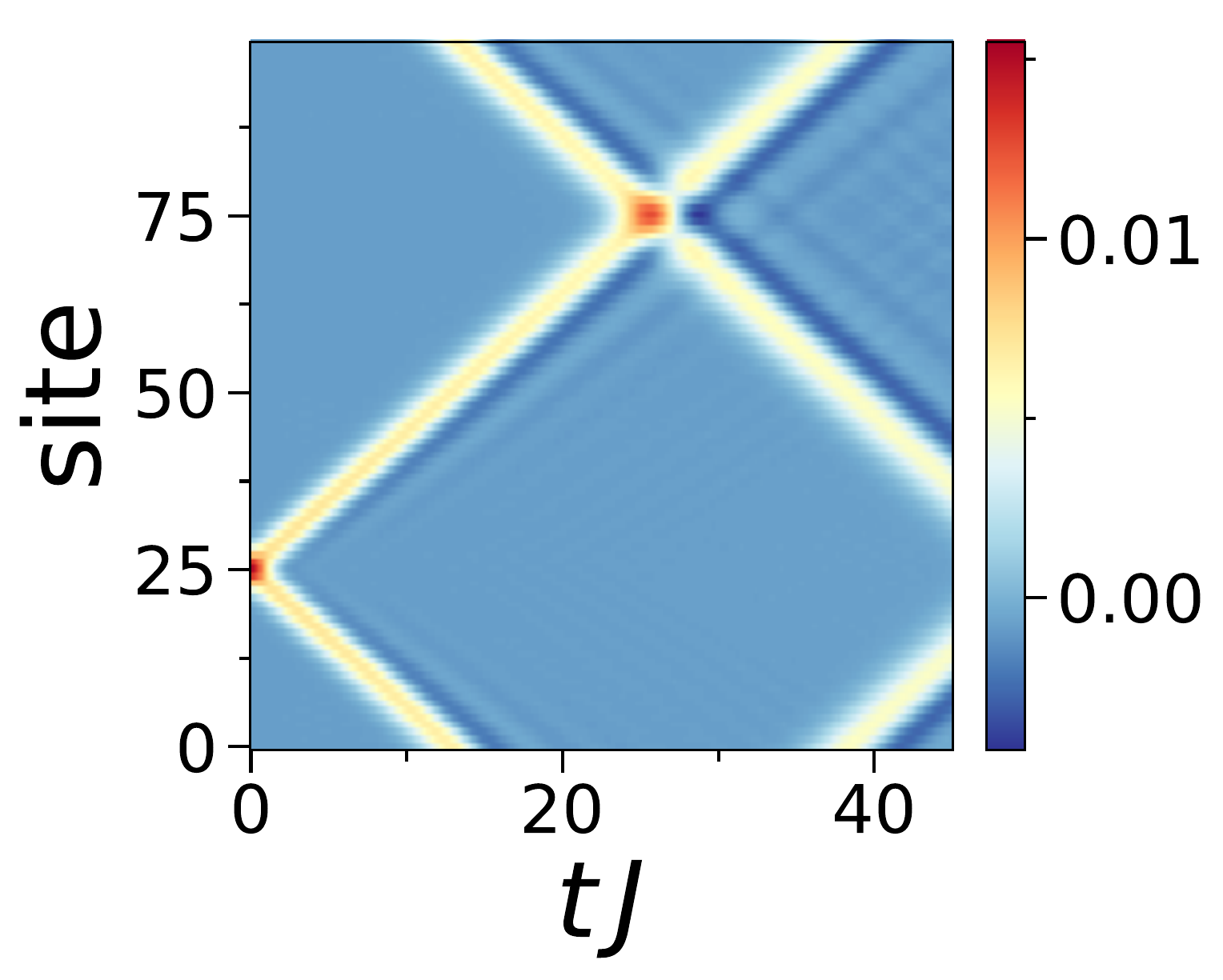}\hfill
	\subfigimg[width=0.24\textwidth]{b}{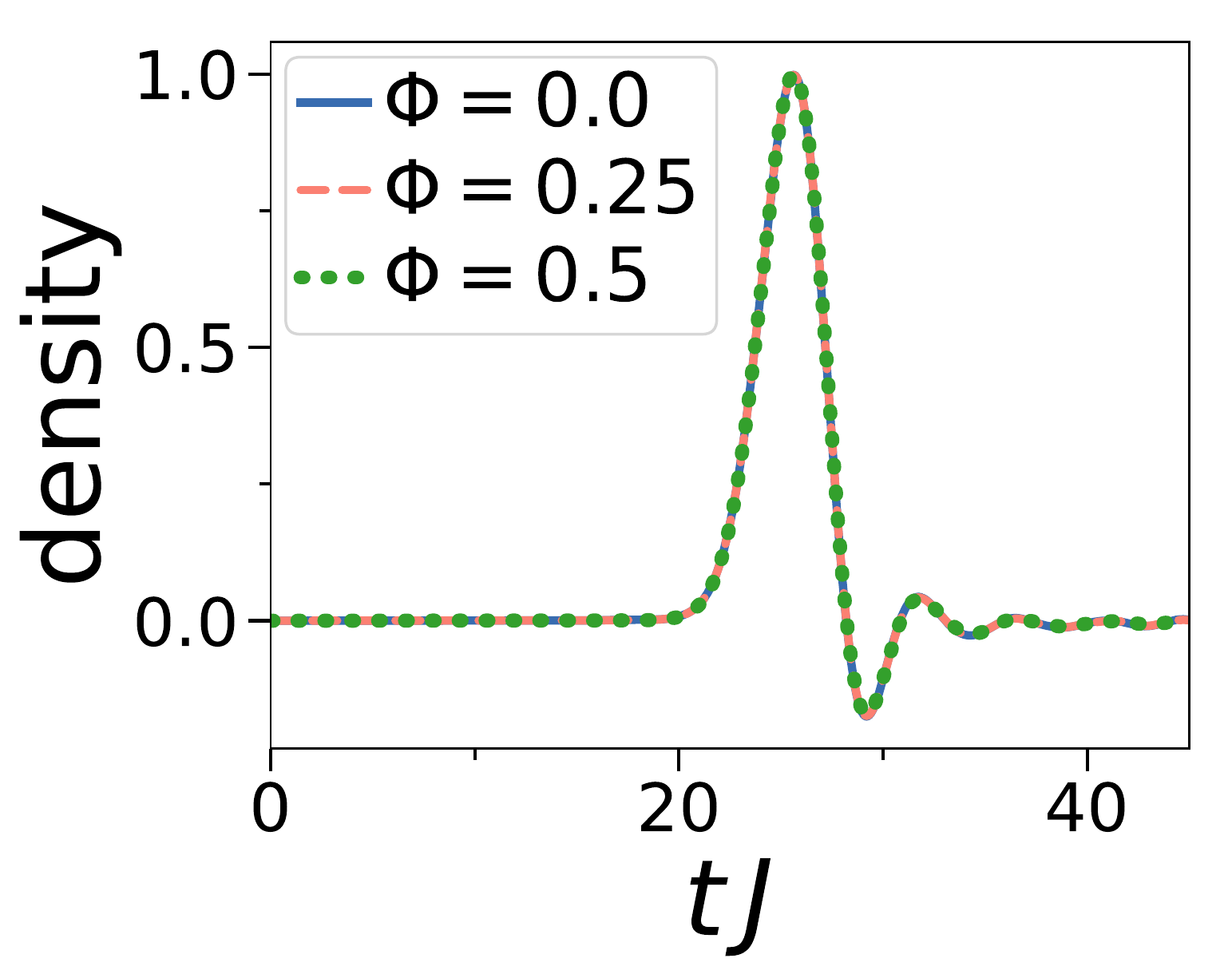}
	\caption{Propagation of a small excitation in a ring without leads with hard-core boson ring or spinless fermions. In this simple configuration, hard-core bosons and spinless fermions can be mapped onto each other. The dynamics are independent of flux and yield the same graph. \idg{a} density in ring for any flux over time. In particular, there is always constructive interference when left and right-moving excitations meet at around site 75. \idg{b} density is averaged between sites 73 and 77 and plotted against time $t$. The ring has length ${L_\text{R}=100}$ with $N_\text{p}=50$ atoms.  The initial potential offset is ${\epsilon_\text{D}=0.1}$ at site 25, with width ${\sigma=2}$. }
	\label{RingOnlyHardcore}
\end{figure}

{\bfseries Bose Hubbard dynamics in isolated ring circuit--}
First, we study the propagation of an excitation in a single ring without leads. The small excitation is created locally by a localized Gaussian potential in the ring, which is switched off at ${t=0}$. The density excitation splits into two, and propagates in left and right winding direction around the ring. When these two waves meet at the opposite side of the ring, they interfere constructively. We find for hard-core bosons and spinless fermions in a ring without leads (they can be mapped exactly onto each other for odd number of particles, and for even number of particles the flux is transformed by ${\Phi\rightarrow\Phi+1/2}$) that the propagation of a small excitation is independent of flux, and they {\it always interfere constructively and there is no Aharonov-Bohm effect} (see Fig.\ref{RingOnlyHardcore}). 

Next, we plot the propagation of an excitation in a ring for the Bose-Hubbard model with finite interaction. Here, we find a slight flux dependence. Close to half-flux, the propagating excitation is dispersing more than without flux. This dispersion effect increases with smaller ring-size. This flux dependence is a finite-size effect of the ring, and vanishes in the limit of many ring sites since boundary effects (like for example AB flux) do not matter in the thermodynamic limit.
\begin{figure}[htbp]
	\centering
	\subfigimg[width=0.24\textwidth]{a}{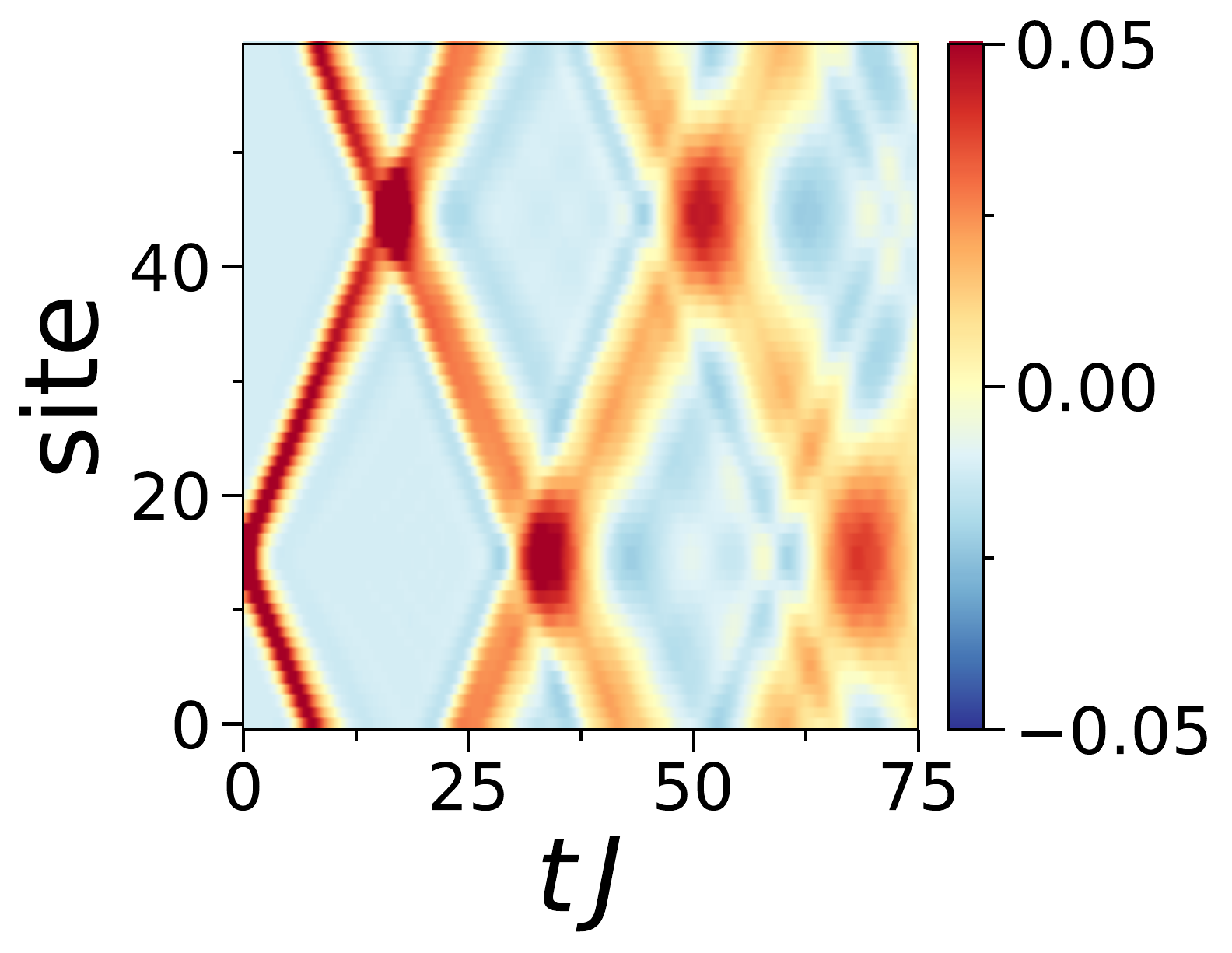}\hfill
	\subfigimg[width=0.24\textwidth]{b}{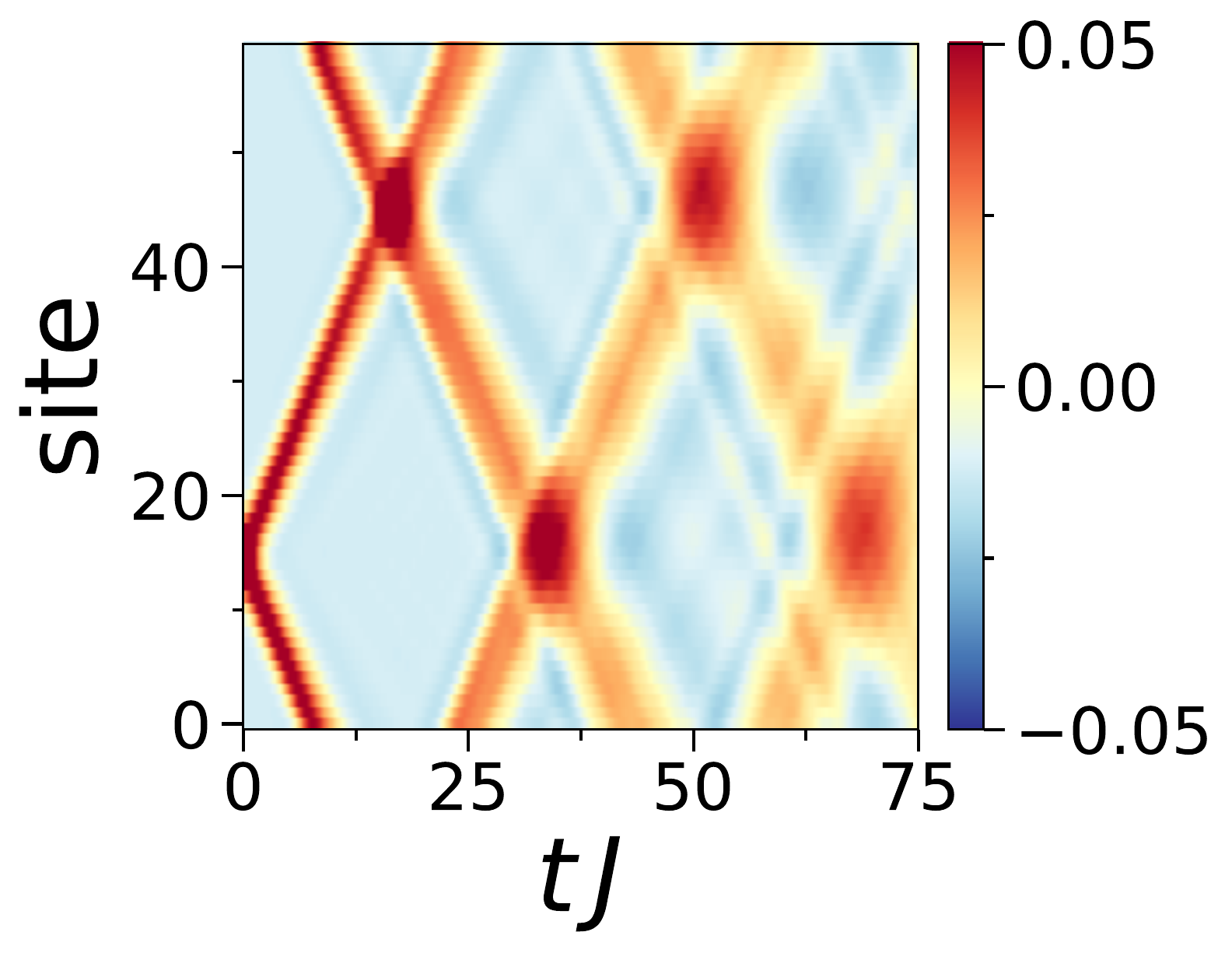}\\
	\subfigimg[width=0.24\textwidth]{c}{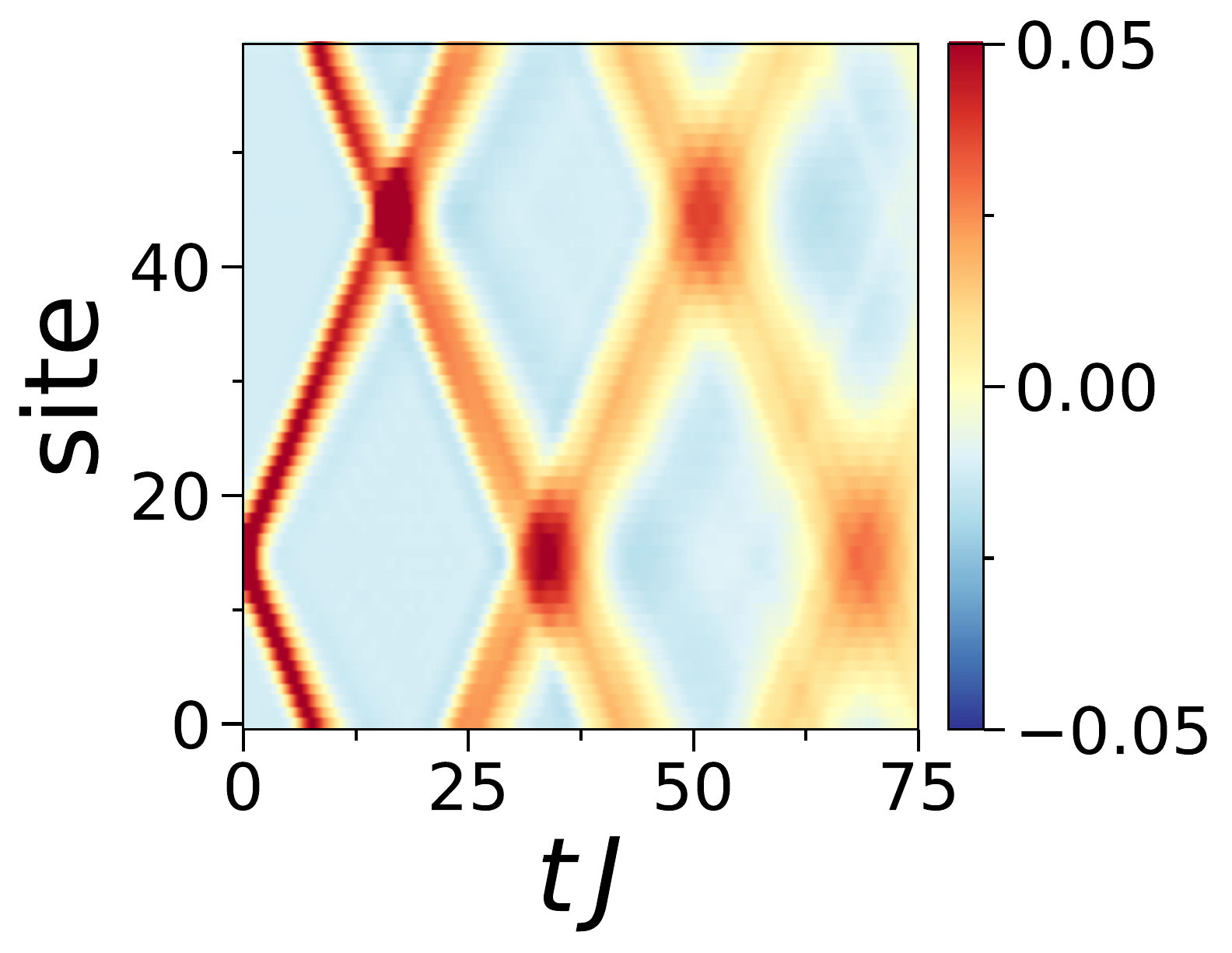}\hfill
	\subfigimg[width=0.24\textwidth]{d}{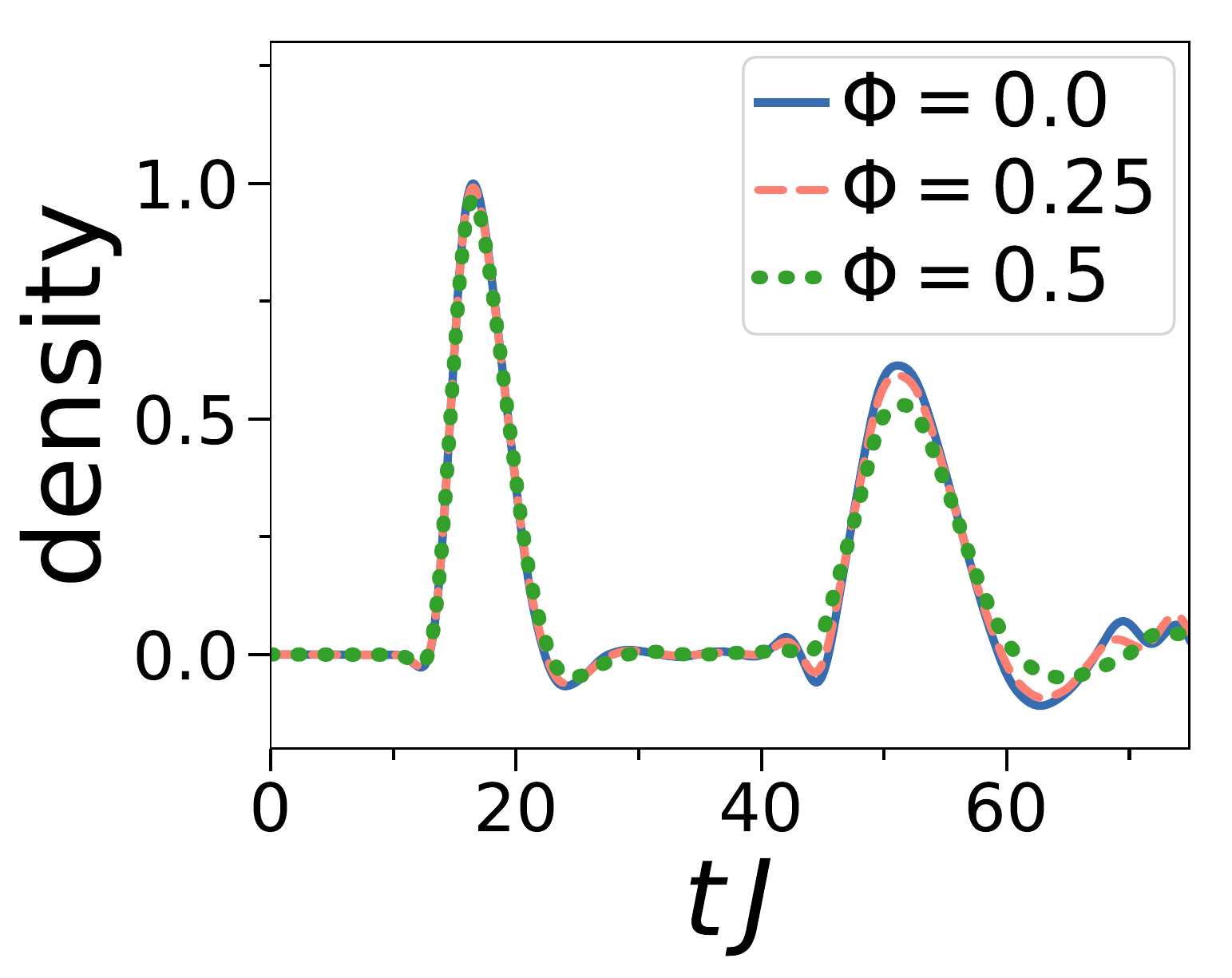}
	\caption{Propagation of a small excitation in a ring without leads for Bose-Hubbard model with ${U=5}$. \idg{a-c} density in ring against time for \idg{a} ${\Phi=0}$, \idg{b} ${\Phi=0.25}$ and \idg{c} ${\Phi=0.5}$.  \idg{d} density averaged between sites 42 and 47 plotted against time $t$. The ring has length ${L_\text{R}=60}$ with $N_\text{p}=30$ atoms.  The initial potential offset is ${\epsilon_\text{D}=0.5}$ at site 15, with width ${\sigma=2}$. The local Hilbertspace is restricted to maximally 4 particles. }
	\label{RingBH}
\end{figure}

However, it is known from past results that a ring attached to leads yields an Aharonov-Bohm effect. When the excitation is injected via leads for spinless fermions (as calculated by B\"uttiker et al.\cite{buttiker1984quantum}), there is destructive interference for at half-flux and the Aharonov-Bohm effect appears. However, in a plain ring without leads, we always find constructive interference. This shows that the interaction with the leads is critical to observe the Aharonov-Bohm effect.

\begin{figure}[htbp]
	\centering
	\subfigimg[width=0.3\textwidth]{}{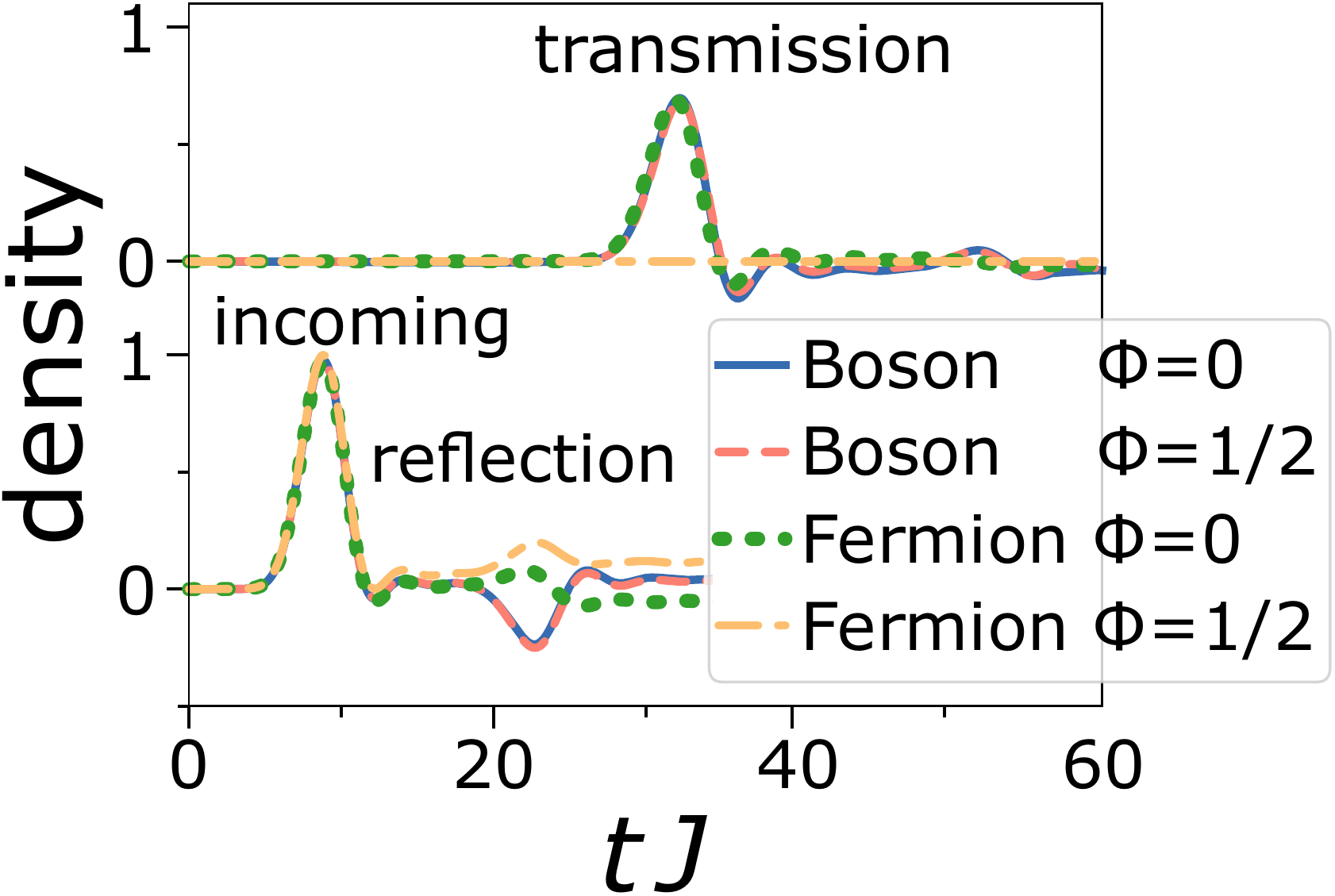}\hfill
	\caption{Propagation of a small excitation in a ring-lead system for hard-core bosons and spinless fermions for zero and half-flux. The source and drain lead has length ${L_\text{S}=L_\text{D}=80}$ and the ring ${L_\text{R}=40}$, the particle number ${N=100}$, strong coupling with ${K=1}$ and ${\epsilon_\text{D}=0.3}$. \densdescr{130}{135}{65}{70} }
	\label{RingFermionBosonComparison}
\end{figure}

%\section{Propagation of excitations in ring-lead system}\label{RingLead}
{\bfseries Nonlinear source-to drain dynamics through Aharonov-Bohm  interferometer.}
Now, we attach two leads to the ring, symmetrically at opposite ends of the ring. We now study the propagation of a small excitation in this  ring-lead system. To proof the difference of spinless fermion and hard-core bosons, we plot the reflected and transmitted density wave for zero and half-flux in Fig.\ref{RingFermionBosonComparison}. For zero flux, we find that the reflected density wave is different for fermions and bosons. For bosons, we find the same characteristic Andreev-like negative reflection peak as seen in the strongly coupled Y-junction. 
The transmission and reflection for hard-core bosons are flux independent. For spinless fermions, there is a transmission for zero flux, while at half-flux we observe zero transmission due to Aharonov-Bohm interference. 

\begin{figure}[htbp]
	\centering
	\subfigimgraised[width=0.24\textwidth]{a}{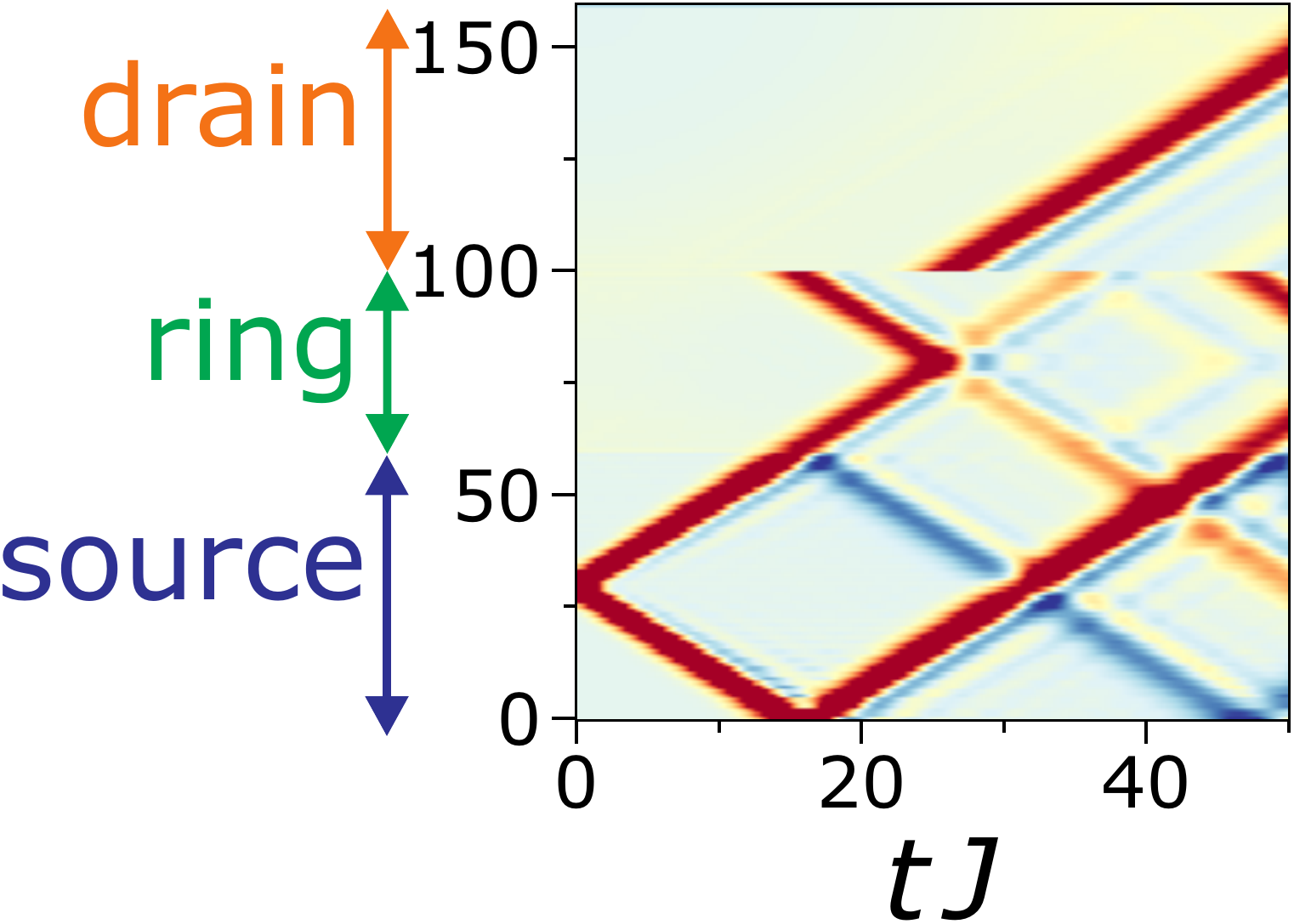}\hfill
	\subfigimg[width=0.24\textwidth]{b}{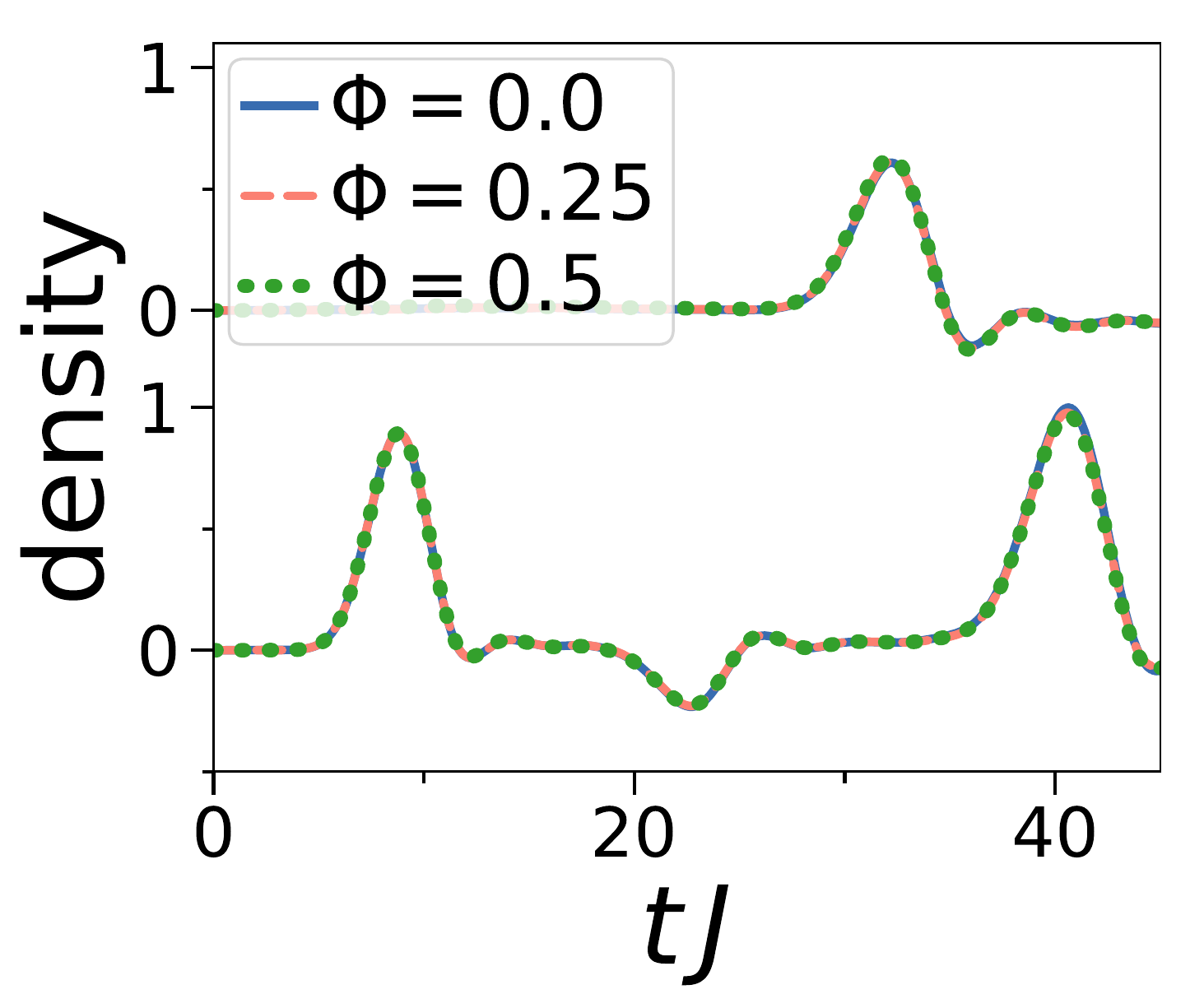}\\
	\subfigimg[width=0.24\textwidth]{c}{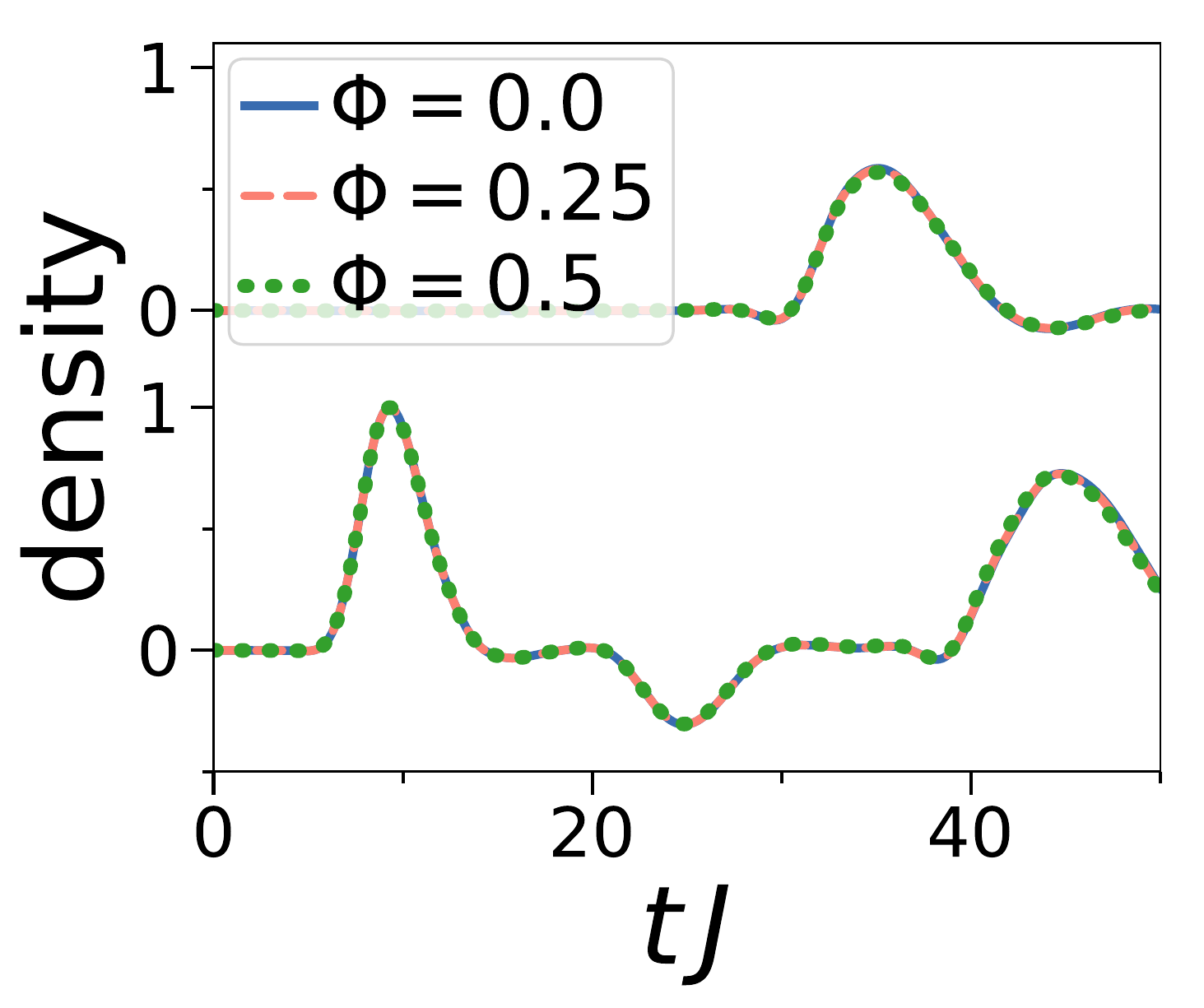}\hfill
	\subfigimg[width=0.24\textwidth]{d}{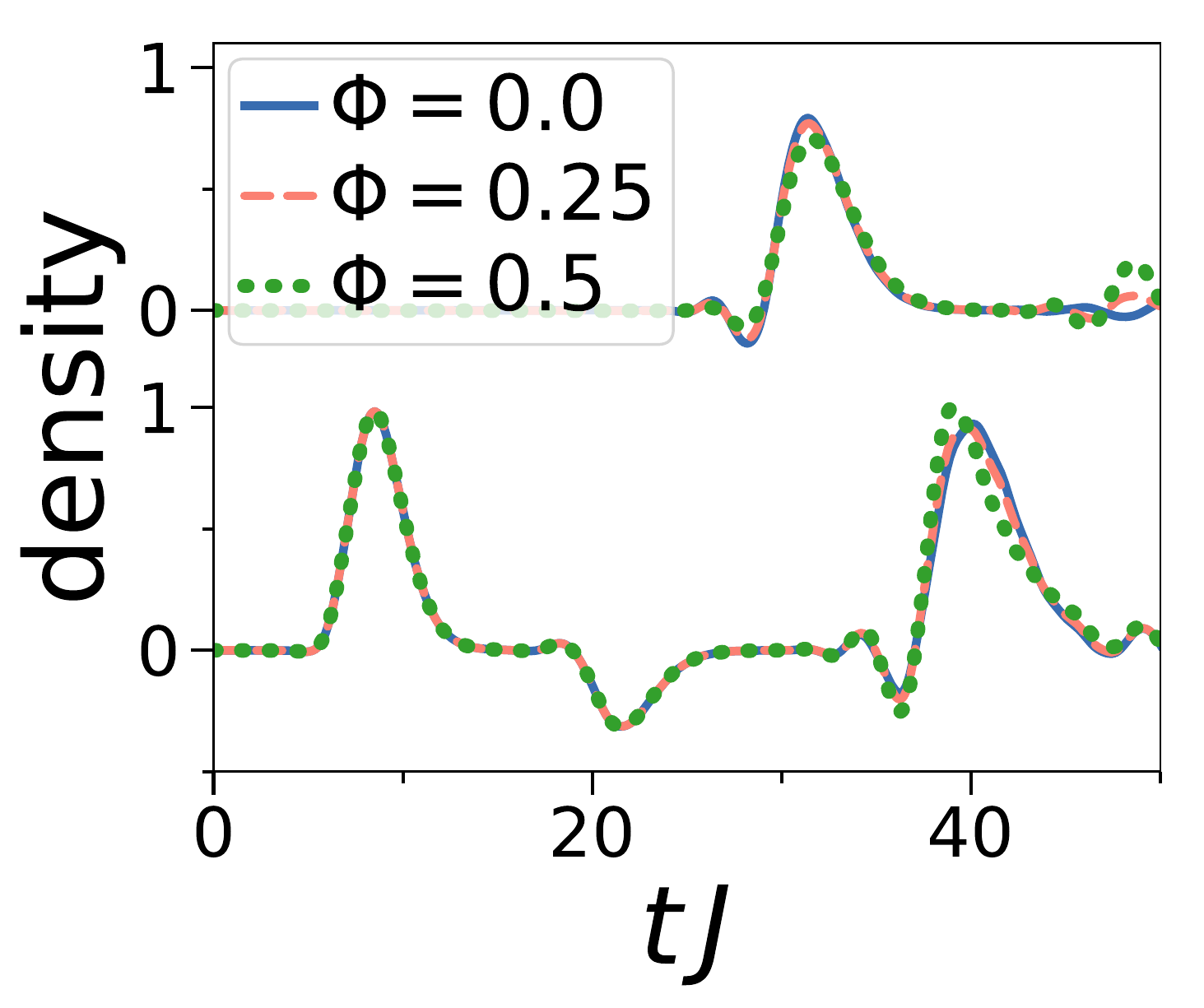}
	\caption{Propagation of small excitation in a ring-lead system for \idg{a,b} hard-core bosons (${U=\infty}$, ${\epsilon_\text{D}=0.3}$),  \idg{c} Bose-Hubbard model (${U=5}$, ${\epsilon_\text{D}=0.5}$) and  \idg{d} discrete Gross-Pitaevskii model (dGPE) with (${g=290}$, ${\epsilon_\text{D}=0.1}$). 
	\idg{a} Full dynamics of density in ring-lead system for hard-core bosons and ${\Phi=0}$. Dynamics of Bose-Hubbard and dGPE look nearly the same.
	 \idg{b,c,d} \densdescr{110}{115}{45}{50} All three models show nearly no flux dependence. The source and drain lead has length ${L_\text{S}=L_\text{D}=60}$, the ring ${L_\text{R}=40}$, the particle number ${N=80}$, ${d=30}$ and ${\epsilon_\text{D}=0.3}$.  The last peak of the lower curve corresponds to a back-reflection from the backward propagating part of the excitation. }
	\label{RingDensityComp}
\end{figure}

Next, we study the system with interacting bosons in more detail.
We show the propagation of a small excitation in the ring-lead system with hard-core bosons, finite $U$ Bose-Hubbard model and dGPE in Fig.\ref{RingDensityComp}a,b,c respectively. (Spinless fermions in Fig.\ref{RingDensityfullFermion} as a reference in the supplemental materials). For interacting bosons in both weak and strongly interacting regime, we find no flux dependence. Both transmission and reflection is independent of flux. %However, spinless fermions show a strong flux dependence, with zero transmission into the drain at $\Phi=0.5$. 

However, the speed of the excitation in the ring for hard-core and finite $U$ Bose-Hubbard depends slightly on flux. \rev{We measure this effect by subtracting the density of the two arms of the ring.} For zero ${\Phi=0}$ and half flux ${\Phi=1/2}$, we find that the excitation velocity for left- and right winding excitation is the same, with a group velocity of ${c=2J}$. However, for ${\Phi=0.25}$, the speed depends on the direction of the excitation. We estimate from our numerical simulations that at this value of flux the speed of the excitation is in forward and backward direction ${c_\pm\approx2J(1\pm1/L_\text{R})}$. This effect is only observed when the excitation is initially prepared in the leads, and then propagates into the ring. The accrued temporal delay between left- and right winding excitation after traveling through the ring is independent of the length of the ring. %However, it is very small and barely visible.
%We also find that the ground state density in the ring is slightly reduced compared to the leads for flux $\Phi=0.5$. 
For the cGPE, we find that the velocity of left- and right winding excitations is different for any non-zero $\Phi$, with ${c_\pm=\sqrt{g n_0}\pm\frac{2\pi\Phi}{L}}$ (as shown in Eq.\ref{EqFluxVelocity} with ${A=\frac{2\pi\Phi}{L}}$). Even for ${\Phi=1/2}$ the velocity is different in left- and right moving excitation in the ring, which is in contrast to what we find for the Bose-Hubbard model. 
The BHM shows macroscopic phase coherence at half-flux, while the cGPE as mean-field theory does not. We suspect that this causes  the difference in velocity.
%From Eq.\ref{EqFluxVelocity}, we find that the direction dependent effect is ${\propto A-v_0}$, where $A$ is the flux and $v_0=-\partial_x\phi(x,t)$ is the change of phase along the ring.  This means that a changing phase along the ring could cancel out the effect of the flux. In the GPE this is not the case as $v_0$ is constant along the ring and therefore the velocity is direction dependent for ${\Phi=1/2}$. 
%However, for the BHM the propagation speed is the same in both directions for ${\Phi=1/2}$. Since the GPE is a limit of the BHM, we may can assume that the relationship between $v_0$, $A$ and velocity also applies for the BHM. Using the same equations, we can argue that the effect of the flux is canceled out by the phase of the condensate for the BHM (${v_0^\text{BHM}=A}$), however not for the GPE (${v_0^\text{GPE}=0}$).
Also, it is known that the GPE is not sufficient to describe interference scenarios for strongly interacting bosons\cite{girardeau2000breakdown,das2002interference}, which may explain the difference between the models.

\begin{figure*}[htbp]
	\centering
	%\subfigure{\includegraphics[width=0.24\textwidth]{{chiralcurrentFluxN1DTwistDataN10000}.pdf}}\hfill
	%\subfigure{\includegraphics[width=0.24\textwidth]{{chiralcurrentTwistN1DTwistN10000}.pdf}}
	\subfigimgraised[width=0.34\textwidth]{a}{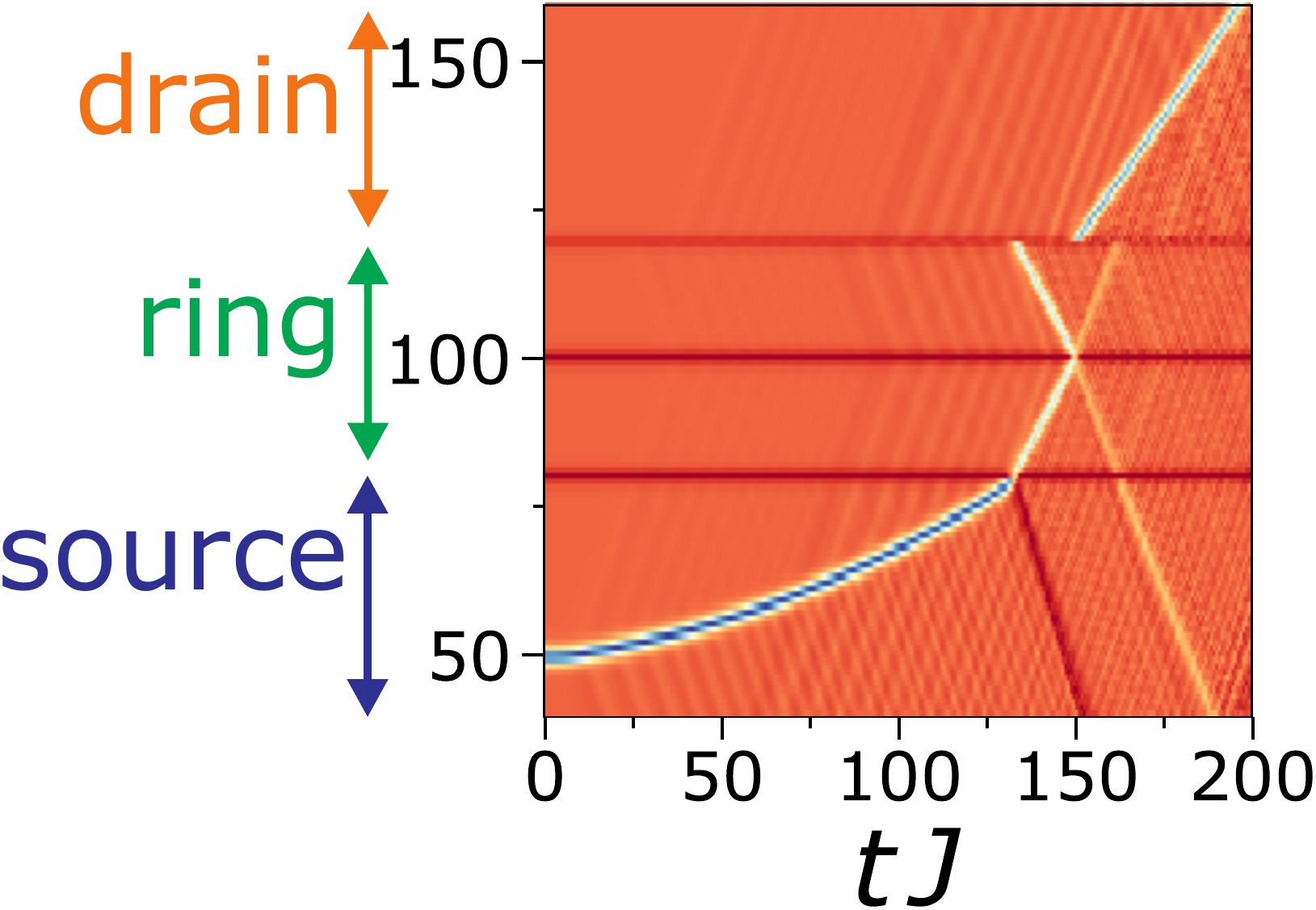}
	\subfigimg[width=0.3\textwidth]{b}{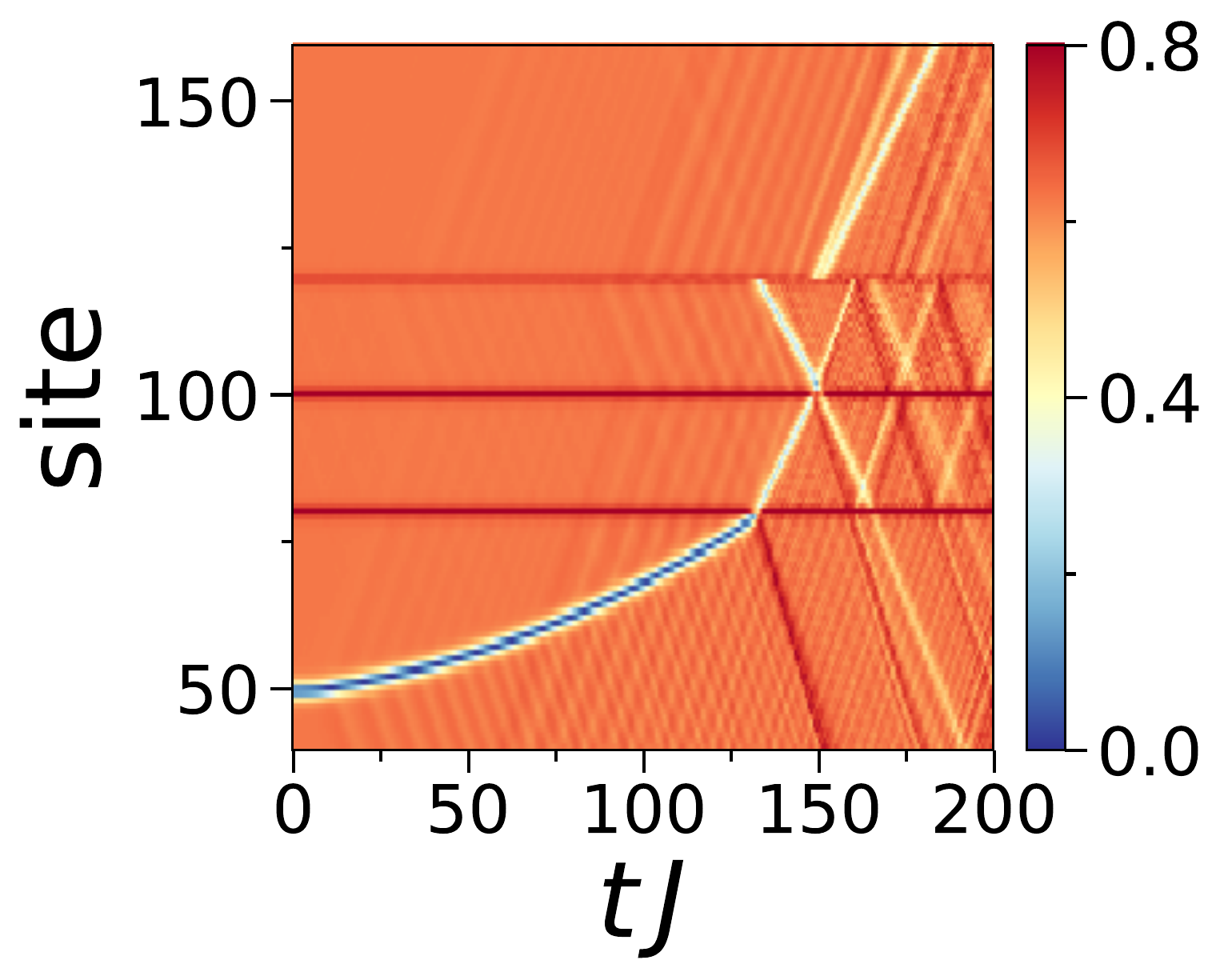}
	\subfigimg[width=0.3\textwidth]{c}{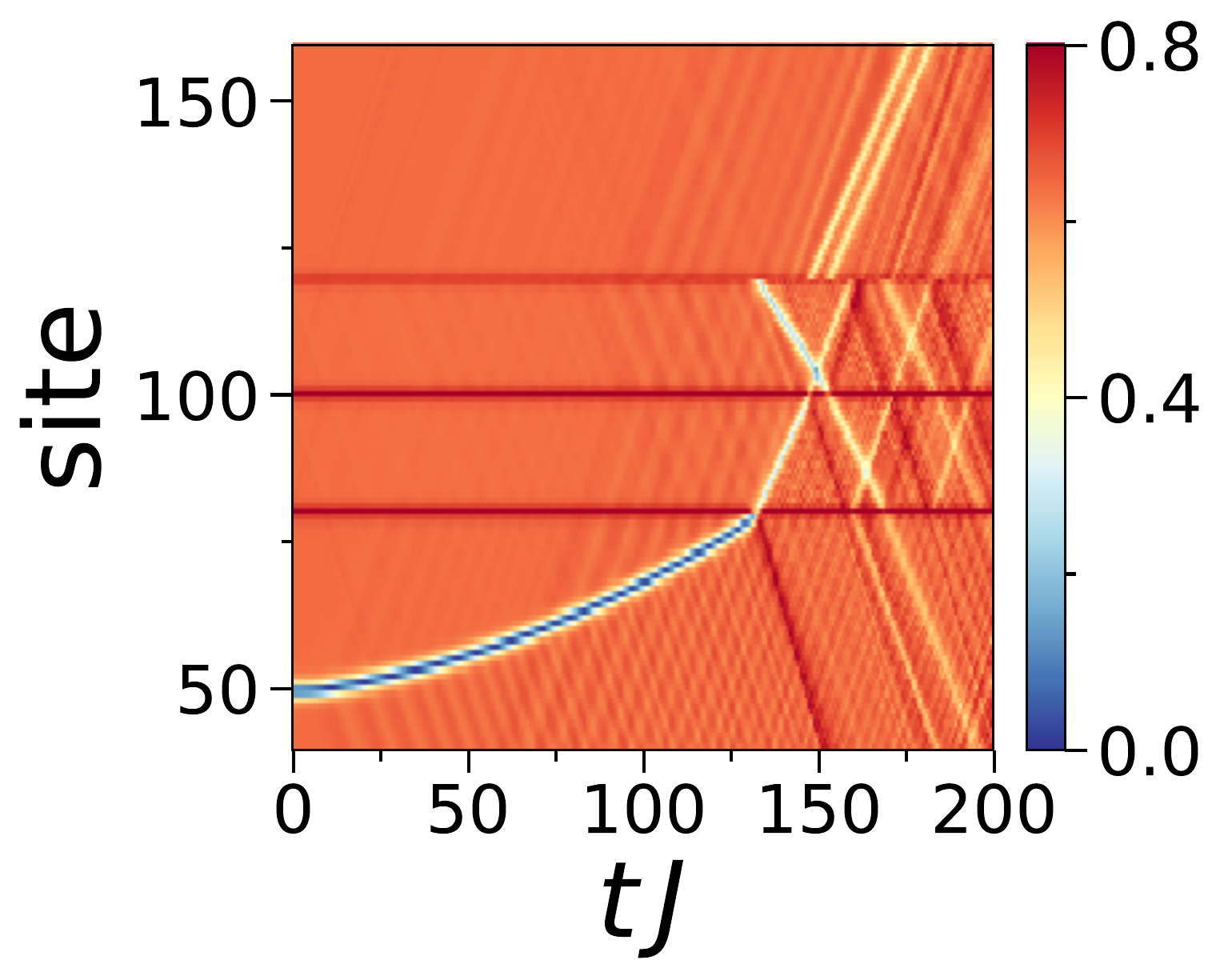}
	\caption{Propagation of a gray soliton in a ring-lead system for the discrete Gross-Pitaevskii model (dGPE). The source and drain lead has length ${L_\text{S}=L_\text{D}=80}$ and the ring ${L_\text{R}=40}$, the interaction ${g=280}$ and ${d=30}$. The flux in the ring is \idg{a} ${\Phi=0}$, \idg{b} ${\Phi=0.25}$, \idg{c} ${\Phi=0.5}$. We create the grey soliton by preparing an initial state with a phaseshift by $\pi$ in the source lead, then evolving that state in imaginary time. }
	\label{RingDensitydarkGPE}
\end{figure*}

Now, we look at large amplitude excitations in the dGPE. It supports solutions which are non-changing in shape, called solitons\cite{pethick2008bose}. For repulsive interaction, they are gray solitons as moving density depression, shown in Fig.\ref{RingDensitydarkGPE}.  The flux changes the velocity and amplitude of left- and right winding solitons in the ring. For zero flux, one gray soliton is transmitted into the drain (Fig.\ref{RingDensitydarkGPE}a). For non-zero flux, two separate gray solitons are transmitted (Fig.\ref{RingDensitydarkGPE}b,c). The grey solitons circulating inside the ring in left- and right winding direction have different amplitude and velocity for non-zero flux. As they reach the drain coupling at different times, the resulting transmission consists of two separate solitons. The difference in left- and right winding velocity is strongly enhanced compared to the small excitation case.

\begin{figure*}[htbp]
	\centering
	%\subfigure{\includegraphics[width=0.24\textwidth]{{chiralcurrentFluxN1DTwistDataN10000}.pdf}}\hfill
	%\subfigure{\includegraphics[width=0.24\textwidth]{{chiralcurrentTwistN1DTwistN10000}.pdf}}
	\subfigimgraised[width=0.28\textwidth]{a}{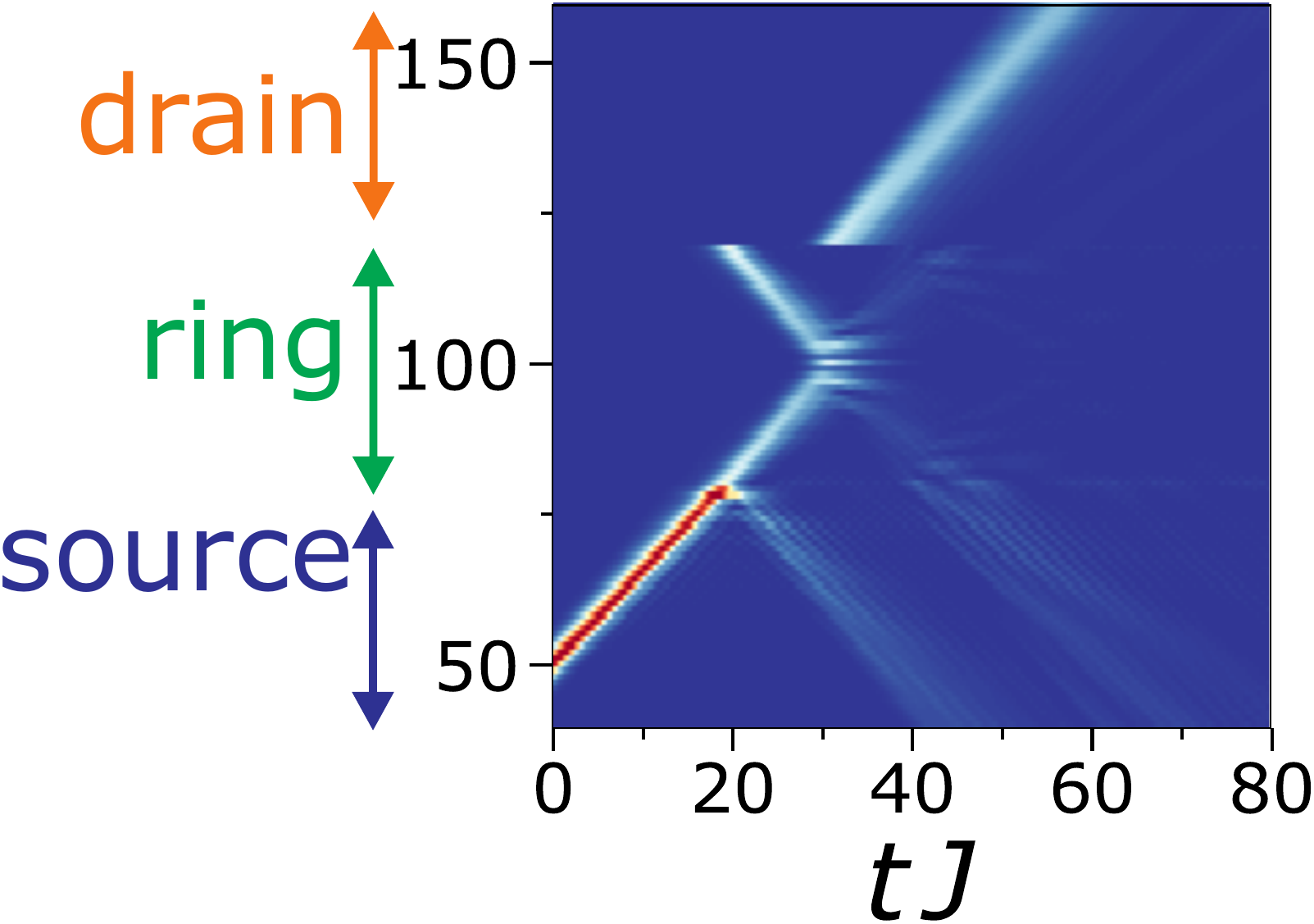}
	\subfigimg[width=0.26\textwidth]{b}{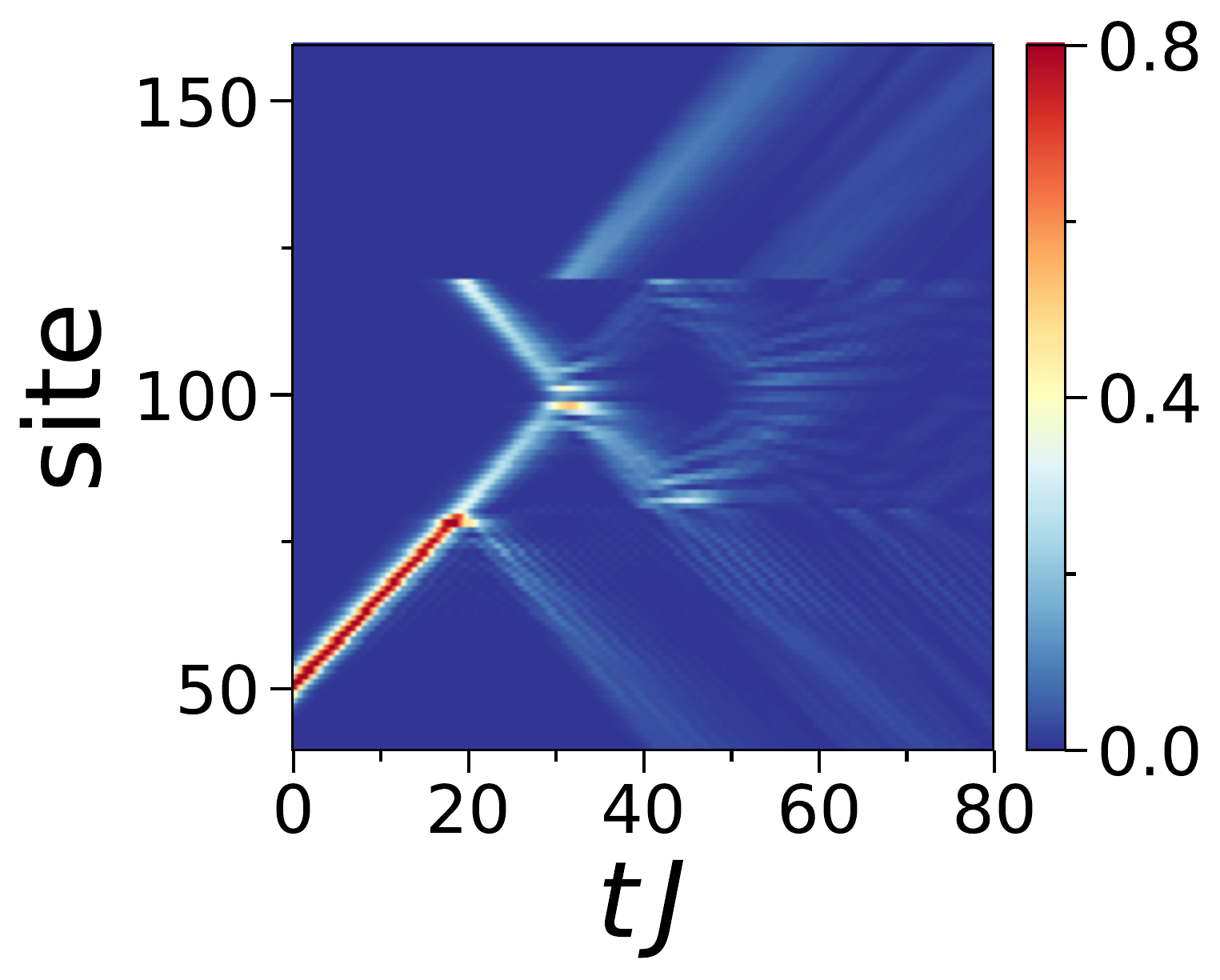}
	\subfigimg[width=0.26\textwidth]{c}{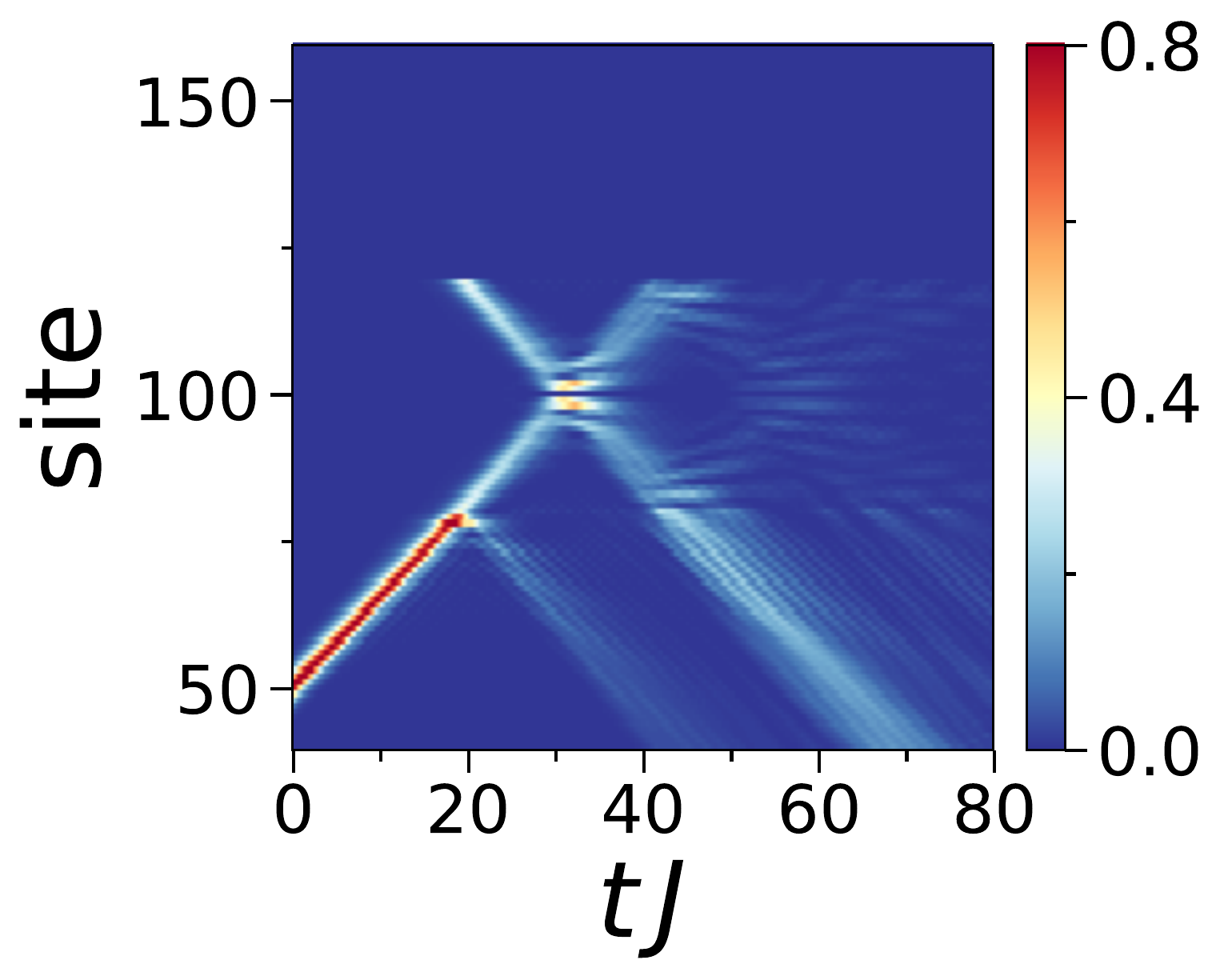}
	\caption{Propagation of a bright soliton in a ring-lead system for the discrete Gross-Pitaevskii model (dGPE). The source and drain lead has length ${L_\text{S}=L_\text{D}=80}$ and the ring ${L_\text{R}=40}$, the interaction ${g=-1}$ and ${d=30}$. The flux in the ring is \idg{a} ${\Phi=0}$, \idg{b} ${\Phi=0.25}$, \idg{c} ${\Phi=0.5}$. We prepare the soliton according to the formula $\Psi(x)\propto\text{sech}((x-L_0)/\sigma)\exp(0.3i\pi x)$, \rev{with $x$ the lattice sites, ${L_0=50}$ the initial soliton position and width ${\sigma=0.01}$. We chose $\sigma$ by varying its value until we achieved a non-dispersive wavepacket.} }
	\label{RingDensitybrightGPE}
\end{figure*}

In the case of attractive interactions, we find non-dispersing wave packets for the dGPE. These bright solitons are plotted in Fig.\ref{RingDensitybrightGPE}. We observe the typical Aharonov-Bohm effect in a ring for bright solitons, with characteristic interference patterns and no transmission at half-flux. 

\begin{figure*}[htbp]
	\centering
	\includegraphics[width=0.75\textwidth]{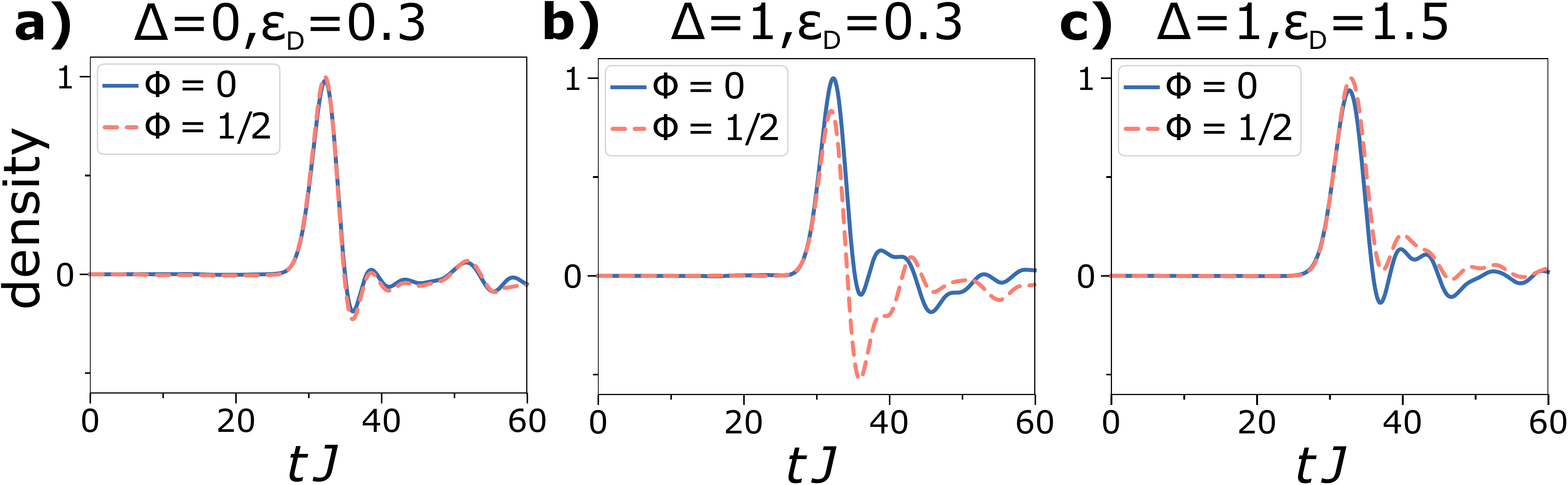}
	\caption{Propagation of density excitation in the drain lead (transmission) in a ring-lead system for hard-core bosons with two impurities of strength $\Delta$. Height of density excitation is controlled by potential offset $\epsilon_{D}$.  \idg{a} $\Delta=0$,${\epsilon_{D}=0.3}$ \idg{b} $\Delta=1$,${\epsilon_{D}=0.3}$ \idg{c} $\Delta=1$,${\epsilon_{D}=1.5}$. The total transmission at time ${tJ=60}$ is tabulated below.
}
	\begin{tabular}{c|c|c} % <-- Alignments: 1st column left, 2nd middle and 3rd right, with vertical lines in between
		&transmission ${\Phi=0}$&transmission ${\Phi=1/2}$\\ \hline
		$\Delta=0$, ${\epsilon_{D}=0.3}$&$0.758$&$0.732$\\ 
		$\Delta=1$, ${\epsilon_{D}=0.3}$ &$0.757$&$0.077$\\
		$\Delta=1$, ${\epsilon_{D}=1.5}$&$0.632$&$0.788$
	\end{tabular}
	\label{TwoImpRingNew}
\end{figure*}

%\section{Propagation of excitations in ring-lead system with impurities}\label{RingImp}
{\bfseries Dynamics through ring condensates interrupted by weak-links.}
In this subsection, we add two impurities symmetrically in the center upper and lower part of the ring. We achieve this by adding this impurity Hamiltonian to the ring Hamiltonian $\mathcal{H}_\text{R,impurity}=\Delta(\nn{a}{L_\text{R}/4}+\nn{a}{3L_\text{R}/4})$. The dynamics for the hard-core Bose-Hubbard model is plotted in Fig.\ref{TwoImpRingNew}. 
As established before, transmission is independent of flux $\Phi$ for no barrier. 
However, it becomes highly flux dependent with barrier ${\Delta=1}$. Here, the density wave that is transmitted into the ring has both positive and negative contributions, which nearly have the same amplitude and thus reduce the net transmission to nearly zero. We observe the same effect for the finite $U$ Bose-Hubbard model (see supplementary materials Sec.\ref{ImpurityBH}).
When the amplitude of the excitation is increased to ${\epsilon_\text{D}=1.5}$, transmission becomes nearly flux-independent again. We find that even with a higher potential barrier in the ring these effects remain. 
We find similar effects for both positive and negative $\Delta$.

%\begin{figure}[htbp]
%	\centering
%	\subfigimg[width=0.24\textwidth]{a}{density0L1DTwoImpNewL40N100l80J1g1f0U0u0V0v0p0P1y1Y1i2d0s1T60t0_008m1M170e3c9c10A77_36HRP7.pdf}\hfill
%	\subfigimg[width=0.24\textwidth]{b}{density0L1DTwoImpNewL40N100l80J1g1f0U0u0V0v0p0P1_5y1Y1i2d0s1T60t0_008m1M170e3c9c10A72_24HRP10.pdf}
%	\caption{Propagation of large excitation (${\epsilon_\text{D}\ge 1}$) in the drain lead (transmission) in a ring-lead system for hard-core bosons with two impurities of strength ${\Delta=1}$ in the ring for different excitation amplitudes $\epsilon_\text{D}$. Other parameters same as in Fig.\ref{TwoImpRing}. \idg{a} ${\epsilon_\text{D}=1}$ \idg{b} ${\epsilon_\text{D}=1.5}$. ${\Phi=0}$ (solid)  ${\Phi=0.25}$ (dashed), ${\Phi=0.5}$ (dots). The transmitted density changes strongly with flux for $\epsilon_\text{D}=1$ and decreases strongly for ${\Phi=0.5}$. However it is nearly independent of $\Phi$ for $\epsilon_\text{D}=1.5$. At half-flux, the transmission depends on the wavepackets amplitude. It blocks transmissions for wavepackets with smaller amplitude, and transmits for larger amplitude. This effect persists for larger ${\Delta>1}$.}
%	\label{TwoImpRingBig}
%\end{figure}

%\section{Quench of ring}\label{RingQuench}
{\bfseries Quench dynamics through an Aharonov-Bohm interferometer.}
In this subsection, we study the non-equilibrium quench dynamics of the ring-lead system. The atoms are initially loaded into the source. We calculate the ground state of atoms in the source, without coupling to the rest of the system. Then, we suddenly switch on the coupling $K$ at ${t=0}$. Then, the atoms flow out into the ring into the drain lead.
We study the resulting dynamics of the density and the expectation value of the current from source to ring, and ring to drain (Eqs.\ref{ringcurrent}) for hard-core bosons in Fig.\ref{RingDensityfullBoson} (Spinless fermions for reference in the supplemental materials in Fig.\ref{RingDensityfullFermion}). 
For hard-core bosons, we find that the dynamics of the current starts from zero, then increases over time, until its dynamics slows down. Source and drain current are initially very different, however they come closer over time. It is reasonable to assume that at longer times they both assume the same value once they reach the steady state. However, we are unable to calculate numerically the dynamics at longer times, as the numerical effort for tDMRG increases drastically for strong quenches.
The drain current is highly flux dependent. We find that the initial dynamics of the drain current is slower for half-flux. The dynamics is characterized by small oscillations, which decrease over time. The current at longer times depends only little on flux.

\begin{figure}[htbp]
	\centering
	%\subfigure{\includegraphics[width=0.24\textwidth]{{chiralcurrentFluxN1DTwistDataN10000}.pdf}}\hfill
	%\subfigure{\includegraphics[width=0.24\textwidth]{{chiralcurrentTwistN1DTwistN10000}.pdf}}
	\subfigimgraised[width=0.24\textwidth]{a}{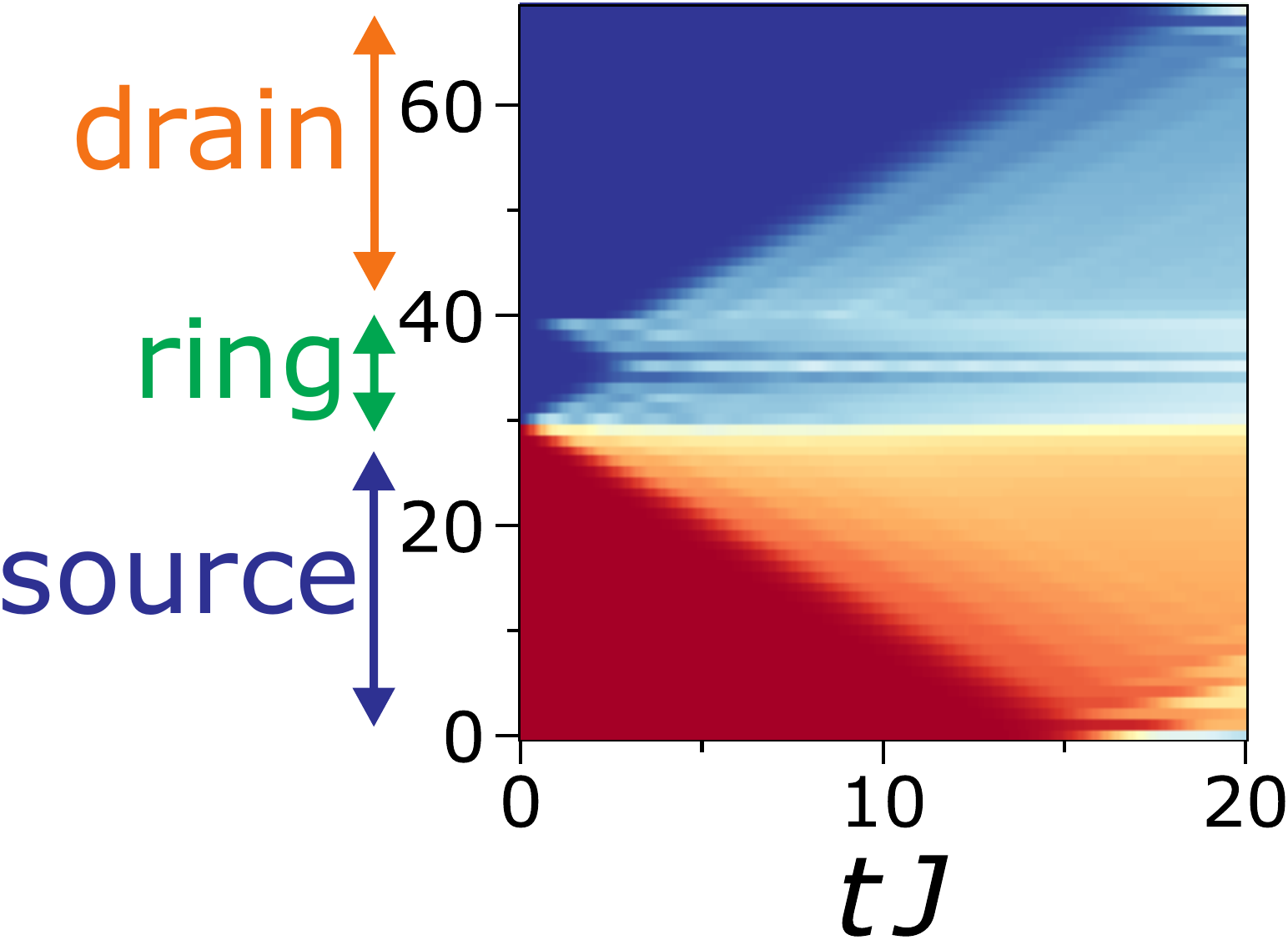}\hfill
	%\subfigimg[width=0.24\textwidth]{d}{densitySD1DL10N15l30J1g1f0U0u0V0v0p0P0i1d0s1T20t0_01m1M1000e3c10A21_3QR1.pdf}\\
	\subfigimg[width=0.24\textwidth]{b}{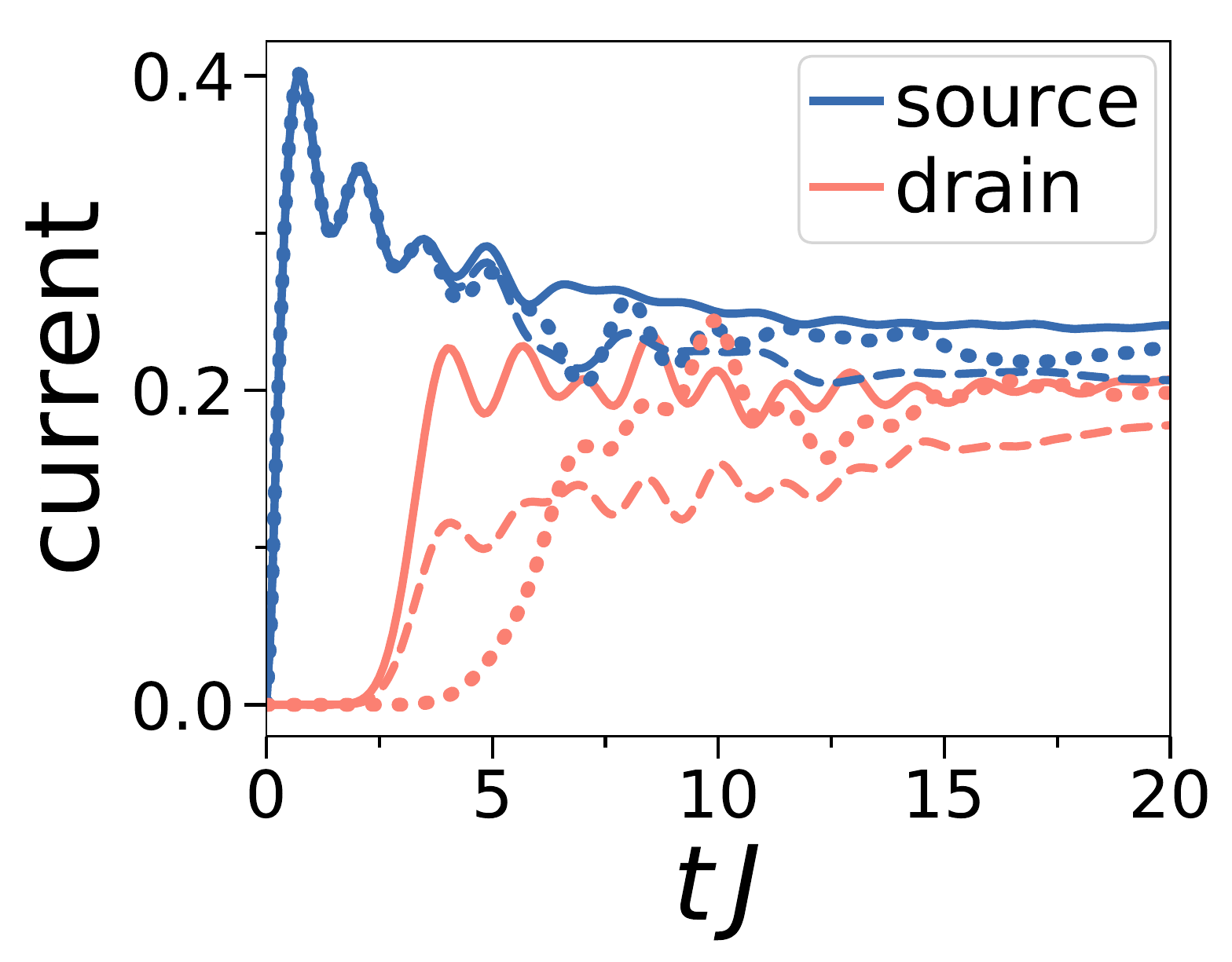}
	\caption{Time evolution of atoms initially loaded into the leads of a ring-lead system for hard-core bosons and strong coupling ${K=1}$. The source and drain lead has length ${L_\text{S}=L_\text{D}=30}$ and the ring ${L_\text{R}=10}$, the particle number ${N=15}$. As an example, we show the dynamics of the density in the full system in \idg{a} ${\Phi=0}$. %\idg{d} The total density in source, ring and drain leads. . 
	\idg{b} Expectation value of current (Eq.\ref{ringcurrent}) from source lead to ring (blue) and ring to drain lead (orange) for different values of flux ${\Phi=0}$ (solid)  ${\Phi=0.25}$ (dashed), ${\Phi=0.5}$ (dots).%(line style same as in \idg{d}).
	}
	\label{RingDensityfullBoson}
\end{figure}

%\begin{figure}[htbp]
%	\centering
%	%\subfigure{\includegraphics[width=0.24\textwidth]{{chiralcurrentFluxN1DTwistDataN10000}.pdf}}\hfill
%	%\subfigure{\includegraphics[width=0.24\textwidth]{{chiralcurrentTwistN1DTwistN10000}.pdf}}
%	\subfigimgraised[width=0.25\textwidth]{a}{densL10N15l30J1g1f0U0u0V0v0p0P0i1d0s1T20t0_01m1M1000e3c10A21_3QR1.pdf}\hfill
%	\subfigimg[width=0.23\textwidth]{b}{densL10N15l30J1g1f0_25U0u0V0v0p0P0i1d0s1T20t0_01m1M1000e3c10A20_6QR2.pdf}
%	\subfigimg[width=0.24\textwidth]{c}{densL10N15l30J1g1f0_5U0u0V0v0p0P0i1d0s1T20t0_01m1M1000e3c10A21_9QR3.pdf}\hfill
%	%\subfigimg[width=0.24\textwidth]{d}{densitySD1DL10N15l30J1g1f0U0u0V0v0p0P0i1d0s1T20t0_01m1M1000e3c10A21_3QR1.pdf}\\
%	\subfigimg[width=0.24\textwidth]{d}{currentSD1DL10N15l30J1g1f0U0u0V0v0p0P0i1d0s1T20t0_01m1M1000e3c10A21_3QR1.pdf}
%	\caption{Time evolution of atoms initially loaded into the leads of a ring-lead system for hard-core bosons and strong coupling ${K=1}$. The source and drain lead has length ${L_\text{S}=L_\text{D}=30}$ and the ring ${L_\text{R}=10}$, the particle number ${N=15}$. The flux in the ring is \idg{a} ${\Phi=0}$, \idg{b} ${\Phi=0.25}$, \idg{c} ${\Phi=0.5}$. %\idg{d} The total density in source, ring and drain leads. . 
%	\idg{d} Expectation value of current from source lead to ring (blue) and ring to drain lead (orange) for different values of flux ${\Phi=0}$ (solid)  ${\Phi=0.25}$ (dashed), ${\Phi=0.5}$ (dots).%(line style same as in \idg{d}).
%}
%	\label{RingDensityfullBoson}
%\end{figure}

\begin{figure*}[htbp]
	\centering
	\subfigimg[width=0.28\textwidth]{a}{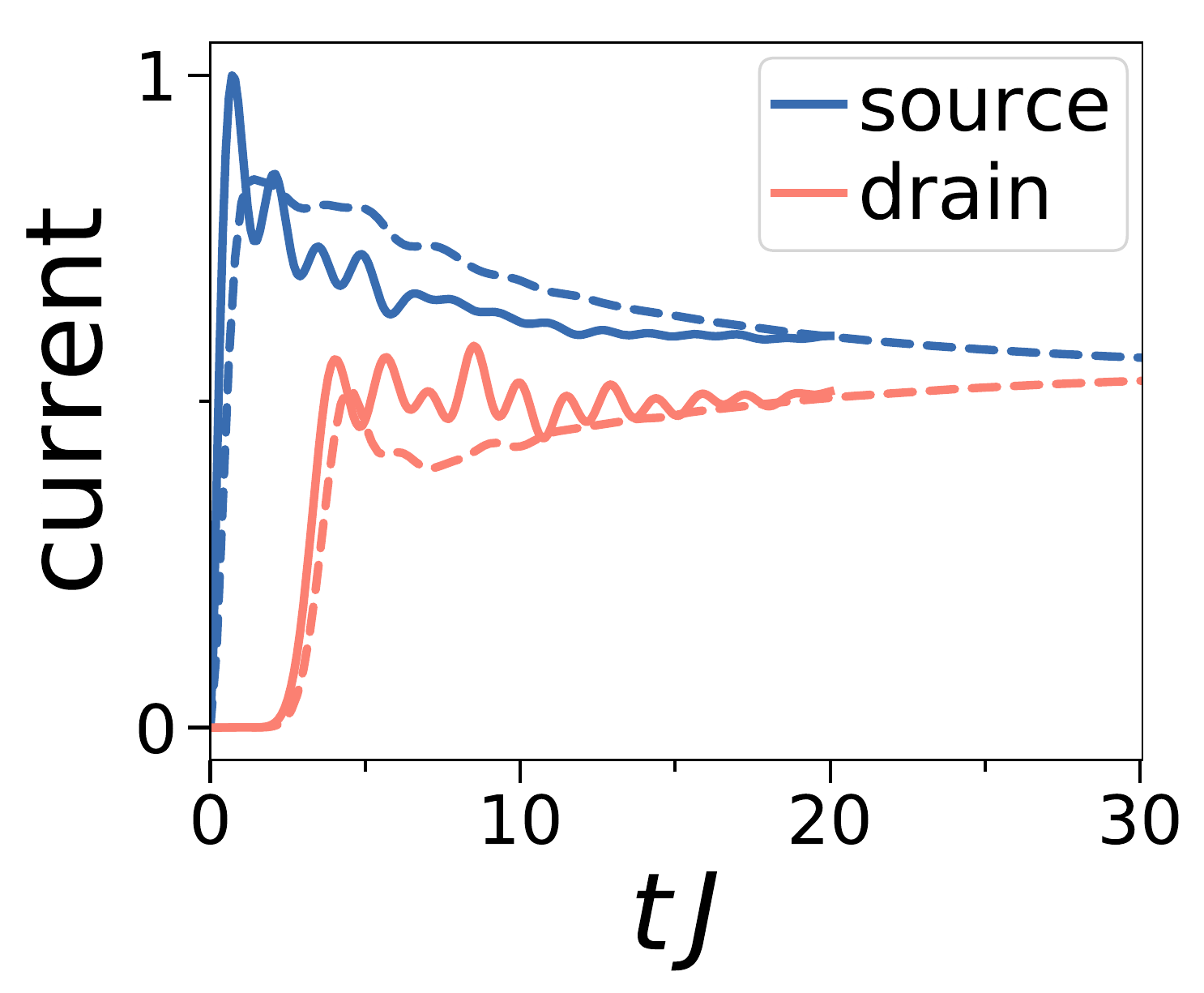}
	\subfigimg[width=0.28\textwidth]{b}{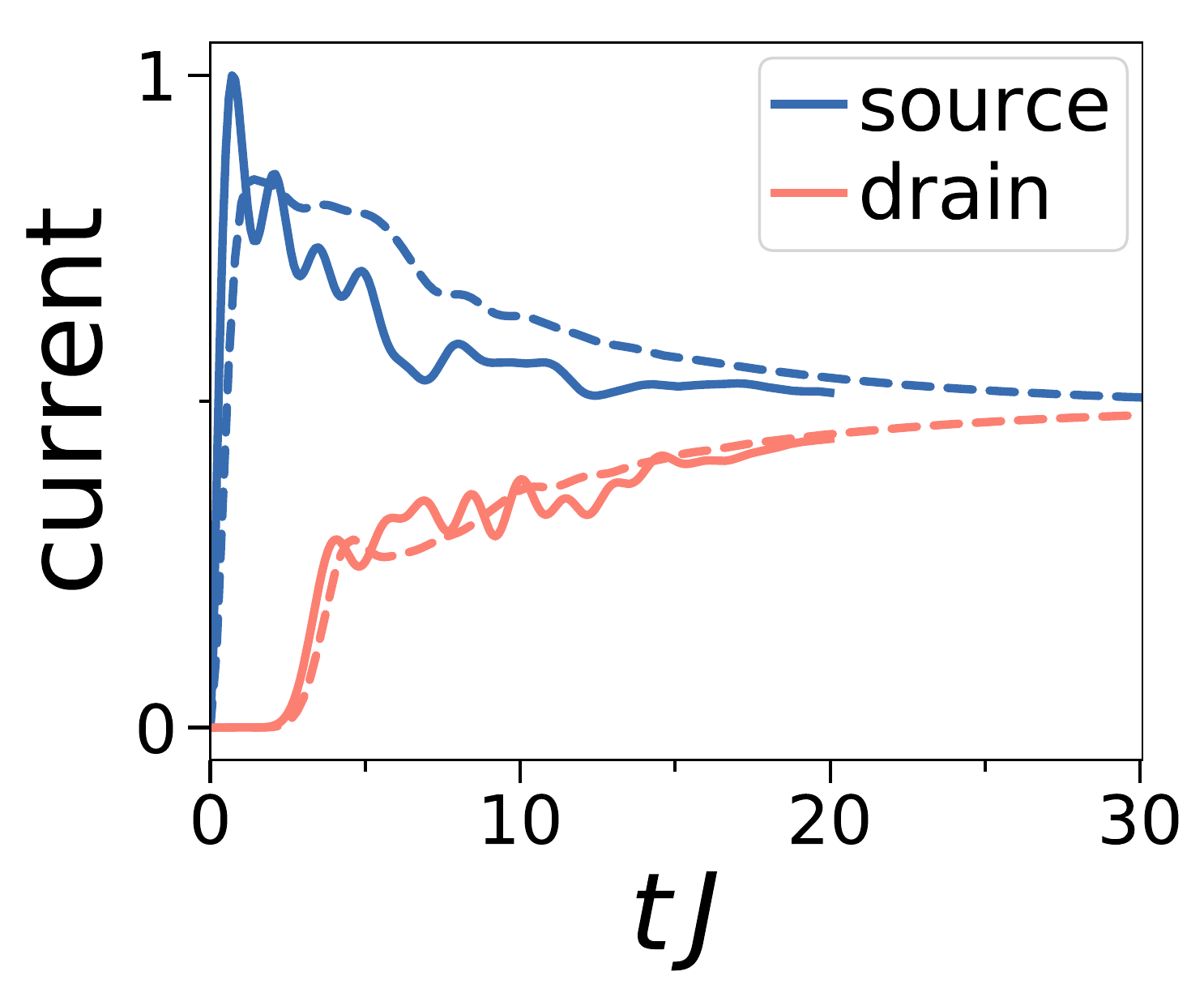}
	\subfigimg[width=0.28\textwidth]{c}{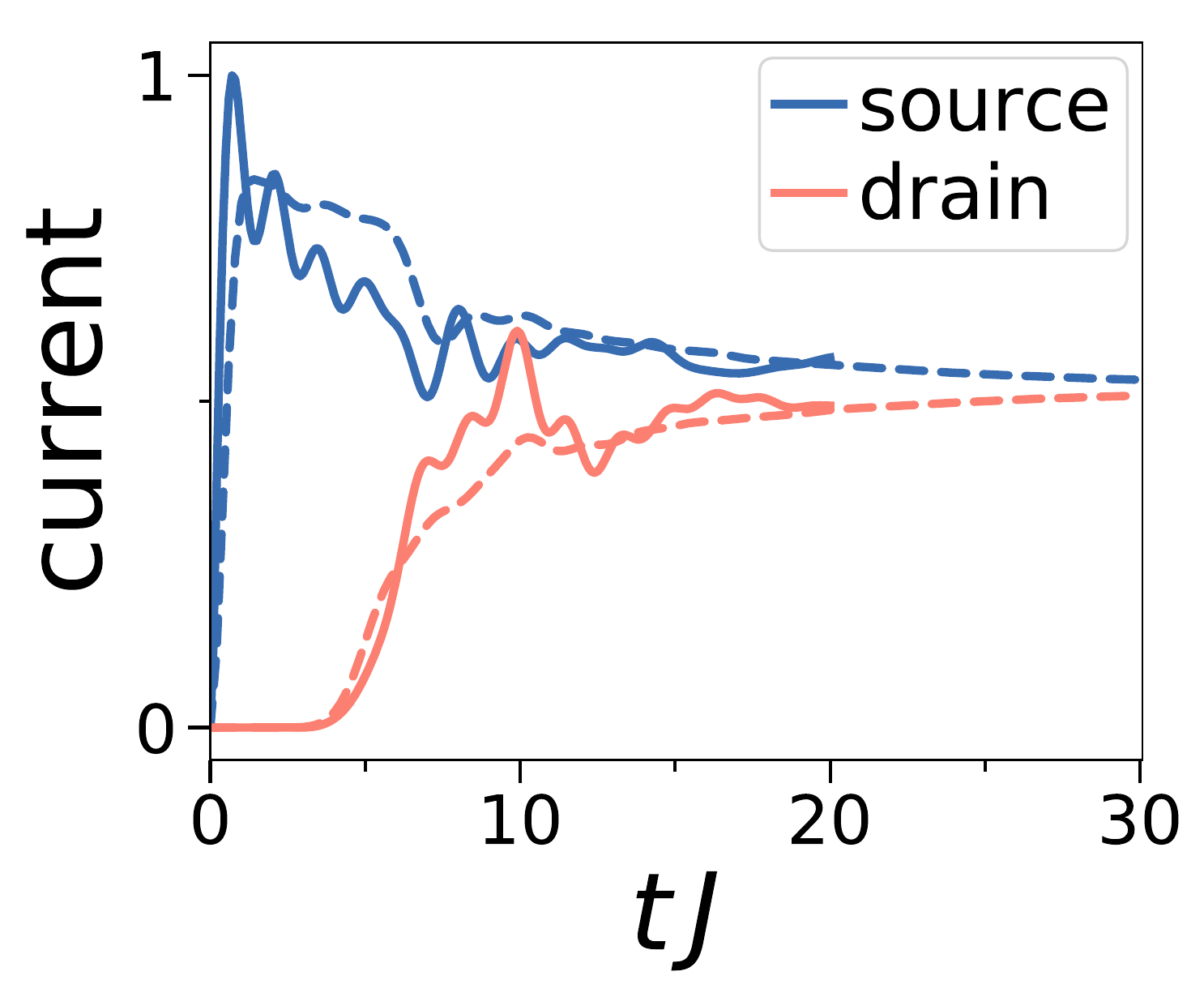}
	\caption{Time evolution of atoms initially loaded into the source lead of a ring-lead system for hard-core bosons and strong coupling ${K=1}$. Comparison of the current with two different methods: Closed system with full modeling of ring and lead (solid) and open system with leads modeled with Markovian bath (dashed). For DMRG, the source and drain lead has length ${L_\text{S}=L_\text{D}=30}$ and the ring ${L_\text{R}=10}$, the particle number ${N=15}$. The atoms are prepared in the ground state of the uncoupled source, then the coupling is switched on instantaneously at ${t=0}$. For the open system approach, the ring has ${L_\text{R}=10}$ sites, the source and drain consist of one site each. Lindblad parameters are ${\Gamma=1.5}$ and ${r=0.65}$. The current from source to ring (blue) and ring to drain (orange) over time for different values of the flux \idg{a} ${\Phi=0}$, \idg{b} ${\Phi=0.25}$, \idg{c} ${\Phi=0.5}$. }
	\label{RingDensityComparison}
\end{figure*}

So far, we simulated the full system including leads as a closed system. We found that with tDMRG we could not reach the steady-state of the quench dynamics. %The reason for that is that for highly excited states the entanglement grows quickly in time, increasing the computation time drastically.
However, we can reach the steady-state by using the open system method as introduced in the quench of the Y-junction in section \ref{YExc} (further see supplementary materials Sec.\ref{opensystem}): We trace out the leads and simulate the dynamics of the interesting subsystem only (the ring).
We compare the results for the two methods for the ring-lead system in Fig.\ref{RingDensityComparison}. We find that the initial dynamics as well as the slow dynamics towards the steady-state agrees well with the tDMRG results. This is surprising since the open system approach uses a strong approximation. \rev{This Born-Markov approximation assumes that the baths are memoryless, not influenced by the system dynamics and the bath-system coupling is weak\cite{breuer2002theory}.}

%\section{Quench of Y-junction}\label{YQuench}

\section{Discussion}
\label{discussion}
We studied the transport dynamics of density excitations propagating through  bosonic systems with two specific configuration: a  Y-junction and a Aharonov-Bohm matter wave interferometer.
We carried out a thorough analysis of the transmission and reflection of the density excitations in terms of the system parameters: particle-particle interaction and coupling between the leads and the systems. The dynamics is analysed with Bose-Hubbard Model (BHM)  and the discrete Gross-Pitaevskii Equation (dGPE). The non-equilibrium dynamics is  also studied by a quench protocol: The state is prepared as the ground state of the source lead; then, the density evolves once the empty ring is put into contact with the leads. 
We point out that, in contrast with the recent study carried out in \cite{haug2017aharonov}, here we model the leads as long $1d$ chains of interacting bosons. Such approach turned out to have important consequences on the possible dynamics that can be established  in the system.

%Andreev-like, oscillating and positive reflections), which are controlled by the coupling strength and the on-site interaction. For large excitations in the Gross-Pitaevskii limit, we find that there i%s a specific coupling strength, at which the type of transmission and reflection changes abruptly. Around this coupling strength, excitations can get trapped in the junction.
%
%For a ring-lead system, we found that the propagation of excitations through the ring is independent of flux and there is no Aharonov-Bohm effect for interacting bosons. The Aharonov-Bohm effect appears when a barrier is introduced into the ring.
%For quenched ring, we find that the dynamics depends on the flux, however the steady-state is independent of flux.

{\bf{Y-junction--}}
For a small density packet passing through a Y-junction, the dynamics governed by the BHM displays  three different regimes of reflection governed by the lead-system coupling of the junction (Fig.\ref{YDensitysmall},\ref{YDensityHubbardsmall}): For strong lead-system coupling, we find a clear Andreev-like reflection with negative density amplitude. For intermediate coupling, we find a reflection with both positive and negative contribution. And for weak coupling, we find a purely positive reflection (no Andreev scattering). \rev{Andreev reflections are also found analytically within the linearized Gross-Pitaevskii wave equations. The total transmitted density is $4/3$ of the incoming density excitation and  the reflected density is $-1/3$. { These numbers come from the linearized dynamics as in Eqs. (\ref{LinEq}), (\ref{transmission}) and match the result for BHM in the strongly coupled regime. }  {\it This demonstrates that the mechanism generating the Andreev reflection is related to the wave nature of excitations on top of the condensate.}
%The total transmitted and reflected density integrated over long times depends on the junction coupling for the Bose-Hubbard model  (Fig.\ref{YDensitysmall}).
However, for weakly coupled systems, the dGPE (Fig.\ref{YDensitysmallGPE}) dynamics yields different results compared to BHM: the dGPE transmission is always $4/3$, which corresponds to the strongly coupled regime, independent of the lead-system coupling. The weakly coupled regime as found in the BHM is absent for dGPE.  This effect is likely to be related to the regime of validity of dGPE, namely in the limit of large number of atoms $\&$ weak interaction. } %This absence of weak coupling matches the prediction of \cite{kane1992transport} that for repulsive bosons (Luttinger liquids with ${g>1}$) barriers are renormalized to zero. However, it remains unclear why the absence is only observed in the dGPE, and not in the BHM. }

%For weakly coupled systems, the dGPE (Fig.\ref{YDensitysmallGPE}) dynamics yields different results compared to BHM: the total transmitted density is always $4/3$  of the incoming density excitation and  the reflected density is fixed to $-1/3$, independently of lead-system coupling strength. These numbers come from the linearized dynamics as in Eqs. (\ref{LinEq}), (\ref{transmission}). }
% For strong coupling, there is 4/3 transmission and -1/3 reflection. Reflection increases and transmission decreases with increasign coupling strength. 
For large negative excitations of the dGPE ('gray soliton' like), we find two regimes of lead-system coupling. For strong coupling, we find that part of the incoming wave is Andreev-like reflected and the other part is transmitted. For weak coupling we find total reflection of the incoming excitation. This behaviour arises because of the interplay between the kinetic energy of the soliton,  the strength of the barrier and the interaction.  The two regimes are separated by a critical value of the lead-system coupling.  Interestingly enough, we find that close to the critical coupling, the gray soliton has a finite residing time in the junction (Fig.\ref{YDensitybigNegGPE}). \rev{Analogous  phenomenon is found in solitons scattering off impurities\cite{krolikowski1996soliton}}.

{\bf{Aharonov-Bohm matter wave interferometer--}}
By creating the excitations inside the ring  of interacting bosons,  {\it without leads},   we find no Aharonov-Bohm interference effect (Fig.\ref{RingOnlyHardcore},\ref{RingBH}). For spinless fermions we find the same results (as they can be mapped to hard-core bosons with the Jordan-Wigner transformation).

For a {\it ring-lead system}, we studied the dynamics in dependence of particle-particle interaction and  Aharonov-Bohm flux through the ring. When leads are attached to the ring and the excitations are incoming via the leads, spinless fermions and hard-core bosons yield different results (because of the leads, the Jordan-Wigner mapping cannot be applied). Interacting bosons in all interaction regimes (hard-core, finite $U$ and dGPE) are independent of flux, however spinless fermions show a clear Aharonov-Bohm effect (Fig.\ref{RingFermionBosonComparison}). This feature confirms that the leads play a crucial role  for  the Aharonov-Bohm interference and agree with the field-theoretic analysis results\cite{tokuno2008dynamics}. 
\rev{Our results indicate  that  bosonic condensate phase is able to cancel the phase shift induced by the Aharonov-Bohm effect and suppress it in interacting bosons.  For fermions, there is no condensation, thus the AB effect is still present.  In support of the argument, we mention that  when the statistics of the particles is intermediate between bosons and fermions it can weaken the flux dependence of the current\cite{haug2017aharonov}. } 

Dynamics based on GPE equation qualitatively confirm the picture provide above. For the cGPE, Eq.\ref{EqFluxVelocity} shows that  the flux has a minor effect on the propagation velocity. The speed of sound becomes dependent on the direction of propagation (clockwise or anti-clockwise) along the ring. However, for small excitations this velocity difference  is small.  
For the dGPE, instead,  with large negative excitations (gray soliton), we find that the transmitted excitations depends on the flux. For non-zero flux, it splits into two parts as the velocity of the soliton depends strongly on the direction of propagation in the ring (Fig. \ref{RingDensitydarkGPE}). 
%This effect could be used to measure the effective flux piercing the ring (in other words, the system can be used as a quantum %detector for rotation).
For large positive excitations in a ring (bright soliton), we observe the Aharonov-Bohm interference effect (Fig.\ref{RingDensitybrightGPE}).

%\rev{The absence of the Aharonov-Bohm effect for interacting bosons may be related to the spontaneous breaking of the $U(1)$ %symmetry by the system condensate. We conjecture that the emerging condensate phase is canceling the Aharonov-Bohm flux.

\rev{The translational symmetry can be broken in the ring by adding two impurity potentials symmetrical into the ring. 
In this way, we study the transport through a  dc-atomtronic-SQUID device\cite{Ryu2013}.
For small excitations with  BHM dynamics on this device, {\it we do find a clear  flux dependence for the transmitted density}(Fig.\ref{TwoImpRingNew}). Indeed, we find that the impurities creates two 'dark solitons' in the ring condensate with a strong density depression (by more than 50\% compared to density in the rest of the ring).  Correspondingly, a strong quantum phase slip occurs across the soliton center  (see for instance\cite{cominotti2014optimal}) that is able to wash out the condensate's phase and  can imprint the Aharonov-Bohm flux onto the density wave. However, we observe this effect only for small excitations, larger excitations do not show the Aharonov-Bohm effect with impurities: large excitations carry enough density such that they can annihilate the dark soliton, causing no phase slips. We found that GPE dynamics is not able to produce such effects.}
\rev{
%Furthermore, the flux dependence disappears for increasing excitation amplitude. 
We observe that, in this regime, for half-flux, this system acts as a nonlinear device that blocks low amplitude density waves and transmits high amplitude ones. }

{\bf{Quench dynamics--}}
For both Y-junction and AB-rings, we find that the approximated  open system approach using Lindblad equations produces similar results as the exact treatment using DMRG (Fig.\ref{YCurrentComparison},\ref{RingDensityComparison}). This is surprising since the open system neglects memory effects of the bath as well as assumes weak coupling of system and bath. 
\rev{The open system approach allows to study the system in a much smaller Hilbert space, making it a powerful tool to study quench problems. It would be interesting to characterize the open system approach for other setup configurations.
Although the quench dynamics is clearly  of different nature compared to  the transport dynamics of a  density packet, we find that the results on the Aharonov-Bohm effect point to the same direction:
any flux dependence vanishes in the steady-state of the current (Fig.\ref{RingDensityfullBoson}, \cite{haug2017aharonov}). However, the quench dynamics of the current itself is flux dependent. }
%For the Y-junction, we find specific oscillations value and oscillations of the current through the junction are  controlled by the junction coupling. 
%An approximated open system approach can reproduce such dynamics . 

%This is an indication that the open system approach is suited to describe non-equilibrium dynamics of cold atoms.
%The same quench procedure within the GPE is highly flux dependent and does not converge towards a steady-state, and instead shows large and long-lived oscillations in density (Fig.\ref{RingDensityfullGPE})
%\section{Conclusion}
%The dynamics of excitations in a X-junction configuration could be explored\cite{girardeau2002theory}. Our numerical methods using DMRG and open system could be applied to the %transport dynamics of fermionic systems\cite{husmann2015connecting}.

We believe that our results could be important for the actual detection of the Andreev scattering and Aharonov-Bohm effect in bosonic cold atom systems. 
Many of the effects we encountered in the specific systems we studied  can be clearly interesting for the realization of new quantum devices and sensors.

\begin{acknowledgments}
	{\it Acknowledgments}. We thank M. Rizzi, D. Feinberg, S. Safaei and J. Helm for enlightening discussions. The Grenoble LANEF framework (ANR-10-LABX-51-01) is acknowledged for its support with mutualized infrastructure. We thank National Research Foundation Singapore and the Ministry of Education Singapore Academic Research Fund Tier 2 (Grant No. MOE2015-T2-1-101) for support. The computational work for this article was partially performed on resources
	of the National Supercomputing Centre, Singapore (https://www.nscc.sg).
\end{acknowledgments}

\bibliography{library}
\appendix

\section{Numerical methods}
\begin{figure}[htbp]
	\centering
	\subfigure{\includegraphics[width=0.49\textwidth]{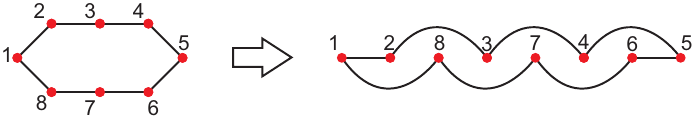}}
	\caption{Sketch of the two different ways of ordering the interaction links of a ring lattice. The lattice sites are connected to the nearest-neighbors only.  The regular numeration of sites is shown on the left. This numeration has the disadvantage that there is a long-range interaction between the first (site 1) and last site (site 8). This long-range interaction inefficient within DMRG. On the right we show the same configuration, however with a different internal numeration, which is used in our simulations. We avoid long-range interaction by re-arranging the links into nearest and next-nearest neighbor links.  }
	\label{DMRGExpl}
\end{figure}

Low-energy static and dynamic problems of quasi one-dimensional systems can be solved efficiently with Matrix-Product state (MPS) and Density matrix renormalization group (DMRG) techniques\cite{white1992density,white2004real}.
The high-dimensional wavefunction can be efficiently numerically represented with MPS in terms of tensors. The Hamiltonian of the system is represented as a Matrix Product operator (MPO). Its numerical efficiency depends on the range of interactions. In our system, we have ring structure, which in a naive implementation requires an interaction between the first and last site of the ring over an interaction length $L_\text{R}$. 
To shorten the interaction length, we reorder the site numbering of the ring part of the Hamiltonian, such that physical neighbors are connected via next-nearest neighbor interactions in the numerical implementation (see Fig.\ref{DMRGExpl}). 

Thus, only short range interactions with at most next-nearest neighbor interaction have to be evaluated in the MPO. A similar procedure is applied for the Y-junctions.
The ground state is calculated by the DMRG sweep method. The time evolution of the state $\ket{\Psi(t)}=\expU{i\mathcal{H}t}\ket{\Psi(0)}$ is calculated by  repeated application of an approximated time evolution operator $\expU{i\mathcal{H}\Delta t}$ with small time steps $\Delta t$ \cite{itensor}.

For the dGPE equation, we solve the equations of motions with a differential equation solver. We find the respective ground state by imaginary time evolution, and then evolve the system in real time.

%%%%%%%
\section{Open systems}
\label{opensystem}
The methods so far presented calculate the full system including the leads. However, the interesting part of our system is usually only a small part of the full system, e.g. for the ring-leads system, we are mainly interested in the dynamics of the ring only.
Thus, we propose an approximation: We trace out the leads, and simulate the dynamics of the interesting subsystem only (e.g. the ring or the junction of the Y-junction)\cite{breuer2002theory,benenti2009charge}. 
We model the bulk of the leads as Markovian baths, which are coupled to the relevant subsystem.  The bath-system coupling is assumed to be weak and within the Born-Markov approximation. The resulting Lindblad master equation is then
\begin{equation}\label{LindbladEq}
\pdif{\rho}{t}=-\frac{i}{\hbar}\left[H,\rho\right]-\frac{1}{2}\sum_m\left\{\cn{L}{m}\an{L}{m},\rho\right\}+\sum_m\an{L}{m}\rho\cn{L}{m}~,
\end{equation}
$\an{L}{m}$ the Lindblad operator which describes the action of the reservoir on subsystem, and $\left[\cdot,\cdot\right]$ ($\left\{\cdot,\cdot\right\}$) is the commutator (anti-commutator). 
The Hamiltonian $H$ now only includes the interesting part of our full system: In case of the ring system, we only keep the ring Hamiltonian and a single lead site of both source and drain, coupled to the ring with strength $K$. For the Y-junction, we keep two sites of each the source and drain leads. The effect of the rest of the leads is contained in the Lindblad operators.
We parameterize the reservoir-lead interaction with the following Lindblad operators
\begin{align*}
L_1={}&\sqrt{\Gamma }\cn{a}{S},\;\; L_2=\sqrt{r \Gamma}\an{a}{S},\;\; L_3=\sqrt{\Gamma }\an{a}{D}\;,
\end{align*}
%\begin{equation}
%f_\text{L}(\epsilon)=\left[1+e^{(\epsilon-\mu_\text{L})/(\kb T)}\right]^{-1}~,
%\end{equation}
with $\an{a}{S}$ ($\an{a}{D}$) the annihilation operator acting on the single site of the source (drain), $\Gamma$ the reservoir-lead coupling and $r$ controls the strength of tunneling back into the source reservoir. On the drain side, we assume that atoms can only tunnel into the drain bath and are unable to come back.

We follow the prescription of \cite{guo2017dissipatively} to reduce the dimension of the Lindblad superoperator $\mathcal{L}$. The size of the superoperator can be reduced when particle number conservation is broken only by the dissipation terms, and is conserved by the Hamiltonian. Then, certain terms of the density matrix are zero in the steady-state, and can be removed from the superoperator. The Master equation is solved with $\partial_t\rho=\mathcal{L}\rho$.

Such a Lindblad formalism describes strong driving, which creates a non-equilibrium steady-state with a high temperature. It is suitable for quenches of the system.

\section{Density wave with positive or negative sign}\label{PosNeg}
\begin{figure}[htbp]
	\centering
	\subfigimg[width=0.235\textwidth]{a}{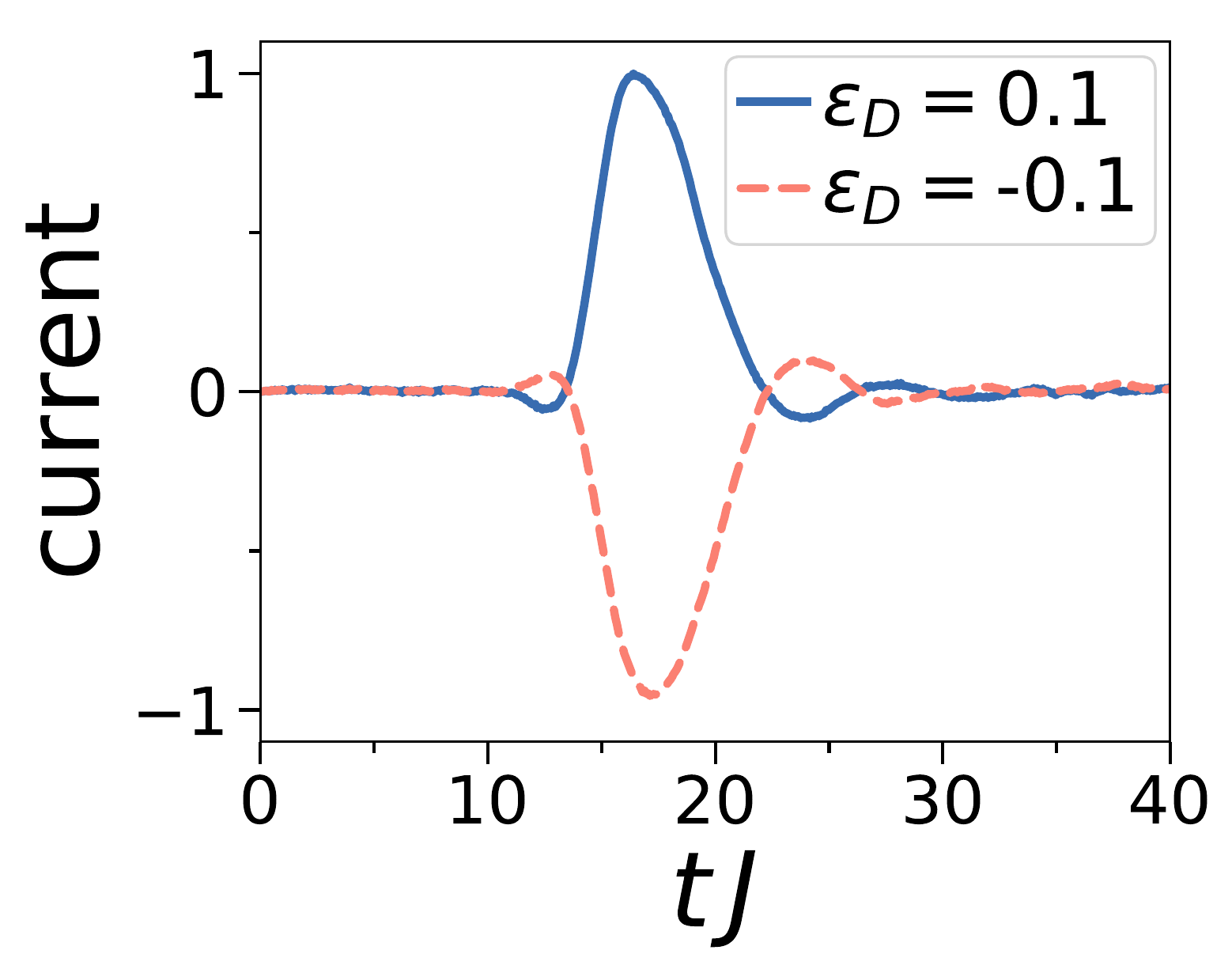}\hfill
	\subfigimg[width=0.245\textwidth]{b}{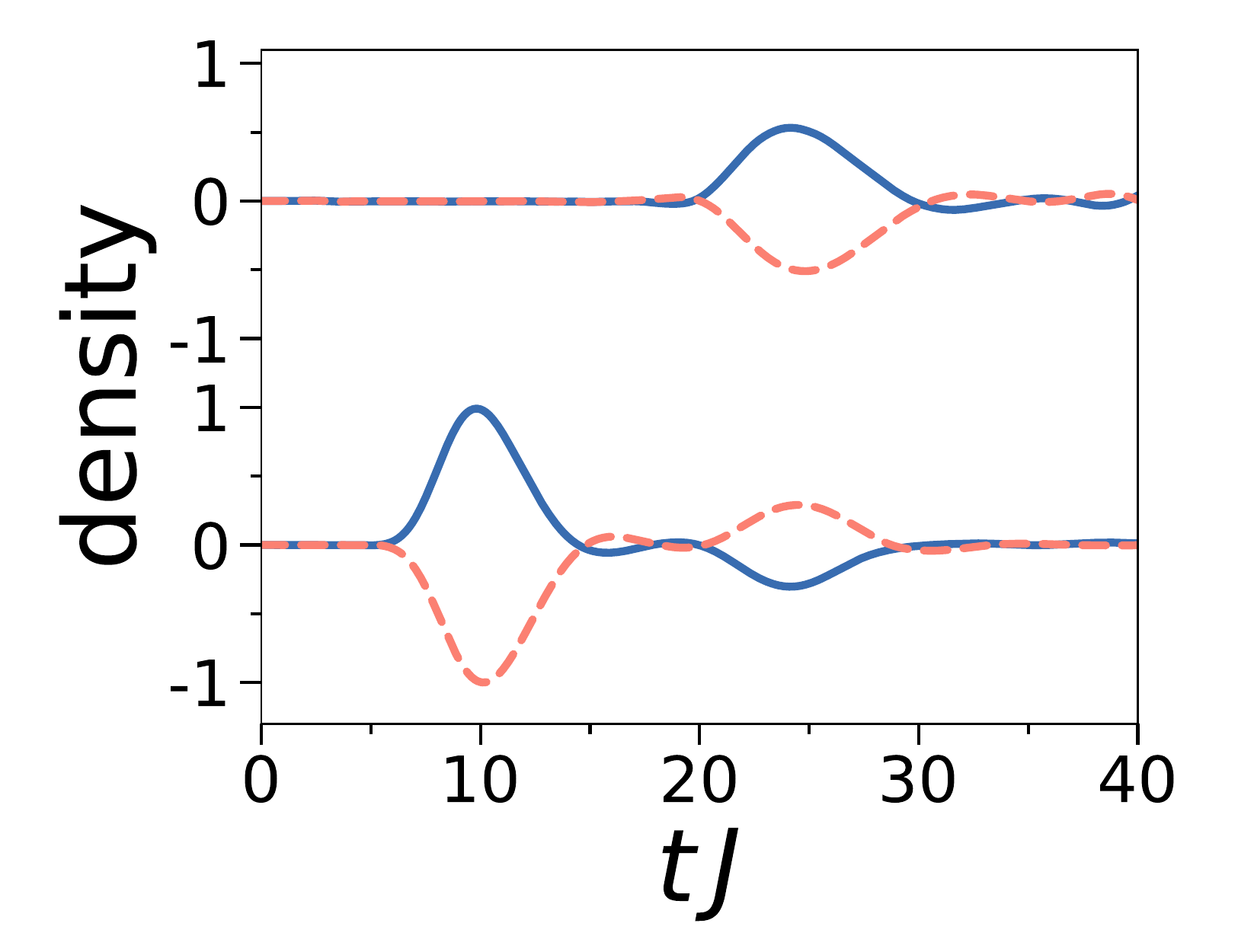}\\
	\subfigimg[width=0.235\textwidth]{c}{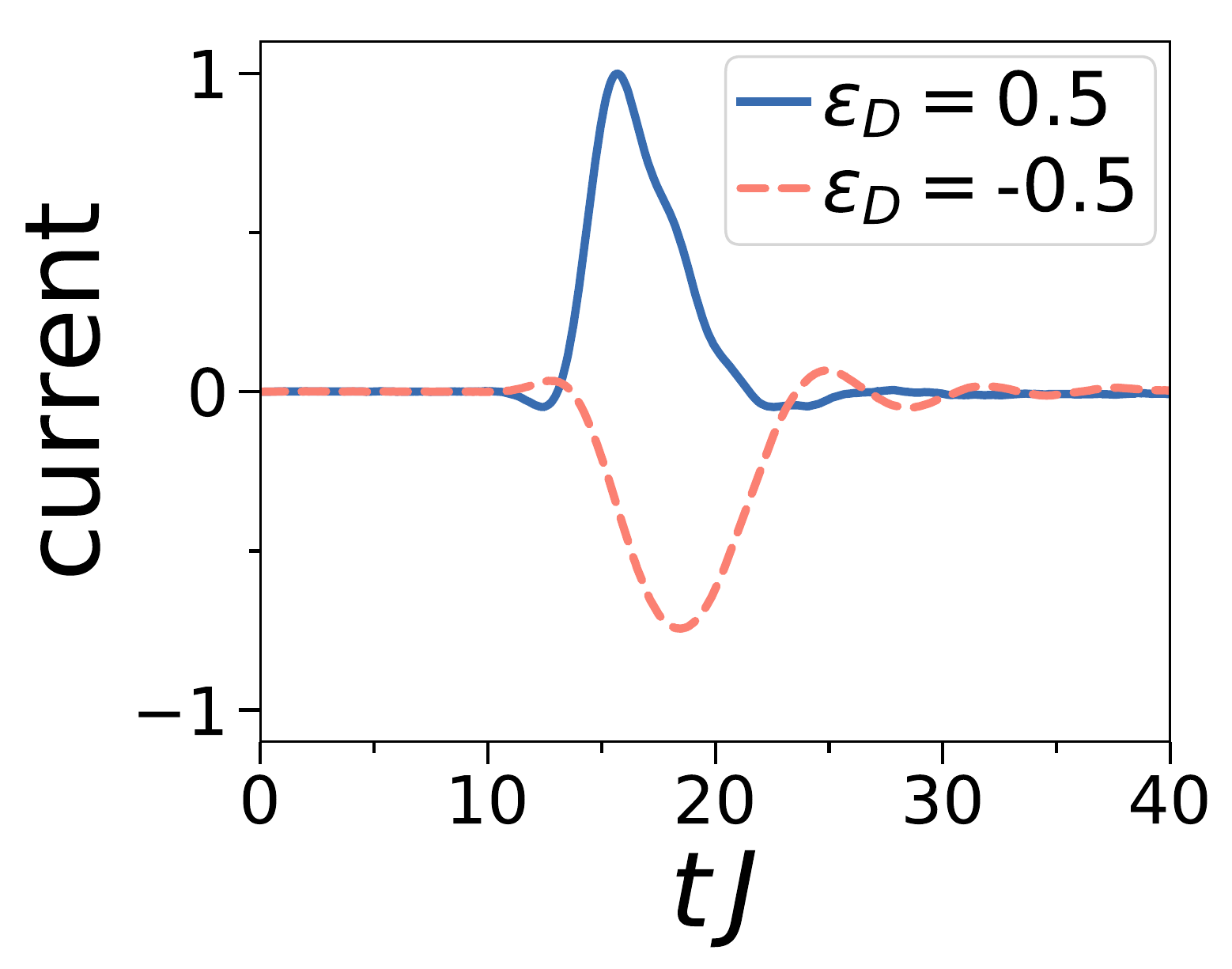}\hfill
	\subfigimg[width=0.245\textwidth]{d}{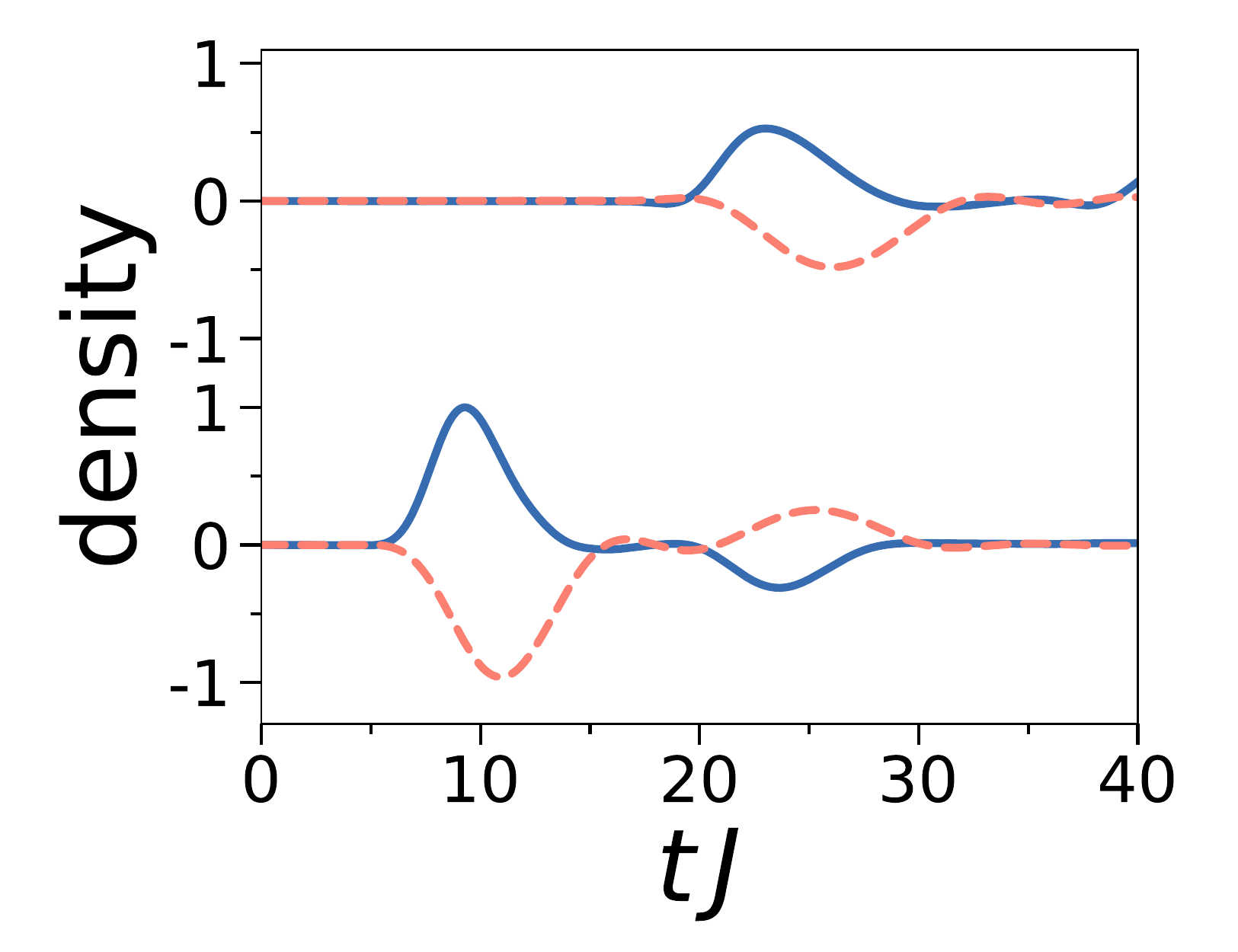}
	\caption{Propagation of excitations with different amplitude height and sign in a Y-junction for the Bose-Hubbard model for ${U=5}$ and ${K=1}$. The source lead has length ${L_\text{S}=80}$, the drain lead each ${L_\text{D}=30}$, the particle number ${N=70}$. The source lead is from site 1 to 80, the first drain lead from 80 to 110, and the second one from site 110 to 140. \idg{a,c} current at the junction \idg{b,d} \densdescr{80}{85}{65}{70} \rev{In all graphs, the solid blue curve describes positive $\epsilon_\text{D}$, and the orange dashed negative $\epsilon_\text{D}$.} In \idg{a,b} we choose a small density excitation amplitude by choosing a small initial potential offset ${\epsilon_\text{D}=\pm0.1}$. We see that the Andreev-like reflection has opposite sign to the initial density excitation. Both positive and negative waves have symmetric profile. 
		In \idg{c,d} we plot a larger ${\epsilon_\text{D}=\pm0.5}$. We see that profiles become asymmetric, and the negative density wave is slower compared to the positive one.
	}
	\label{YDensityHubbardsmalllarge}
\end{figure}

Next, we investigate the effect of sign and amplitude height of the density excitation using the finite $U$ Bose-Hubbard model. 
In Fig.\ref{YDensityHubbardsmalllarge}, we plot both the density and the current for small and large initial potential offsets $\epsilon_\text{D}$ with different signs. 
For small amplitudes, there is no significant difference in the propagation between positive and negative excitation amplitudes. The Andreev-like reflection has always opposite sign to the incoming wave (for an incoming negative wave, the Andreev reflection is positive). 
For larger amplitudes, we find that the speed of the wave depends on the amplitude. Positive amplitude excitations are faster than negative ones. We can explain this with the relation between speed of sound $c$ and density $n_0$ in a superfluid condensate ${\propto\sqrt{n_0}}$: the speed of sound increases with density and thus is larger for higher density amplitude. There are also small differences in the temporal shape of the transmitted and reflected waves. However, the overall dynamics and transmission/reflection properties remain independent of sign and amplitude.

\section{Ring-lead dynamics with two barriers for Bose-Hubbard model}\label{ImpurityBH}
\begin{figure}[htbp]
	\centering
	\subfigimg[width=0.26\textwidth]{}{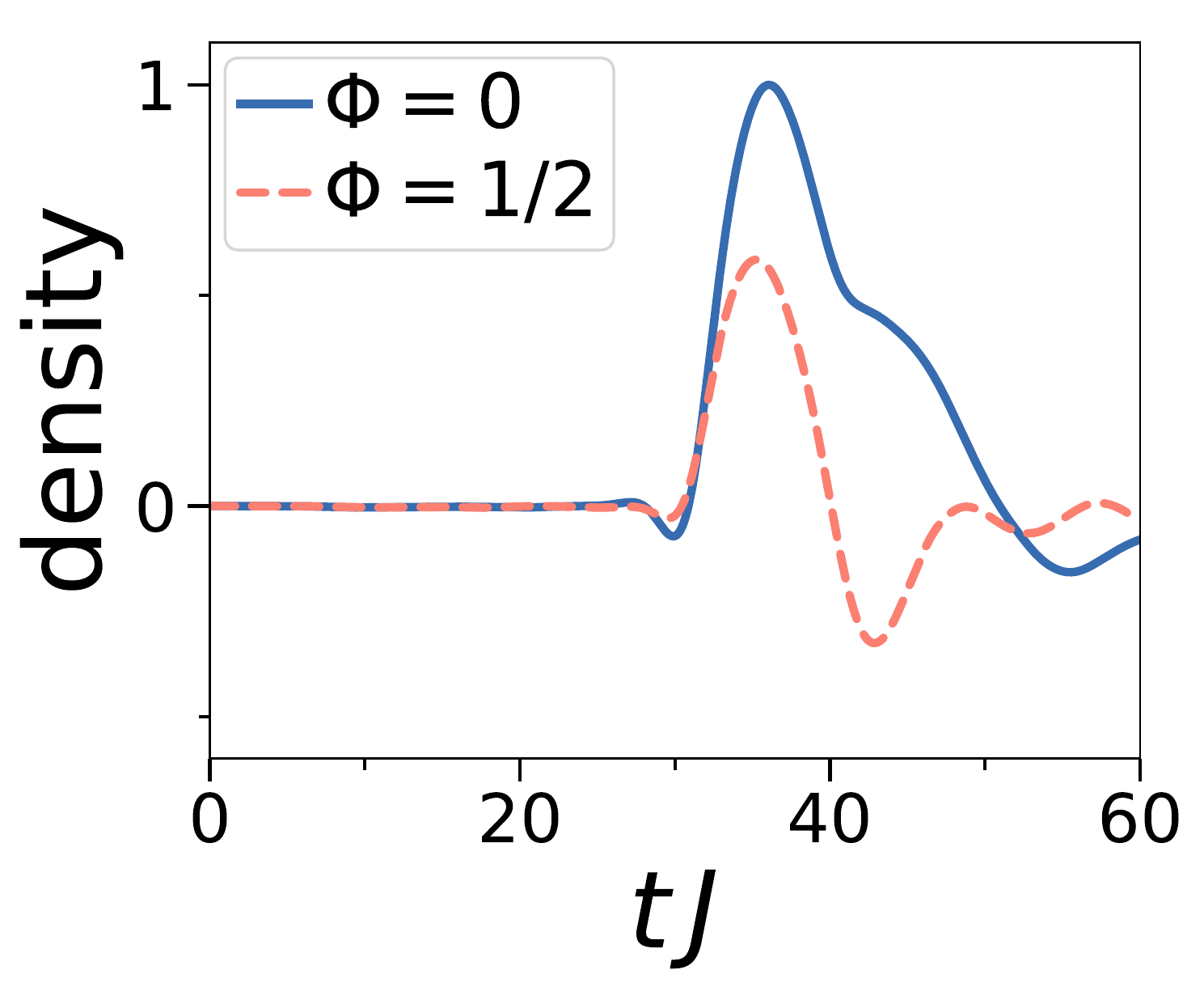}
	\caption{Propagation of small excitation (${\epsilon_\text{D}=0.5}$) in the drain lead in a ring-lead system for Bose-Hubbard model (${U=5}$) with two impurities in the ring with height ${\Delta=3}$. Other parameters same as in Fig.\ref{TwoImpRingNew}. The transmitted density wave changes with $\Phi$. For ${\Phi=0}$ a positive density enters the drain, while for ${\Phi=1/2}$, it is a density oscillation with nearly equal positive and negative amplitude, and nearly no net density enters the drain. The total transmission at time ${tJ=60}$ is 0.855 (${\Phi=0}$) and 0.119 (${\Phi=1/2}$).}
	\label{TwoImpRingBH}
\end{figure}

In Fig.\ref{TwoImpRingBH}, we check the effect of potential barriers for the finite $U$ Bose-Hubbard model and small density excitations. We find that the transmitted wave is dependent on flux for high enough barrier $\Delta$, showing the same behavior as for hard-core bosons for small excitation amplitudes. Due to the renormalization of potential barriers with on-site interaction $U$, higher values of ${\Delta=3}$ are required compared to the hard-core model to observe the effect.

\section{Spinless fermions}
\begin{figure*}[htbp]
	\centering
	%\subfigure{\includegraphics[width=0.24\textwidth]{{chiralcurrentFluxN1DTwistDataN10000}.pdf}}\hfill
	%\subfigure{\includegraphics[width=0.24\textwidth]{{chiralcurrentTwistN1DTwistN10000}.pdf}}
	\subfigimg[width=0.24\textwidth]{a}{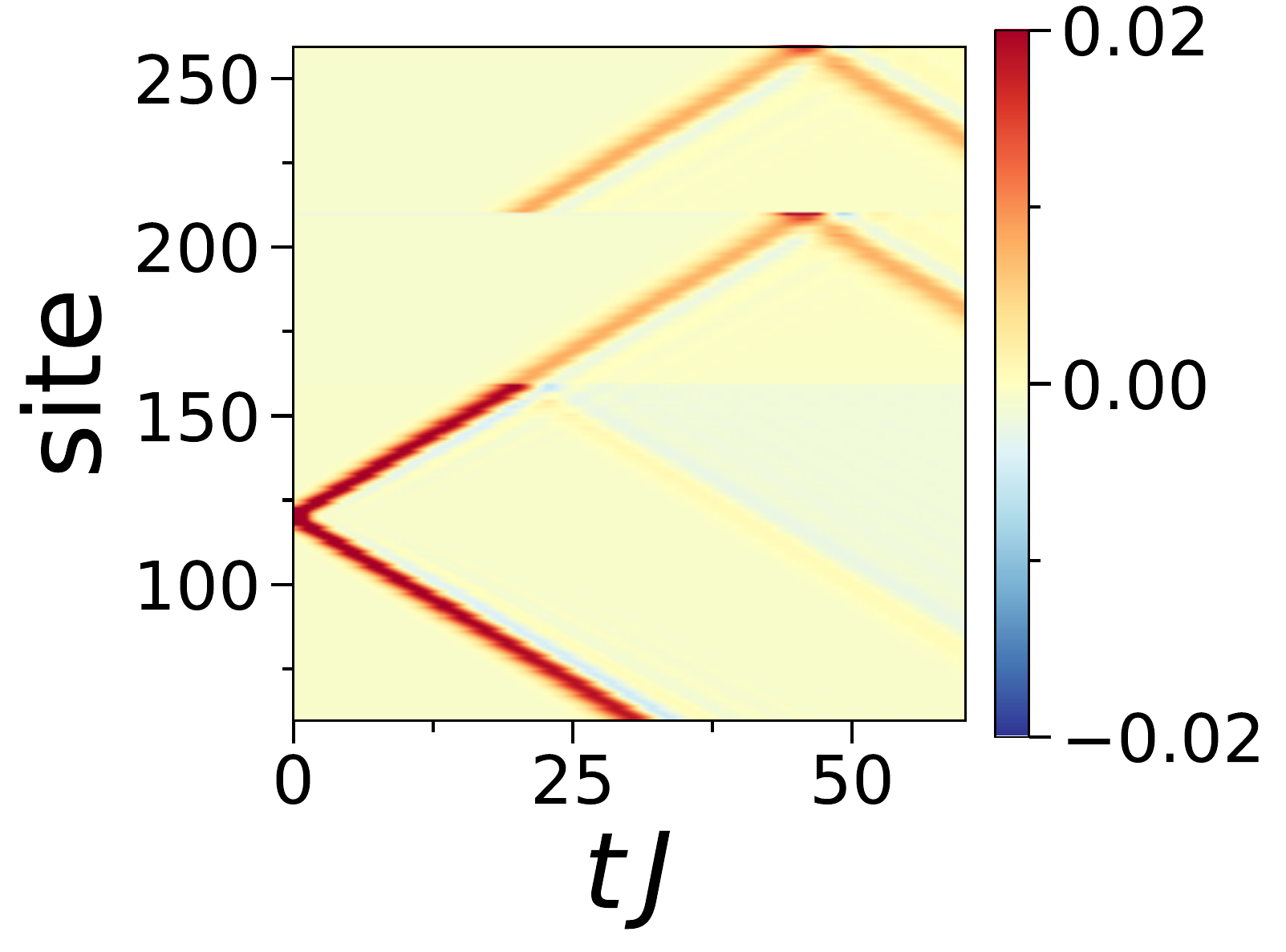}
	\subfigimg[width=0.24\textwidth]{b}{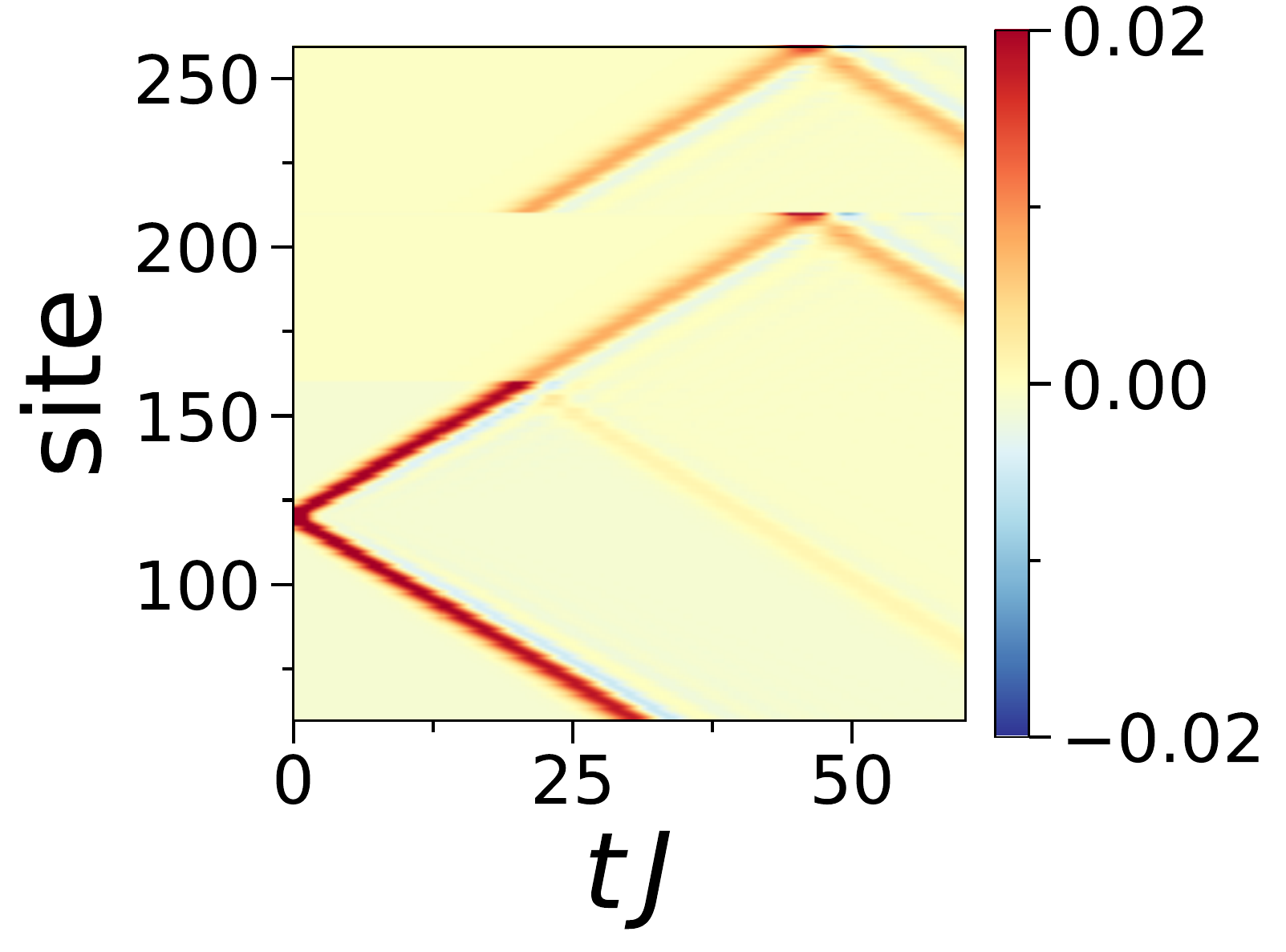}
	\subfigimg[width=0.24\textwidth]{c}{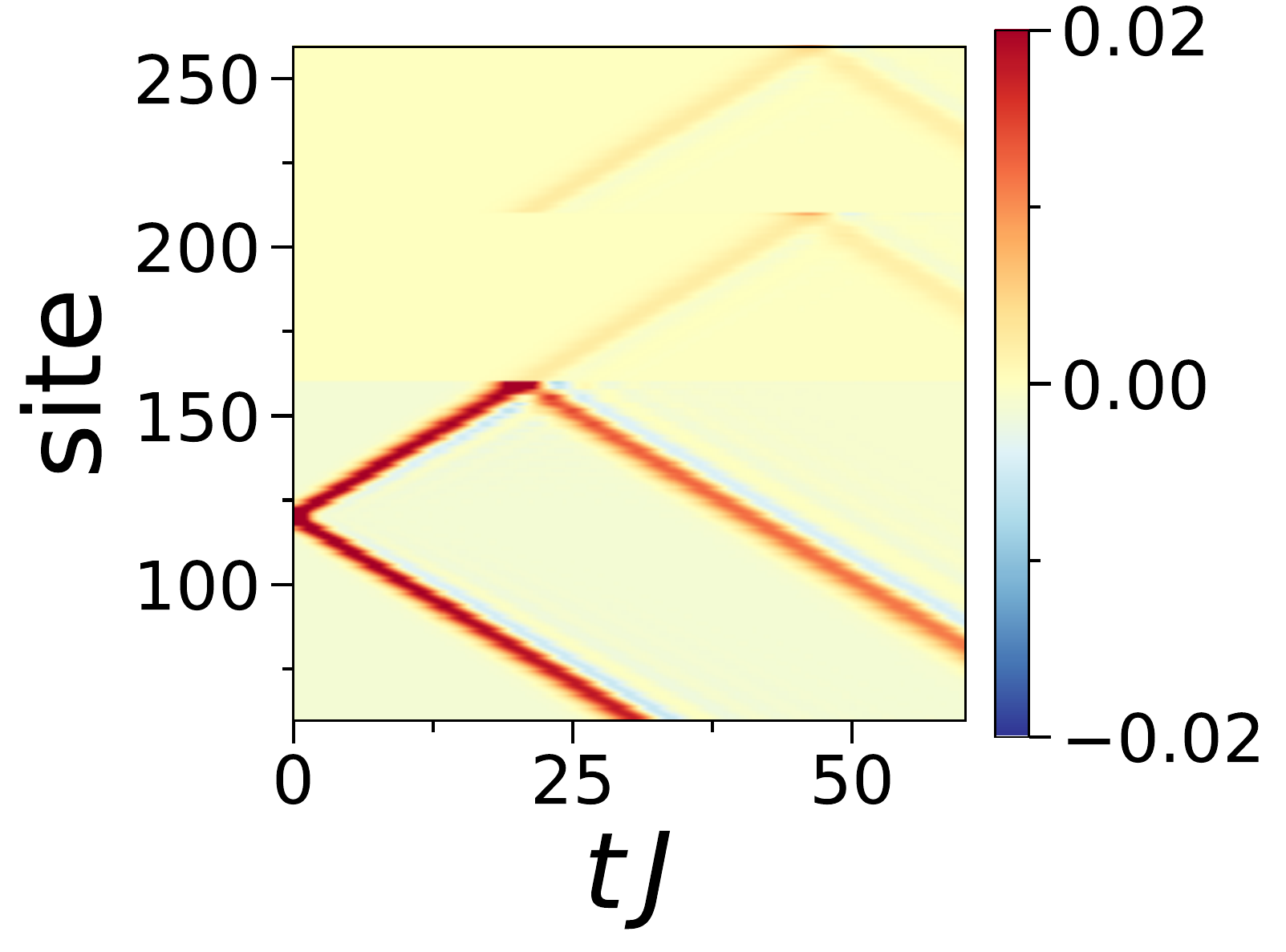}\\
	\subfigimg[width=0.24\textwidth]{d}{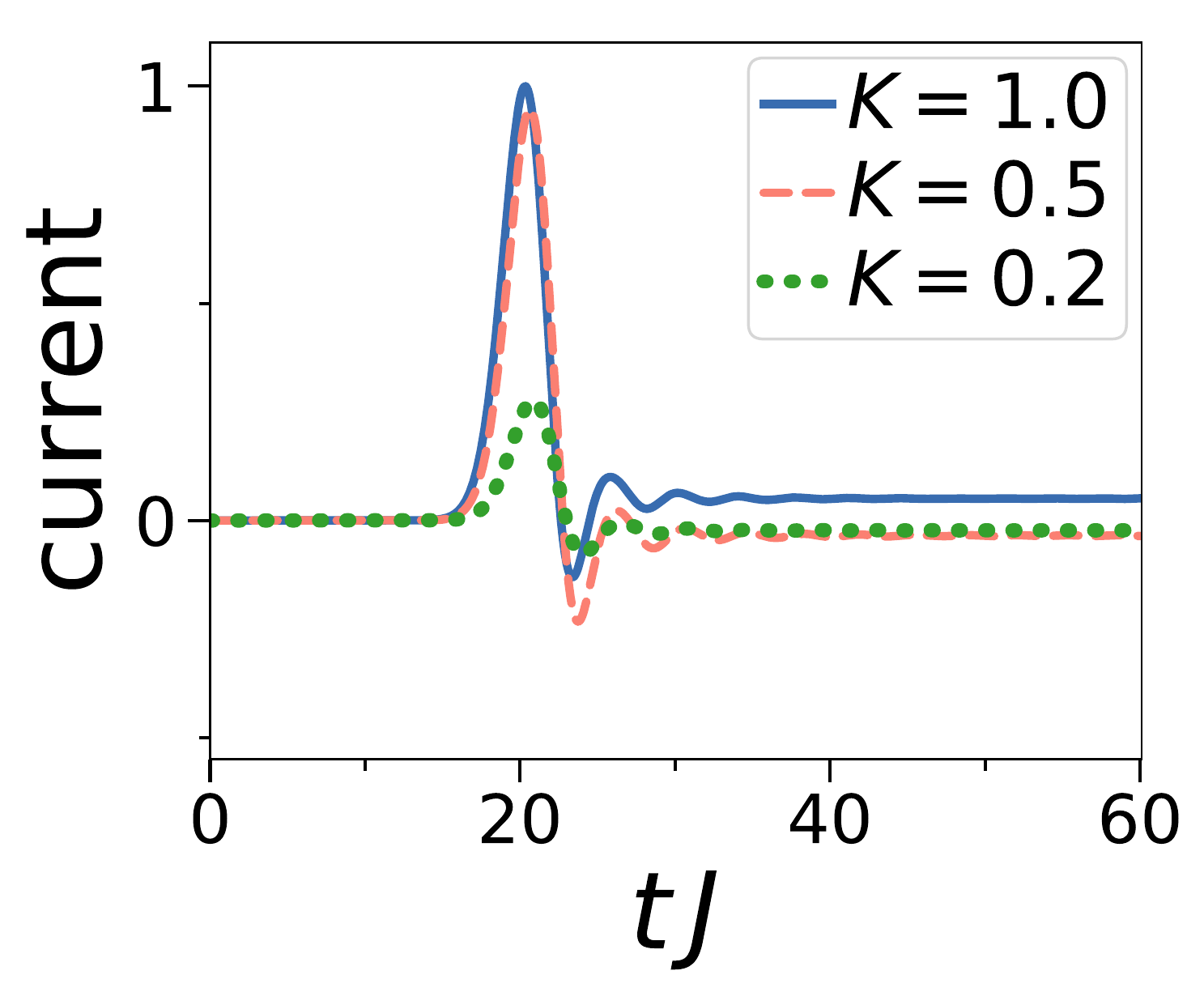}
	\subfigimg[width=0.24\textwidth]{e}{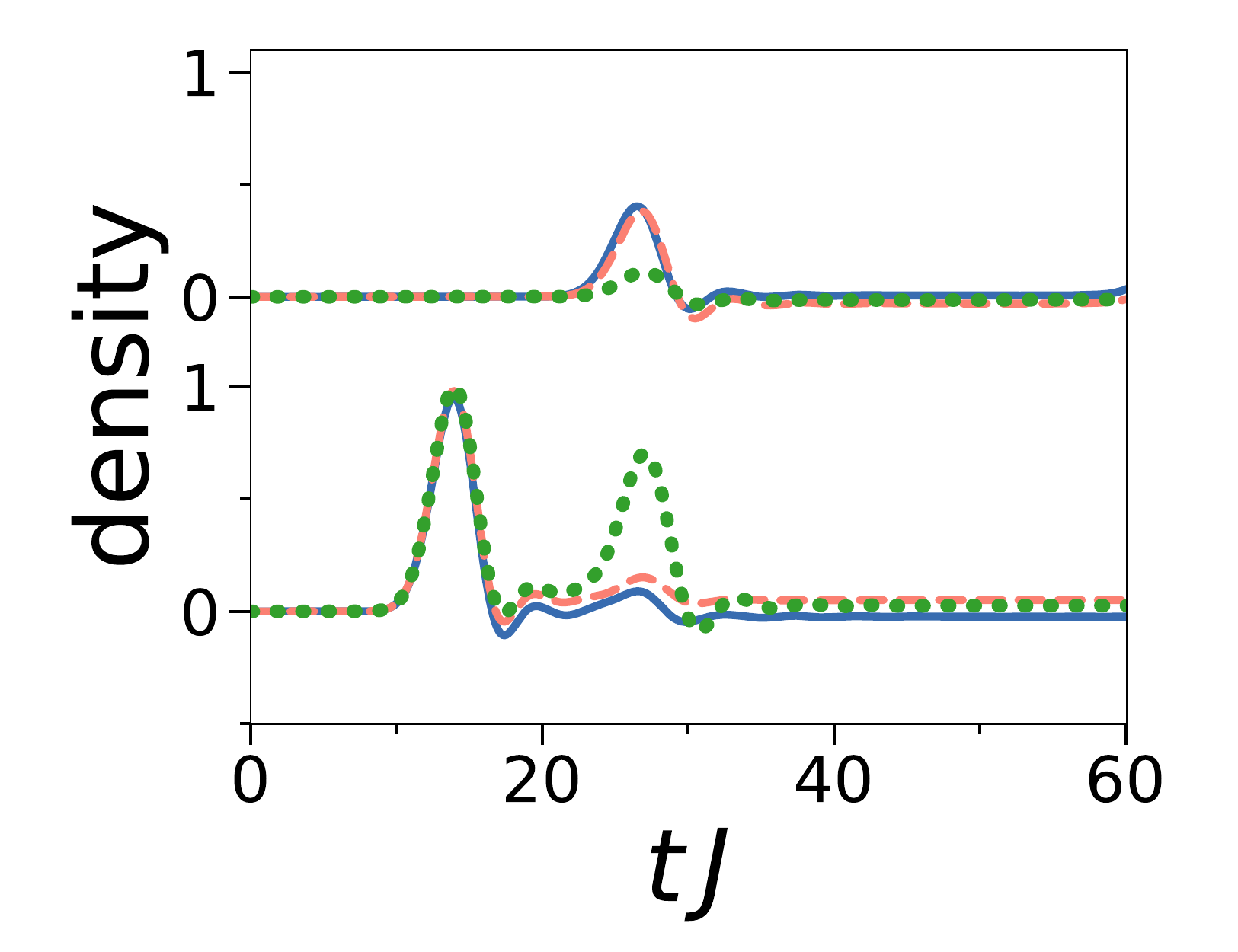}
	\caption{Propagation of small excitation in a Y-junction for the spinless fermion model. The source lead has length ${L_\text{S}=160}$, the drain lead each ${L_\text{D}=50}$, the particle number ${N=130}$, ${d=40}$ and ${\epsilon_\text{D}=0.3}$. The source lead is from site 1 to 160, the first drain lead from 160 to 210, and the second one from site 210 to 260. The coupling at the junction (site 160) is \idg{a} ${K=1}$, \idg{b} ${K=0.5}$, \idg{c} ${K=0.2}$. \idg{d} current at the junction \idg{e} \densdescr{170}{175}{145}{150} For ${K=1}$ (solid) we observe a wave with positive and negative amplitude, ${K=0.5}$ (dashed) nearly no reflection amplitude, ${K=0.2}$ (dots) a large positive reflection amplitude.
	The table below shows the transmission and reflection coefficients, calculated at ${t=31/J}$ with Eq.\ref{EqInc}-\ref{EqRef} (${t_\text{in}=15}$, ${a=30}$). \rev{There is a small, non-zero negative background current for ${K=1}$. This is a side-effect of our procedure to generate the density excitation, and is much more pronounced for fermions than for hard-core boson (further details in the main text). This current causes the small negative reflection coefficient. We avoid this background current as much as possible by evaluating the transmission and reflection coefficient at early times (${t=31/J}$), right after the source-drain current has rung down. }%\rev{The transmission for ${K=1}$ is greater one and we see a negative total transmission. However, the reflection density clearly lacks the negative Andreev-reflection wavepacket seen for bosons and thus is not associated with Andreev-reflection. Instead, the negative reflection is caused by a small, but long-lived current into the drain. This current is caused by the initial atom imbalance: We prepare an initial wavepacket in the source, and thus have slightly more atoms initially in the source than in the drain. The negative current we see is the result of atom equilibration between source and drain. It can be eliminated by a small additional constant offset potential in the source. }
	}
	\begin{tabular}{l|*{3}{c}}
	& ${K=1}$ & ${K=0.5}$ & ${K=0.2}$ \\
	\hline
	%transmission & $1.201$ & $0.673$ & $0.135$   \\
	%reflection   & $-0.201$ & $0.327$ & $0.865$  
	%transmission & $1.042$ & $0.777$ & $0.205$   \\
	%reflection   & $-0.042$ & $0.223$ & $0.795$ 
	transmission & $1.061$ & $0.772$ & $0.199$   \\
	reflection   & $-0.061$ & $0.228$ & $0.801$  
	\end{tabular}
	\label{YDensitysmallFermion}
\end{figure*}

\begin{figure}[htbp]
	\centering
	%\subfigure{\includegraphics[width=0.24\textwidth]{{chiralcurrentFluxN1DTwistDataN10000}.pdf}}\hfill
	%\subfigure{\includegraphics[width=0.24\textwidth]{{chiralcurrentTwistN1DTwistN10000}.pdf}}
	\subfigimg[width=0.24\textwidth]{a}{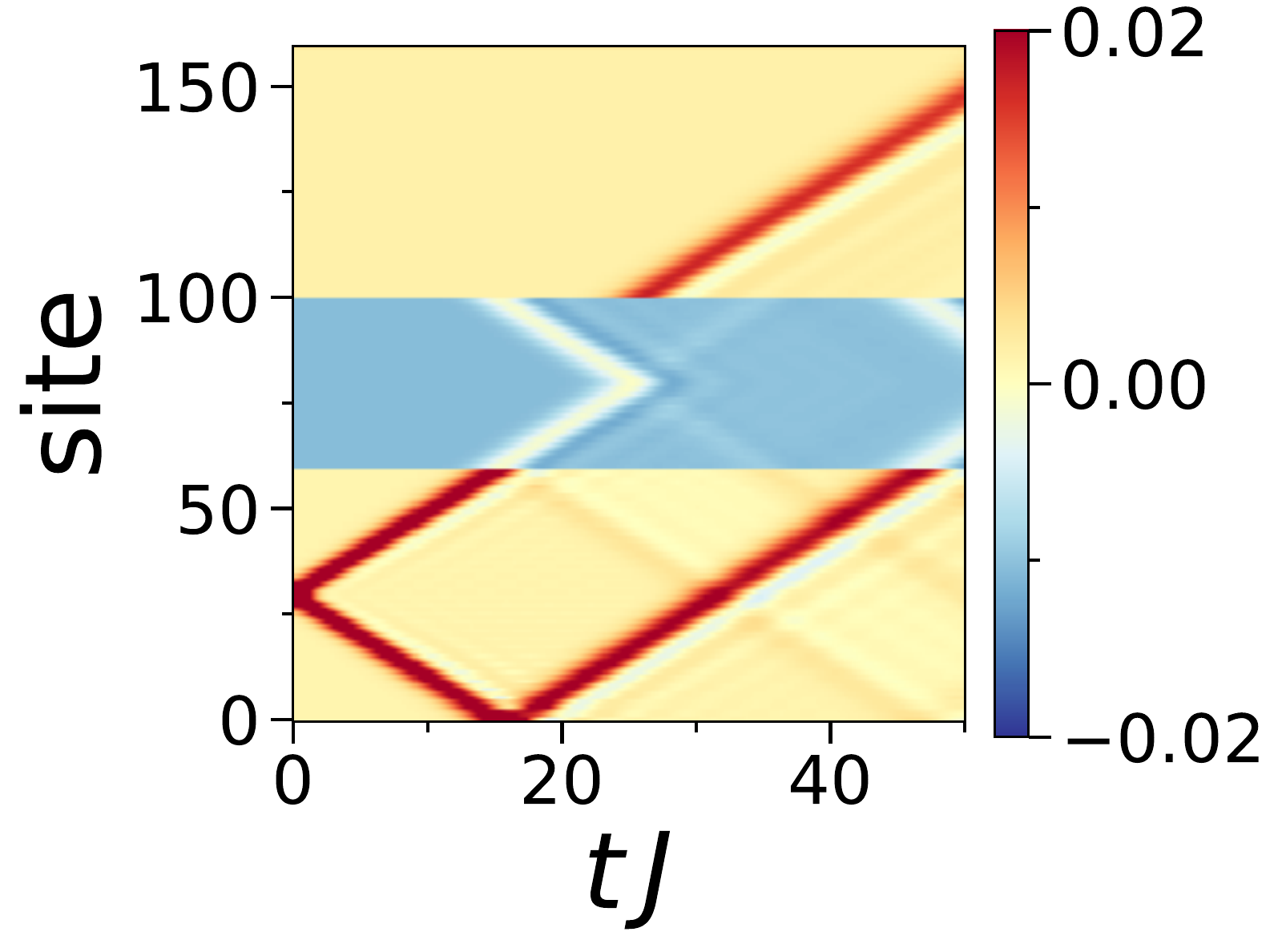}\hfill
	\subfigimg[width=0.24\textwidth]{b}{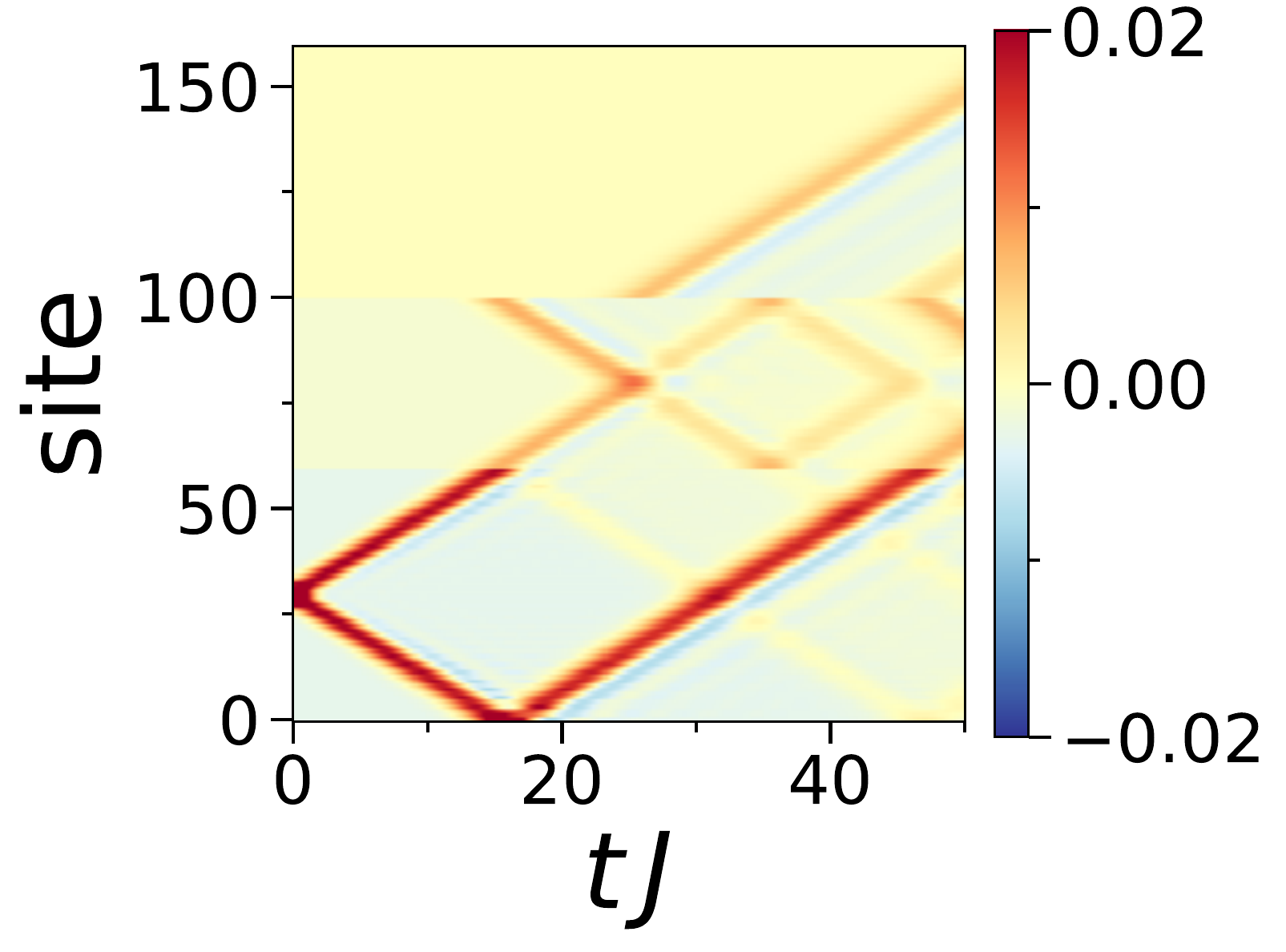}
	\subfigimg[width=0.24\textwidth]{c}{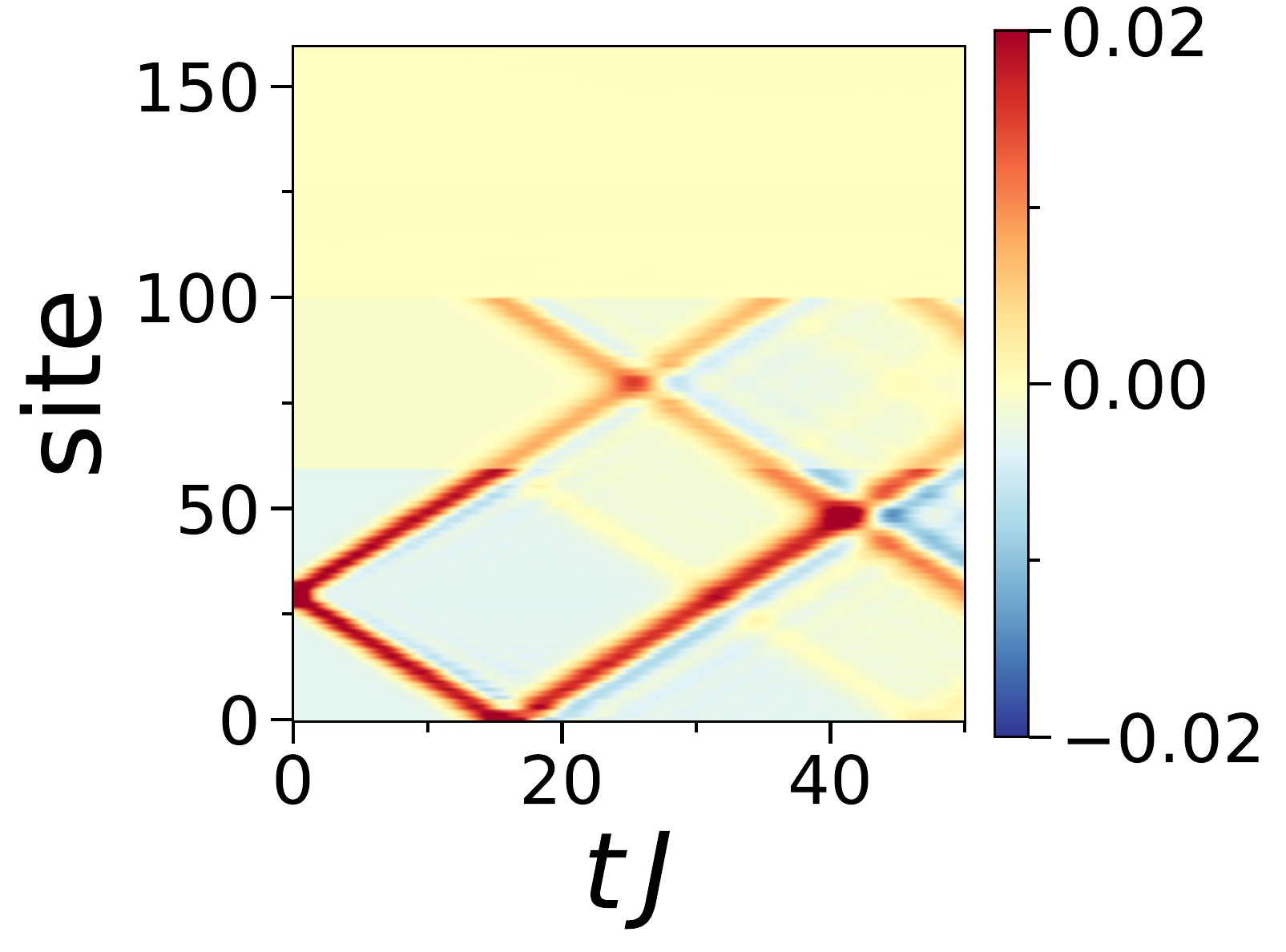}\hfill
	\subfigimg[width=0.24\textwidth]{d}{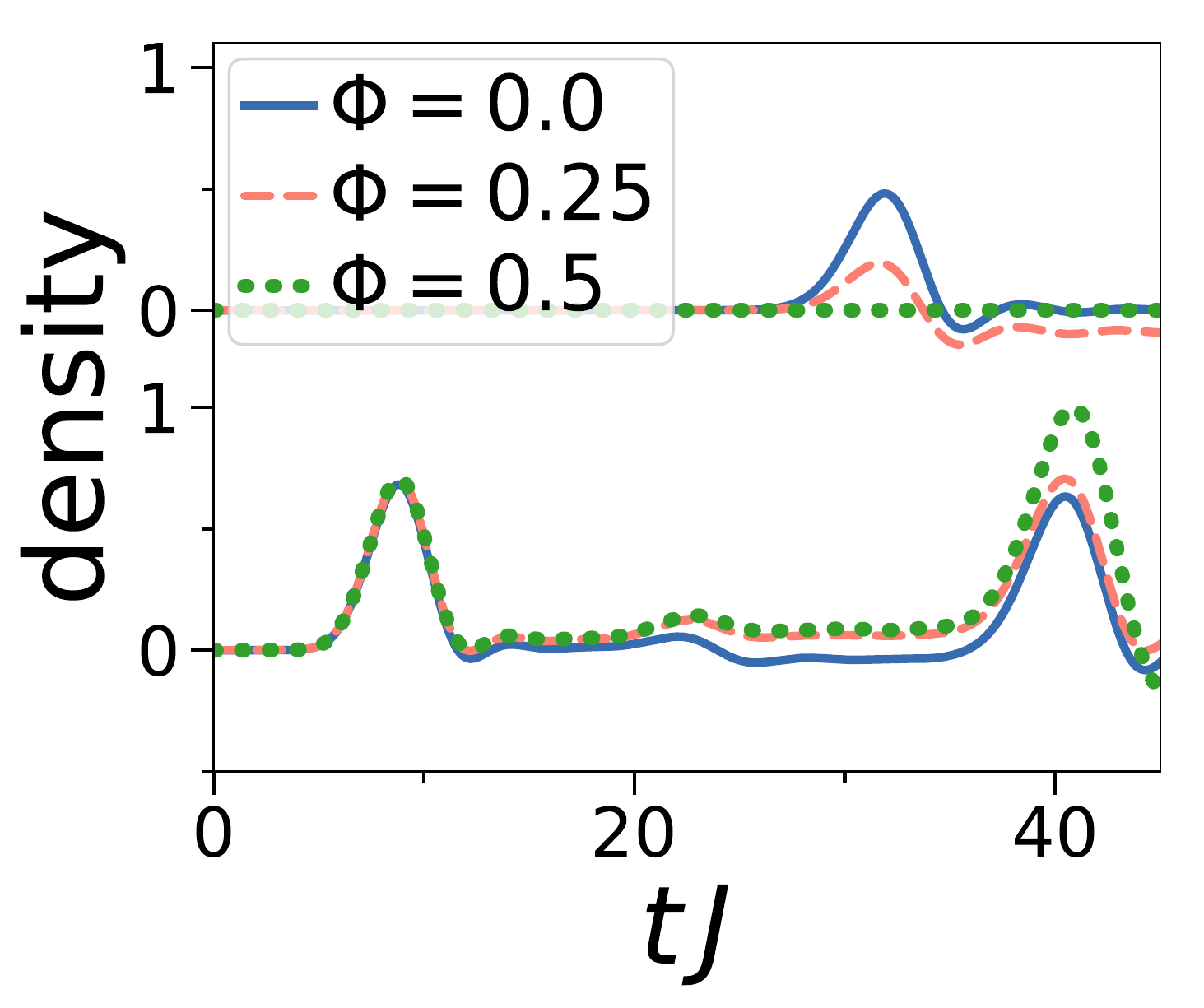}
	\caption{Propagation of small excitation in a ring-lead system for spinless fermions. The source and drain lead has length ${L_\text{S}=L_\text{D}=60}$, the ring ${L_\text{R}=40}$, the particle number ${N=80}$ and initial excitation ${\epsilon_\text{D}=0.3}$. The flux in the ring is \idg{a} ${\Phi=0}$, \idg{b} ${\Phi=0.25}$, \idg{c} ${\Phi=0.5}$. \idg{d} \densdescr{110}{115}{45}{50}  }
	\label{RingDensitysmallFermion}
\end{figure}

\begin{figure}[htbp]
	\centering
	%\subfigure{\includegraphics[width=0.24\textwidth]{{chiralcurrentFluxN1DTwistDataN10000}.pdf}}\hfill
	%\subfigure{\includegraphics[width=0.24\textwidth]{{chiralcurrentTwistN1DTwistN10000}.pdf}}
	\subfigimg[width=0.24\textwidth]{a}{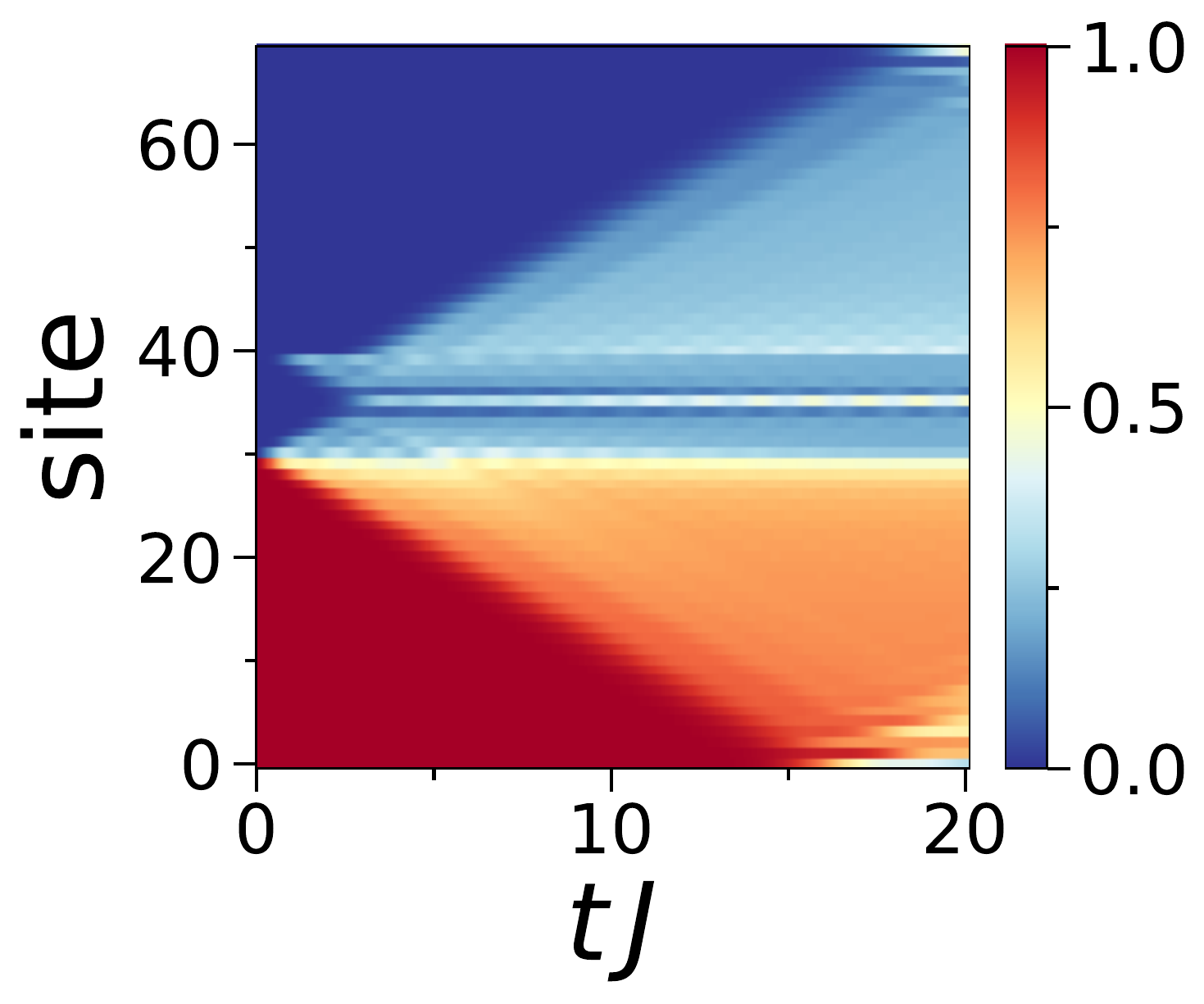}\hfill
	\subfigimg[width=0.24\textwidth]{b}{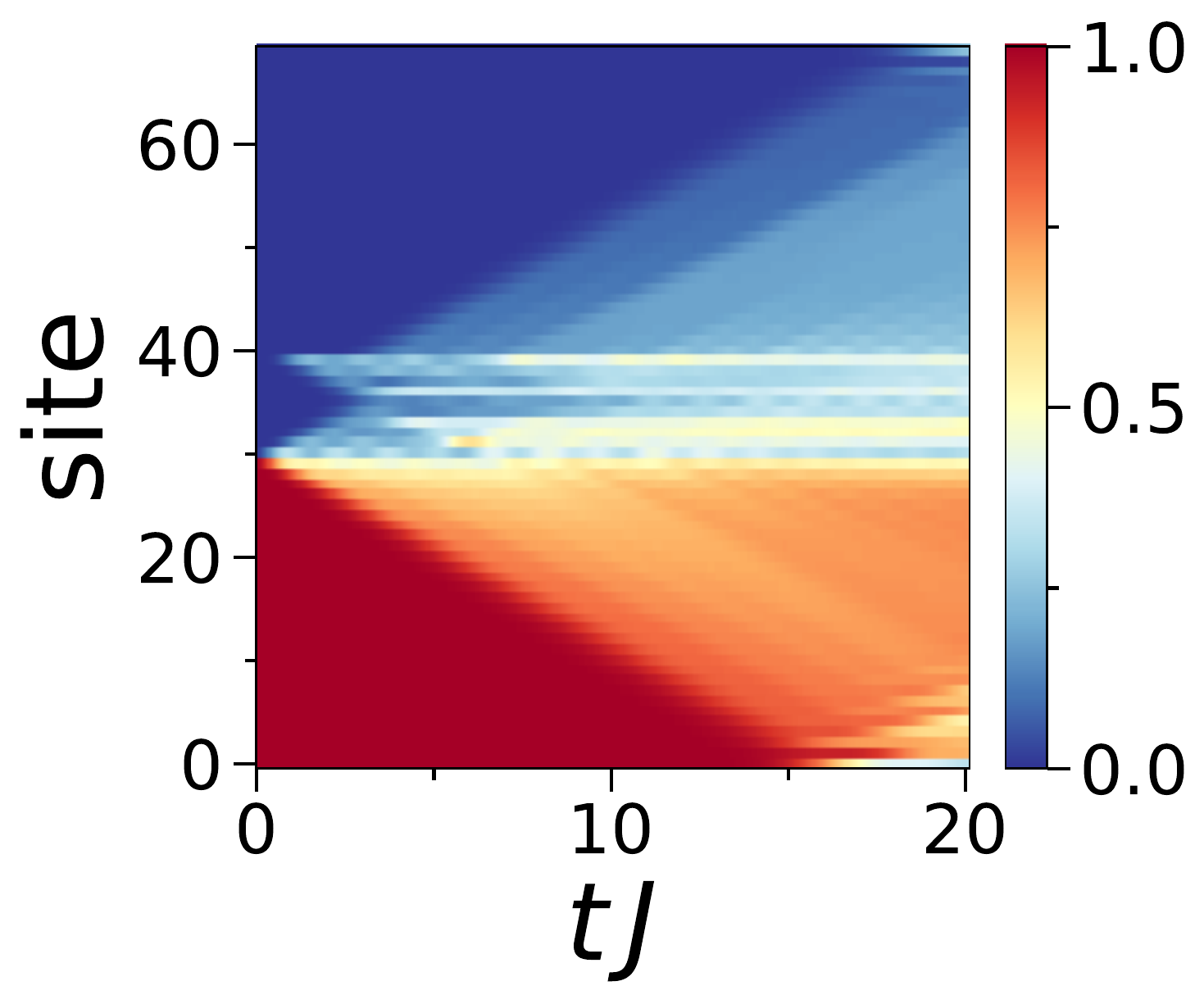}
	\subfigimg[width=0.24\textwidth]{c}{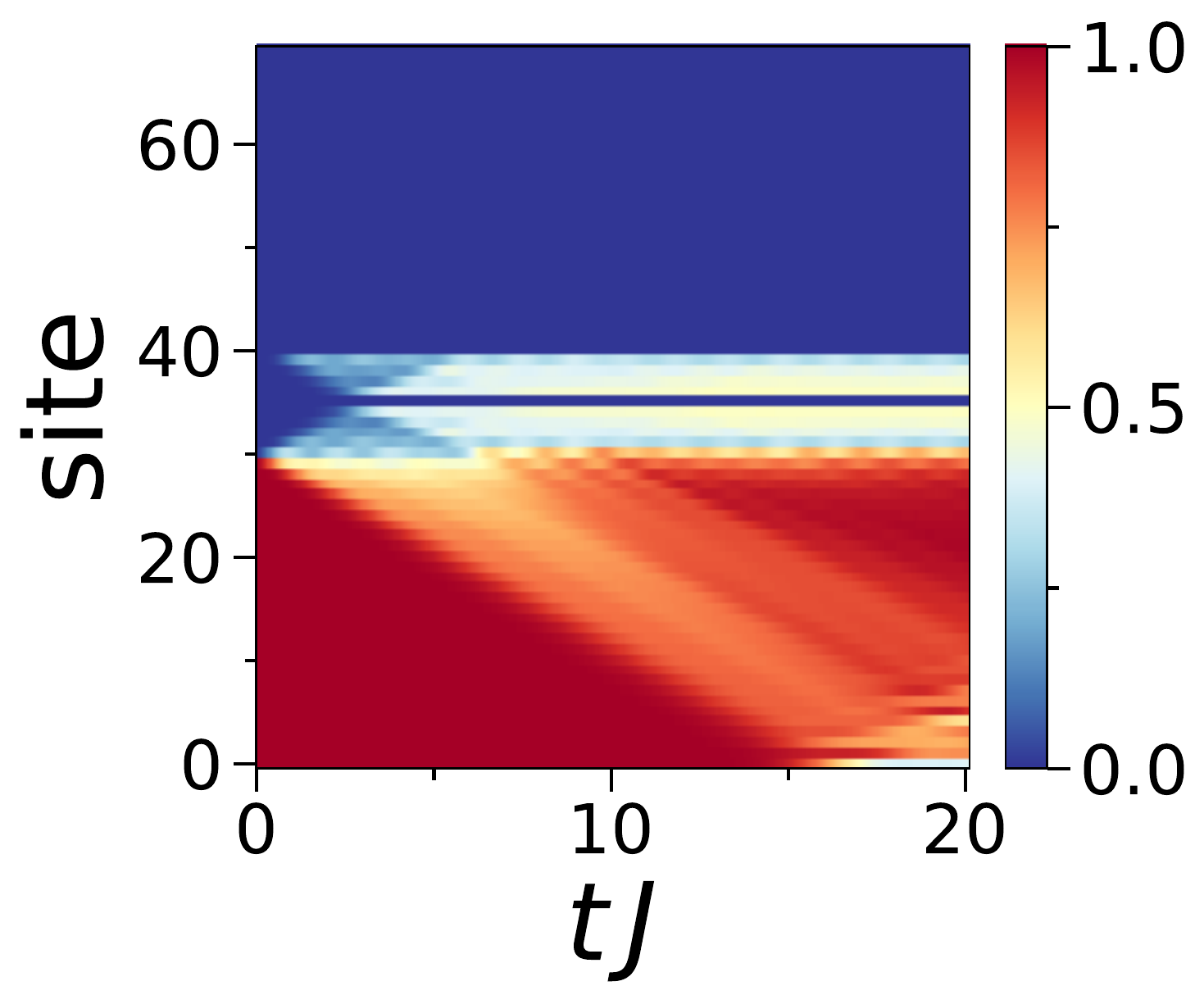}\hfill
	\subfigimg[width=0.24\textwidth]{d}{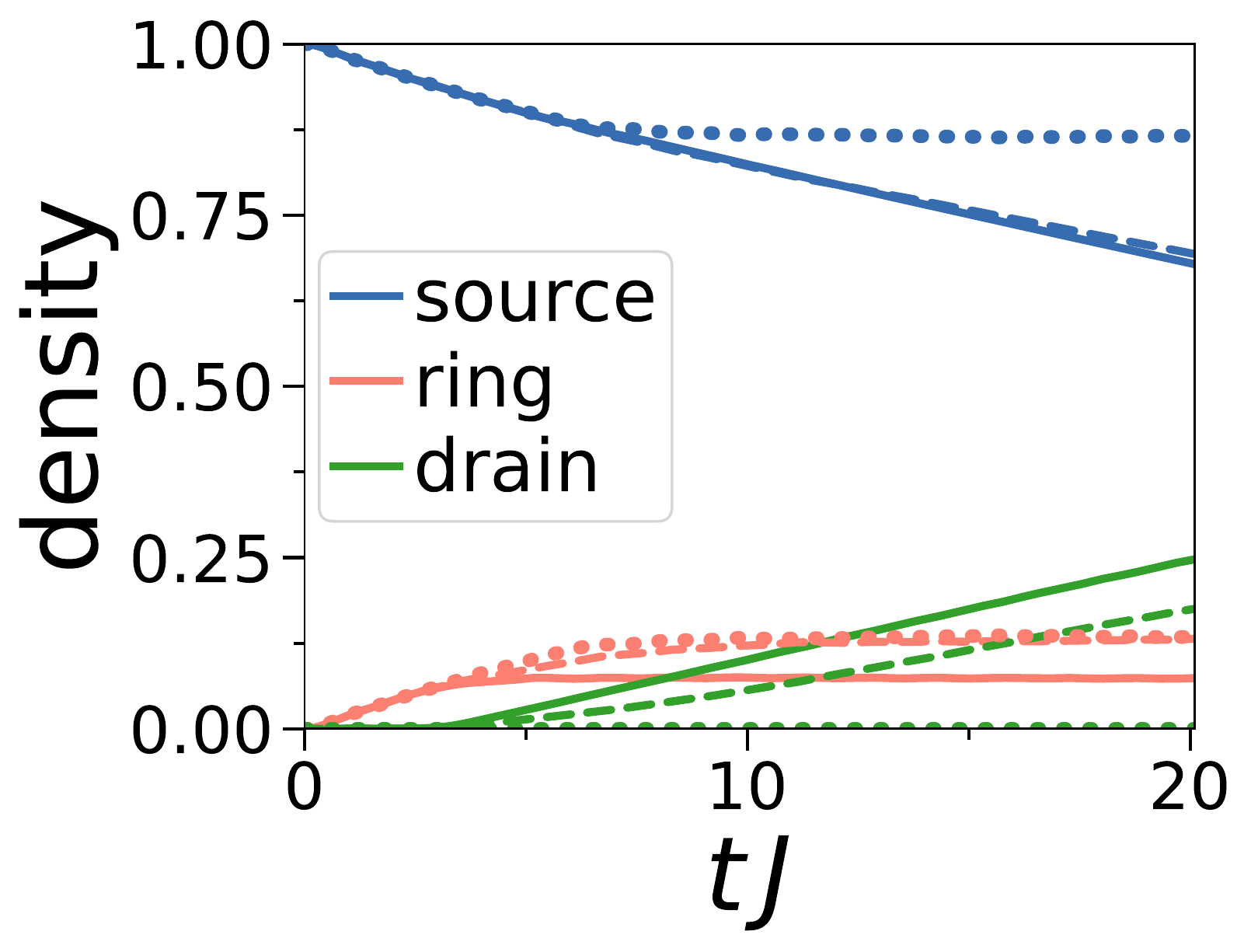}\\
	\subfigimg[width=0.3\textwidth]{e}{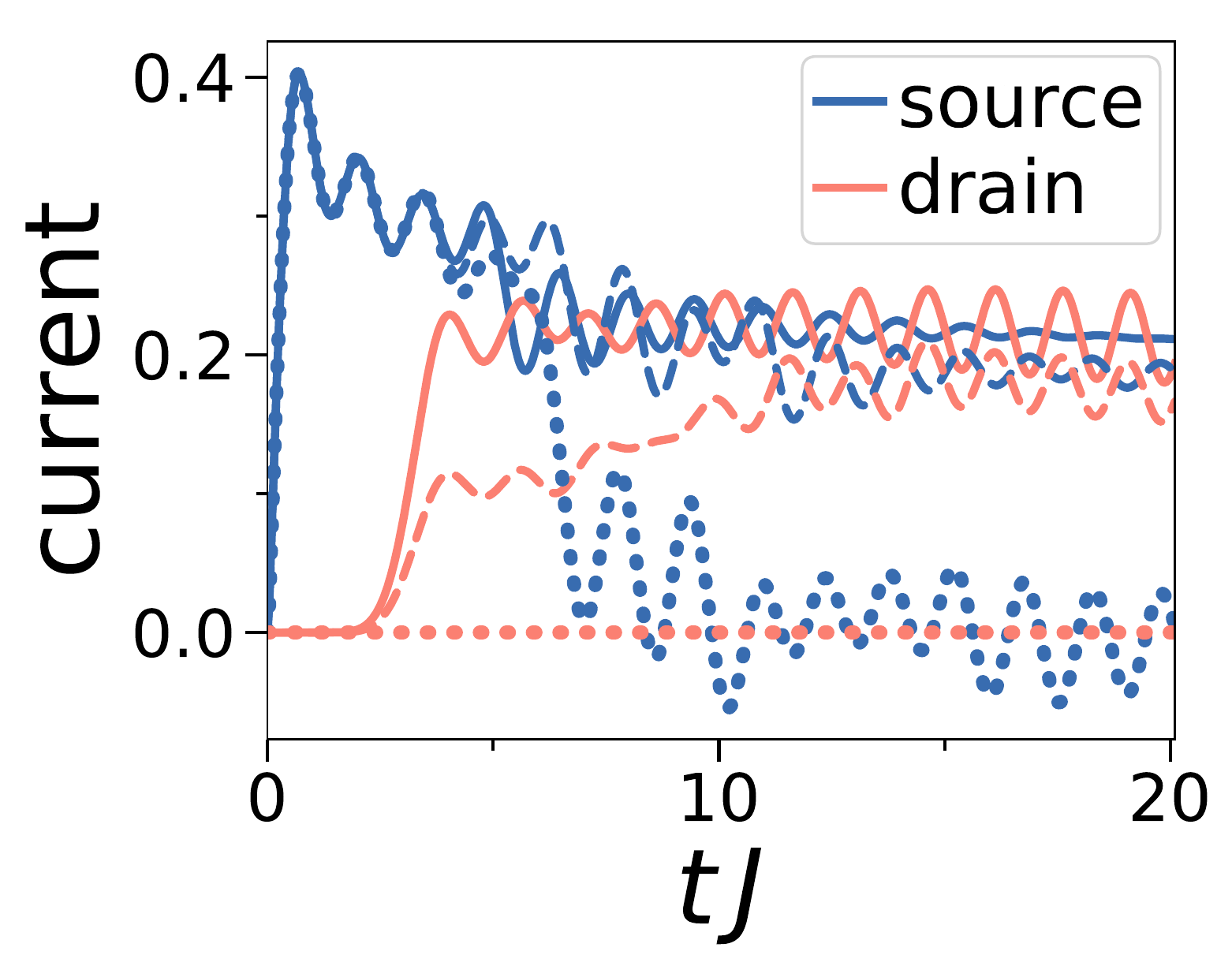}
	\caption{Time evolution of atoms initially loaded into the leads of a ring-lead system for spinless fermions. The source and drain lead has length ${L_\text{S}=L_\text{D}=30}$ and the ring ${L_\text{R}=10}$, the particle number ${N=15}$. The flux in the ring is \idg{a} ${\Phi=0}$, \idg{b} ${\Phi=0.25}$, \idg{c} ${\Phi=0.5}$. \idg{d} The total density in ring, source and drain leads. ${\Phi=0}$ (solid)  ${\Phi=0.25}$ (dashed), ${\Phi=0.5}$ (dots). \idg{e} Current through source and drain junction.}
	\label{RingDensityfullFermion}
\end{figure}

In this section we provide results for the case of non-interacting spinless fermions as a reference.
In Fig.\ref{YDensitysmallFermion} we show the propagation of an excitation in a Y-junction. 
In Fig.\ref{RingDensitysmallFermion} we show the propagation of an excitation in a ring. Here, we find the regular Aharonov-Bohm effect, with destructive interference and no transmission at half-flux.
In Fig.\ref{RingDensityfullFermion} we show the quench of non-interacting fermions which are initially prepared in the source lead. The fermions expand through the ring to the drain lead. The dynamics heavily depends on flux. For half-flux the fermions cannot reach the drain due to destructive interference.
\end{document}